\documentclass[12pt,a4paper]{article}

\usepackage{latexsym,amsfonts,amsmath,amssymb,graphics,bbm}
\usepackage{pifont}    
\usepackage{dsfont}
\usepackage{nccmath}   
\usepackage[dvips]{graphicx}

\usepackage{mathrsfs}  
\usepackage{bbm} 
\usepackage{slashed}   
\DeclareMathAlphabet{\mathpzc}{OT1}{pzc}{m}{it}

\newcommand{\vphi}{{\pmb{\phi}}}
\newcommand{\vom}{{\pmb{\om}}}
\newcommand{\vpsi}{{\pmb{\psi}}}

\newcommand{\mpo}{m^2_{V(\la)}}
\newcommand{\proj}{\bP(\la)}

\newcommand{\rmt}{{\scriptscriptstyle\top}} 

\newcommand{\kA}[0]{}

\newcommand{\kD}[0]{}
\newcommand{\kE}[0]{}

\newcommand{\kG}[0]{}
\newcommand{\kH}[0]{}

\newcommand{\kK}[0]{}
\newcommand{\kL}[0]{}
\newcommand{\kM}[0]{}
\newcommand{\kN}[0]{}
\newcommand{\kO}[0]{}
\newcommand{\kP}[0]{}
\newcommand{\kQ}[0]{}
\newcommand{\kR}[0]{}
\newcommand{\kS}[0]{}
\newcommand{\kT}[0]{}
\newcommand{\kU}[0]{}
\newcommand{\kV}[0]{}
\newcommand{\kW}[0]{}
\newcommand{\kX}[0]{}
\newcommand{\kY}[0]{}
\newcommand{\kZ}[0]{}

\newcommand{\kBB}[0]{}
\newcommand{\kCC}[0]{}
\newcommand{\kDD}[0]{}

\newcommand{\kFF}[0]{}
\newcommand{\kGG}[0]{}
\newcommand{\kHH}[0]{}

\newcommand{\kJJ}[0]{}
\newcommand{\kKK}[0]{}
\newcommand{\kLL}[0]{}

\newcommand{\kNN}[0]{}
\newcommand{\kOO}[0]{}
\newcommand{\kPP}[0]{}
\newcommand{\kQQ}[0]{}

\newcommand{\kSS}[0]{}
\newcommand{\kTT}[0]{}
\newcommand{\kUU}[0]{}

\newcommand{\kWW}[0]{}
\newcommand{\kXX}[0]{}
\newcommand{\kYY}[0]{}
\newcommand{\kZZ}[0]{}

\newcommand{\kAAA}[0]{}
\newcommand{\kBBB}[0]{}
\newcommand{\kCCC}[0]{}
\newcommand{\kDDD}[0]{}
\newcommand{\kEEE}[0]{}
\newcommand{\kFFF}[0]{}

\newcommand{\kHHH}[0]{}
\newcommand{\kIII}[0]{}
\newcommand{\kJJJ}[0]{}
\newcommand{\kKKK}[0]{}
\newcommand{\kLLL}[0]{}
\newcommand{\kMMM}[0]{}
\newcommand{\kNNN}[0]{}
\newcommand{\kOOO}[0]{}

\newcommand{\kQQQ}[0]{}
\newcommand{\kRRR}[0]{}
\newcommand{\kSSS}[0]{}
\newcommand{\kTTT}[0]{}
\newcommand{\kUUU}[0]{}
\newcommand{\kVVV}[0]{}

\newcommand{\kXXX}[0]{}
\newcommand{\kYYY}[0]{}

\newcommand{\kAAAA}[0]{}
\newcommand{\kBBBB}[0]{}
\newcommand{\kCCCC}[0]{}
\newcommand{\kDDDD}[0]{}
\newcommand{\kEEEE}[0]{}
\newcommand{\kFFFF}[0]{}
\newcommand{\kGGGG}[0]{}
\newcommand{\kHHHH}[0]{}

\newcommand{\kJJJJ}[0]{}

\newcommand{\kLLLL}[0]{}

\newcommand{\kNNNN}[0]{}
\newcommand{\kOOOO}[0]{}
\newcommand{\kPPPP}[0]{}

\newcommand{\kRRRR}[0]{}
\newcommand{\kSSSS}[0]{}
\newcommand{\kTTTT}[0]{}
\newcommand{\kUUUU}[0]{}

\newcommand{\kWWWW}[0]{}
\newcommand{\kXXXX}[0]{}
\newcommand{\kYYYY}[0]{}

\newcommand{\kAAAAA}[0]{}
\newcommand{\kBBBBB}[0]{}
\newcommand{\kCCCCC}[0]{}
\newcommand{\kDDDDD}[0]{}

\newcommand{\koment}[1]{}      
\newcommand{\nothing}[1]{}

\newcommand{\anti}[1]{\overline{#1}}

\newcommand{\ve}[1]{\mathbf{#1}}

\newcommand{\derp}[2]{\frac{\partial #1}{\partial #2}} 
\newcommand{\derf}[2]{\frac{\delta #1}{\delta #2}}     

\newcommand{\dd}[0]{{\rm d}}

\newcommand{\pa}[0]{{\partial}}

\newcommand{\re}[0]{{\rm Re}}

\newcommand{\diag}{{\rm diag}}
\newcommand{\hb}{\hbar}
\newcommand{\id}[0]{\mathds{1}}

\newcommand{\refer}[1]{(\ref{#1})}

\newcommand{\nn}[0]{\nonumber}

\newcommand{\tC}[0]{{\widetilde{C}}}

\newcommand{\tG}[0]{{\widetilde{G}}}

\newcommand{\cA}[0]{{\mathcal{A}}}
\newcommand{\cB}[0]{{\mathcal{B}}}
\newcommand{\cC}[0]{{\mathcal{C}}}

\newcommand{\cF}[0]{{\mathcal{F}}}
\newcommand{\cG}[0]{{\mathcal{G}}}
\newcommand{\cH}[0]{{\mathcal{H}}}

\newcommand{\cJ}[0]{{\mathcal{J}}}

\newcommand{\cL}[0]{{\mathcal{L}}}

\newcommand{\cN}[0]{{\mathcal{N}}}
\newcommand{\cO}[0]{{\mathcal{O}}}
\newcommand{\cP}[0]{{\mathcal{P}}}

\newcommand{\cR}[0]{{\mathcal{R}}}
\newcommand{\cS}[0]{{\mathcal{S}}}

\newcommand{\cV}[0]{{\mathcal{V}}}

\newcommand{\cX}[0]{{\mathcal{X}}}

\newcommand{\cZ}[0]{{\mathcal{Z}}}

\newcommand{\ca}[0]{{\mathpzc{a}}}
\newcommand{\cb}[0]{{\mathpzc{b}}}

\newcommand{\ch}[0]{{\mathpzc{h}}}

\newcommand{\sA}[0]{{\mathscr{A}}}
\newcommand{\sB}[0]{{\mathscr{B}}}
\newcommand{\sC}[0]{{\mathscr{C}}}
\newcommand{\sD}[0]{{\mathscr{D}}}

\newcommand{\sF}[0]{{\mathscr{F}}}
\newcommand{\sG}[0]{{\mathscr{G}}}
\newcommand{\sH}[0]{{\mathscr{H}}}

\newcommand{\sL}[0]{{\mathscr{L}}}
\newcommand{\sM}[0]{{\mathscr{M}}}

\newcommand{\sQ}[0]{{\mathscr{Q}}}
\newcommand{\sR}[0]{{\mathscr{R}}}

\newcommand{\sT}[0]{{\mathscr{T}}}
\newcommand{\sU}[0]{{\mathscr{U}}}

\newcommand{\vA}[0]{{\mathbf{A}}}

\newcommand{\vG}[0]{{\mathbf{G}}}

\newcommand{\vZ}[0]{{\mathbf{Z}}}

\newcommand{\va}[0]{{\mathbf{a}}}
\newcommand{\vb}[0]{{\mathbf{b}}}
\newcommand{\vc}[0]{{\mathbf{c}}}

\newcommand{\vh}[0]{{\mathbf{h}}}

\newcommand{\vk}[0]{{\mathbf{k}}}

\newcommand{\vvp}[0]{{\mathbf{p}}}

\newcommand{\bA}[0]{{\mathbb{A}}}

\newcommand{\bC}[0]{{\mathbb{C}}}

\newcommand{\bG}[0]{{\mathbb{G}}}

\newcommand{\bL}[0]{{\mathbb{L}}}

\newcommand{\bP}[0]{{\mathbb{P}}}

\newcommand{\bR}[0]{{\mathbb{R}}}
\newcommand{\bS}[0]{{\mathbb{S}}}

\newcommand{\bV}[0]{{\mathbb{V}}}

\newcommand{\bZ}[0]{{\mathbb{Z}}}

\newcommand{\al}{\alpha}
\newcommand{\be}{\beta}
\newcommand{\de}{\delta}
\newcommand{\De}{\Delta}
\newcommand{\ga}{\gamma}
\newcommand{\Ga}{\Gamma}

\newcommand{\Si}{\Sigma}

\newcommand{\la}{\lambda}
\newcommand{\La}{\Lambda}
\def\th{\theta} 
\newcommand{\Th}{\Theta}
\newcommand{\ep}{\epsilon}

\newcommand{\cphi}{\varphi}
\newcommand{\vep}{\varepsilon}
\newcommand{\om}{\omega}
\newcommand{\Om}{\Omega}
\newcommand{\ze}{\zeta}

\newcommand{\YF}[0]{Y}     

\newcommand{\TS}[0]{\mathcal{T}}       
\newcommand{\TF}[0]{{ \mathfrak{f}}}

\newcommand{\tig}[0]{{\widetilde{\Gamma}}}

\newcommand{\volel}[1]{\!\! {{\rm d}^4 #1 }}

\newcommand{\volfour}[1]{\!\! \frac{{\rm d}^4 #1}{\left(2\pi\right)^4}}

\newcommand{\intt}[3]{\int{{\rm d}^{#1} #2 \text{ } #3}}

\newcommand{\bracket}[1]{\left(#1\right)}
\newcommand{\matri}[1]{\left[#1\right]}
\newcommand{\VEV}[1]{\left<#1\right>}

\newcommand{\mind}[3]{{{#1}}^{#2}_{\phantom{#2}{#3}}}

\begin{document}
\koment{{{\bf  !!! COMMENT !!! }}}

\pagestyle{empty}
\begin{flushright}
\end{flushright}
\vspace*{5mm}

\begin{center}

{\large {\bf LSZ-reduction, resonances and non-diagonal propagators: 
gauge fields}}
\vspace*{1cm}

{Adrian Lewandowski 
\footnote{Email: lewandowski.a.j@gmail.com}
}\\

\vspace{1cm}
{{\it
Albert Einstein Center for Fundamental Physics,\\ 
Institute for Theoretical Physics, University of Bern, \\
Sidlerstrasse 5, CH-3012 Bern, Switzerland 
}}

\vspace*{1.7cm}
{\bf Abstract}
\end{center}
\vspace*{5mm}
\noindent

We analyze in the Landau gauge mixing of bosonic fields in gauge theories
with exact and spontaneously broken symmetries, extending to this case
the Lehmann-Symanzik-Zimmermann (LSZ) formalism of the asymptotic fields.
Factorization of residues of poles (at real and complex values of the
variable $p^2$) is demonstrated and a simple practical prescription for
finding the ``square-rooted" residues, necessary for calculating $S$-matrix
elements, is given. The pseudo-Fock space of asymptotic (in the LSZ sense)
states is explicitly constructed and its BRST-cohomological structure is 
elucidated. Usefulness of these general results, obtained by investigating
the relevant set of Slavnov-Taylor identities, is illustrated
on the one-loop examples of the $Z^0$-photon mixing in the Standard
Model and the $G_Z$-Majoron mixing in the singlet Majoron model.

\vspace*{1.0cm}

\vspace*{0.2cm}

\vfill\eject
\newpage

\pagestyle{plain}

\tableofcontents 

\section{Introduction}

Mixing of fields is a common feature of many quantum field theory models.
For example, 
scalar fields mix in many extended models of the Higgs sector
of the Standard Electroweak Theory; already in the Standard Model (SM)
one has to do 
with mixing of vector fields (the $B_\mu$ and the $W^3_\mu$ 
fields - known as the photon-$Z^0$ mixing) and with mixing of fermions (the 
Cabibbo-Kobayashi-Maskawa mixing), see e.g. \cite{WeinT2}.
In tree level calculations, the mixing is removed by appropriate redefinions
of the fields but in higher orders  it reappears and extraction of 
$S$-matrix elements from Green's functions 
requires addressing this problem. 
Moreover, some 
particle states identified at the tree level become, when the
loop corrections are included, unstable (resonances) and the structure of 
the Fock space of true asymptotic states of the model is usually (even
in the perturbative expansion) different 
than the Fock space of 
the corresponding non-interacting theory.  

In general, the proper way of extracting $S$-matrix elements is provided by 
the Lehmann-Symanzik-Zimmermann (LSZ) asymptotic approach which basically 
consist of analyzing the pole structure of the relevant two-point functions
of the fields which mix, and reconstructing on this basis the Fock space of 
the true asymptotic states. 
Yet, to the best of our knowledge, in the case of field mixing in general gauge 
theories this has never been analyzed in details. 

In the simplest case of mixing of several scalar 
fields $\phi^i$ (which we take to be real, that is Hermitian operators) 
the (connected) 
two point Green's function (propagator) can in general be written in the form
\eq{\label{Eq:PolePart} 
\langle T(\phi^k(x)\phi^j(y))\rangle
=
\int\volfour{p}\,\, e^{-i\,p\cdot(x-y)}
\bigg\{
\!\sum_\ell 
\, 
\ze^k_{S[\ell]} 
\,
\frac{i}{p^2-m^2_{S(\ell)}}
\,
\ze^j_{S[\ell]}
+
\left[\!\!\begin{array}{c}
\text{non-pole}\\ \text{part}
\end{array}\!\!\right]
\!\bigg\}
\,.
}
Factorization of the residues of 
poles, some of which occur at real and other at complex values of $p^2$, 
is a well-known property \cite{Stapp}. 
\footnote{Factorization of residues at real 
poles follows also from formal manipulations \cite{WeinT1,Chank} that is, 
from inserting the complete set of asymptotic states between the field 
operators in the left hand side of (\ref{Eq:PolePart}). The $\ze^k_{S[\ell]}$ 
factors are then simply equal 
$\langle 0|\phi^k(0)|\mathbf{p},\ell\rangle$, 
where 
$|0\rangle$ 
is the true vacuum of the theory and 
$|\mathbf{p},\ell\rangle$ are the states of a single spin 0 particles labeled 
by $\ell$.}
The factors 
$\ze^k_{S[\ell]}$ associated with poles at real values
$m^2_{S(\ell)}$ of $p^2$ are crucial for 
obtaining correctly normalized (i.e. consistent with unitarity) transition 
amplitudes between initial and final states involving stable particles. 
More precisely, the Cutkosky-Veltman rules guarantee  \cite{Velt}
that the $S$-matrix is unitary provided 
{\it i)} asymptotic (free) fields appearing in the LSZ-reduction 
formula for the $S$-operator 
(see e.g. \cite{Bec} and the formula \refer{Eq:S} below) are normalized so as 
to reproduce the behavior of  
the corresponding 
(full) two-point functions near  
the poles associated with stable particles, and
{\it \it ii)} 
poles at complex values of $p^2$ are associated with no asymptotic
   states (fields), i.e. unstable particles contribute to the $S$-matrix only through
   the internal lines.
Thus, the asymptotic field 
$\vphi^j$ associated with $\phi^j$ has the form   
$\vphi^j=\sum{}^{\prime}_{\ell} 
\zeta^j_{S[\ell]}  
\Phi^{\ell}$,
where $\Phi^{\ell}$ are canonically normalized free scalar fields 
constructed out of the annihilation and creation operators of a spin 0 
particle of mass $m_{S(\ell)}$, and the summation runs over indices $\ell$
labeling only real poles (this is indicated by the prime). 

The first LSZ based extraction of 
$S$-matrix elements (disregarding the unstable character of the particles)
in the presence of mixing of two scalar fields (neutral CP even components of 
the two Higgs doublets of the Minimal Supersymmetric Standard Model) 
was presented in \cite{CHPR}. 
The mixing of $W^3_\mu$ and $B_\mu$ gauge fields of the Standard Model was 
studied in this framework in \cite{Stuart}, and factorization of residues of 
the pole at $p^2=0$ (corresponding to the photon) and at a complex value of 
$p^2$ (corresponding to the unstable $Z^0$ boson) was \kAAAAA explicitly 
demonstrated. 
The LSZ approach to fermionic mixing was presented first in the context of 
leptogenesis in \cite{PilafRL1,PlumOld} where factorization of residues of the 
poles corresponding to unstable Majorana neutrinos was demonstrated
(see also \cite{PilafUnd,PlumNew}). 
While asymptotic (in and out) particle states corresponding to poles in 
(\ref{Eq:PolePart}) at complex values of $m^2_{S(\ell)}$ do not, strictly
speaking, exist, the factors $\zeta^j_{S[\ell]}$ associated with such poles
are, nevertheless, useful in studying properties of resonances 
as shown in \cite{Stapp, Stuart, PilafRL1, PlumOld}.

Clearly, treatment of mixing of
scalar fields is essential in studies of multifield Higgs sectors of various
extensions of the SM. 
Similarly, the mixing of vector fields is 
a typical
feature of theories (e.g. GUT models)
based on gauge symmetry groups higher than the group
$SU(3)_c\times SU(2)_L\times U(1)_Y$ of the SM. 
In view of the ubiquity of mixing of fields, the problem of formulating 
an optimal prescription for computing the coefficients $\ze^k_{S[\ell]}$ 
parameterizing the residues in \refer{Eq:PolePart} have  gained in recent 
years a renewed interest. 
Mixing of three scalar fields was analyzed only recently in the paper
\cite{Fuchs}, in which the factorization property \refer{Eq:PolePart} was 
demonstrated and explicit formulae for the coefficients $\ze^k_{S[\ell]}$ were 
given. The results were applied to the neutral Higgs sector of the MSSM; 
it was shown that 
cross-sections obtained neglecting the non-pole part in Eq. 
\refer{Eq:PolePart} agree to a good accuracy with the cross-sections based on 
the full propagators.  
Analysis of a generic mixing of $n$ fermionic fields was recently given in 
\cite{Kniehl-1,Kniehl-2,Kniehl-3}.\footnote{In the context of field mixing 
we should mention also the paper \cite{AOKI} in which the possibility 
of imposing on-shell renormalization conditions in systems with 
mixed scalar, vector and fermionic fields was studied. Mixing of fermions
treated in this approach  was re-examined in \cite{Kaloshin}   
with the aid of special parametrization of the propagator. \kBBBBB}

In \cite{OnFeRu} we have reconsidered general mixing of scalar and fermionic 
fields, simplifying and generalizing prescriptions for calculating the 
$\ze$ factors available in the cited literature. In the case of fermions
we have analyzed in details poles corresponding to 
the arbitrary system of Weyl fields,  
obtaining prescriptions for 
Dirac-type and Majorana-type spin 1/2 particles (or resonances) 
as special cases. 
Our approach is closest  
in spirit to the one of \cite{Kniehl-1,Kniehl-2,Kniehl-3}. 
There are, however, some differences. Firstly, we followed the philosophy of 
keeping the renormalization scheme as general as possible. In particular, we 
did not impose any concrete renormalization conditions on the two-point 
functions.
Second, we offered a technical improvement in comparison with the analyses of 
\cite{Stapp,Kniehl-1,Kniehl-2,Kniehl-3}, where the cofactor matrix 
of one-particle irreducible two-point functions at the pole
was used to get the formulae for $\ze$. 
In contrast, the factors $\ze$ in our approach are expressed directly 
in terms of properly normalized eigenvectors of certain ``mass-squared 
matrices", so that the case of degenerated eigenvalues is naturally covered by 
our prescription. Thus, the prescription for finding $\ze$ proposed in 
\cite{OnFeRu} can be considered a direct generalization of the standard 
procedure for finding tree-level mass eigenstates. 

The purpose of this paper is to extend the approach of \cite{OnFeRu} to the
mixing of $n$ vector fields in a general gauge field theory. The complication 
characteristic for gauge fields in general, and their mixing in particular, is 
the presence of unphysical degrees of freedom which contribute to residues of
poles of the two-point functions but do not correspond to physical particles.
To properly identify the $\ze$ factors corresponding to
physical particles (or resonances) we 
perform a careful analysis of the relevant set of Slavnov-Taylor identities and 
explicitly construct the asymptotic (in the LSZ sense) vector and scalar
fields. We also demonstrate that, in case of 
generic mixing,
the unphysical components of these asymptotic fields 
create (out of the 
vacuum) states which combine into the Kugo-Ojima 
quartet representations of the BRST algebra \cite{KugoOjima} 
what is essential for unitarity of the $S$-matrix \cite{Bec,KugoOjima}. 

We have decided to restrict this study to the Landau gauge,
since this gauge offers some practical advantages: it is 
Lorentz covariant, renormalization group invariant and provides the simplest 
way of calculating the effective potential (see e.g. \cite{Martin-3} for 
a recent determination of the three-loop effective potential of scalar fields
in a general renormalizable model in the $\overline{{\rm MS}}$ scheme). A 
dedicated analysis of the Landau gauge case is justified also at the technical 
level: firstly, the Nakanishi-Lautrup auxiliary fields cannot be integrated 
out in this gauge. Secondly, in the Landau 
gauge the would-be Goldstone bosons 
produce poles at $p^2=0$ in the 
propagators of system of scalar fields and a prescription is necessary to
properly identify the associated $\ze$ factors in situations where there
are also poles at $p^2=0$ corresponding to physical massless Goldstone bosons
of spontaneously broken global symetries. It is here that our approach of 
Ref.  \cite{OnFeRu}, covering also the case of degeneracies, becomes 
particularily advantageous; combined with the additional symmetry of
the Faddeev-Popov sector of the Landau gauge action \cite{BPS:anti-ghost-eq},
it allows for unambiguous identification of the $\ze_{S[\ell]}$ factors 
corresponding to physical massless spin 0 particles.

Analysis of mixing of vector fields in a non-Landau $R_\xi$ gauge 
requires in principle only a few minor changes\footnote{ In general 
(non-Landau) $R_\xi$ gauges
vector fields mix beyond the tree-level with the scalar ones giving rise to 
nonvanishing mixed vector-scalar propagators \cite{Mi1,Mi2,Mi3,Mi4}.~Therefore 
the asymptotic vector fields can create/annihilate 
also physical spin 0 particles. One thus needs a prescription for an additional 
(component of) 
eigenvector $\ze$ which determines the contribution of physical 
scalar field mode to the asymptotic vector field.} 
and will be given elsewhere. 

The paper is organized as follows. In Section \ref{Sec:Sub:Scalars} we 
summarize (and reformulate in a slightly more convenient way) the prescription 
of \cite{OnFeRu} for extracting the factors $\ze^j_{S[\ell]}$ from Green's 
functions of scalar fields. In Sec. \ref{Sec:Sub:Prescr-Vectors} this 
prescription is 
directly generalized to the case of mixing of vector fields without 
presenting the detailed  
structure of the asymptotic fields, so that the reader interested 
in practical aspects of the procedure, that is in extracting the factors
$\ze_{V}$ necessary for computing elements of the 
$S$-matrix with 
spin 1 particles in asymptotic states, is not distracted by technicalities.
The prescription for properly identifying the factors $\ze^j_{S[\ell]}$ 
corresponding to physical Goldstone bosons is also given. 
To illustrate the main points on some examples we first give in Section 
\ref{Sec:Res} general one-loop formulae for all possible (in the Landau gauge) 
self-energies of the system of vector and scalar fields of a general 
renormalizable model in the $\anti{{\rm MS}}$ scheme,\footnote{General 
formulae for fermionic self energies are collected in \cite{OnFeRu}.
All these formulae were obtained using the naive prescription for the 
$\gamma^5$ matrix.} 
cultivating in this way
the long tradition of providing ready-to-use general formulae, see e.g. 
\cite{Weinberg}, \cite{Gross-Wilczek}, \cite{Jack-Osborn}, 
\cite{Mach-Vaughn}, \cite{Martin}, \cite{Martin-3}, \cite{Luo}, 
\cite{ChLM}.  
Using these general results we reconsider in Section 
\ref{Sec:SubSec:Examples:Z-A-mixing} the $Z$-photon mixing from the point of
view of the asymptotic LSZ approach. The problem of properly identifying the 
would-be Goldstone modes and the true Goldstone bosons is 
illustrated in Sec. \ref{Sec:SubSec:Examples:Maj} on the example of 
the Singlet Majoron Model \cite{Majo}.
The technicalities: 
the analysis of 
Slavnov-Taylor identities, construction of the asymptotic fields and related 
issues, which 
constitute in fact the main results of the paper, are relegated to Section 
\ref{Sec:Proof}.

We end this introduction by summarizing our notation and conventions.
In most of formulae indices are suppressed and the matrix 
multiplication is understood. The summation convention is used only when an 
upper index is contracted with a lower one; whenever ambiguities may arise, 
sums are explicitly displayed. The Minkowski metric has the form 
\eq{\nn
\eta=[\eta_{\mu\nu}]=\diag(+1,-1,-1,-1)\,.
}  
Our convention for Fourier transform of fields is summarized by the
formulae
\begin{eqnarray}
\label{Eq:Mom-Pos-Rel}
\cF(x)=\int\,\volel{l} ~\! e^{-i l x} \hat{\cF}(l) 
\qquad\Rightarrow\qquad
\derf{}{\cF(x)}=\int\volfour{l} ~\! e^{i l x}~\!
\derf{}{\hat{\cF}(l)}\,.
\end{eqnarray}
We assume that the model in question has already been renormalized 
in an arbitrary renormalization scheme consistent with the gauge symmetry. 
Thus, all fields and correlation functions are considered as renormalized ones. 
\footnote{
In particular, we assume that the finite counterterms have been adjusted, if necessary, 
so as to restore the Slavnov-Taylor identities for the gauge symmetry
(see e.g. \cite{Bon}  for a discussion in the context of dimensional 
regularization with the consistent 't~Hooft-Veltman-Breitenlohner-Maison 
prescription for $\gamma^5$). 
}
The (renormalized) one-particle irreducible (1PI) effective action $\Gamma$
is the generating functional for renormalized 1PI Green's functions.  
For instance, the two-point function of scalar fields $\phi^j$ is given by 
\begin{eqnarray}
\left.\derf{}{\hat{\phi}^j(p)}
\derf{}{\hat{\phi}^k(p')}~\!\Gamma\matri{\phi,\ldots}\right|_{0}
=(2\pi)^4 \delta^{(4)}(p'+p)\,
\widetilde{\Gamma}^{\phantom{kj}}_{kj} (p',p)~\!.
\phantom{aaaaa}\label{Eq:Notacja1}
\end{eqnarray}
The functional derivatives (which act always from the left) in 
\refer{Eq:Notacja1} are taken at the ``point'' at which \emph{all} fields 
vanish (this is indicated by the vertical bar with the subscript 0). 
In particular, we always assume that the scalar fields have already been 
shifted if necessary, so that they have vanishing vacuum expectation values 
(VEVs); in other words  we assume that
\eq{\label{Eq:TadCondFormal}
\left.\derf{\Gamma\matri{\phi,\,\ldots}}{\phi^i(x)}\right|_{0}=0.
}
\vskip0.2cm

\section{Practical prescriptions}\label{Sec:Prescription}

\subsection{Mixing of scalar fields} \label{Sec:Sub:Scalars}

 We start by recapitulating the prescription, formulated in \cite{OnFeRu},
for the pole part of the propagator of a system of scalar fields $\{\phi^j\}$
(which, without loss of generality are assumed to be all 
Hermitian with vanishing VEVs).
The 1PI two-point function of such a system of
scalars is of the general form
\eqs{\label{Eq:Gamma2-scalar} 
\widetilde{\Ga}_{k j}(-p,p)
=\Big[
p^2\id-M^2_S(p^2)
\Big]_{k j}
\,,
}
with a symmetric matrix $M^2_S(s)=M^2_S(s)^\rmt\equiv(M^{\rm tree}_S)^2+\Si_S(p^2)$.
Inverting the matrix $\widetilde{\Ga}_{k j}(-p,p)$ 
we get the  matrix of propagators 
\eq{\label{Eq:G2-scalar} 
\widetilde{G}^{\,k j}(p,-p)
=
i 
\Big[ \big(p^2\id-M^2_S(p^2)\big)^{-1} \Big]^{k j}
\,.\
}
The poles of (\ref{Eq:G2-scalar}) are at values  
$p^2=m_{S({\ell})}^2$ which are solutions to the following equation \kA 
\eq{\label{Eq:GapEq-0-scalar}
\det(s\mathds{1}-M^2_S(s))\Big|_{s=m_{S(\ell)}^2}=0\,.
}
Let the vectors $\ze_{S[\ell_1]}\,,\ze_{S[\ell_2]}\,,\ldots,$ 
form a basis of the eigenspace of the matrix $M^2_S(m_{S(\ell)}^2)$
corresponding to its eigenvalue\footnote{
In this notation 
$m_{S(\ell)}^2\neq m_{S(\ell')}^2$ for $\ell\neq\ell'$. Notice that, in general, 
eigenvalues of the matrix $M^2_S(m_{S(\ell)}^2)$ other than $m_{S(\ell)}^2$ are 
not solutions of (\ref{Eq:GapEq-0-scalar}) and are irrelevant for the 
problem of mixing.}  
$m_{S(\ell)}^2$
\eq{\label{Eq:xi-Eig-GENERAL-scalars}
M^2_S(m_{S(\ell)}^2)\,  \ze_{S[\ell_r]} 
= m_{S(\ell)}^2\,
\ze_{S[\ell_r]}\,,
}
obeying the following normalization/orthogonality conditions 
\koment{str.MPW:91} \eq{\label{Eq:norm-cond-GENERAL-scalars}
\ze_{S[\ell_r]}^{\,\, \rmt} \,  
\left[
\mathds{1} - M^2{}^{\prime}_{\!\!\!\!S\, } (m_{S(\ell)}^2)
\right]\,
\ze_{S[\ell _q]}
=\delta_{r q}\,, 
}
in which
$
M^2_S{}^{\prime}(s) 
\equiv 
\dd M^2_S(s)/\dd s\,.
$
As shown in \cite{OnFeRu}, 
the propagator \refer{Eq:G2-scalar} takes then the form\footnote{
In order to ensure that the 
propagator takes on the simple form
\refer{Eq:D-as-GENERAL-scalars-indi},
one has to assume that each generalized eigenvector (see e.g. \cite{Axler})   
\kE of $M^2_S(m_{S(\ell)}^2)$ associated with the eigenvalue $m_{S(\ell)}^2$ is an 
ordinary eigenvector  (that is, in the Jordan basis the block of the 
matrix $M^2_S(m_{S(\ell)}^2)$ corresponding to its eigenvalue $m_{S(\ell)}^2$
is diagonal). 
There is no need to investigate other (unphysical) generalized  
eigenspaces of $M^2_S(m_{S(\ell)}^2)$; 
in particular 
as a whole the matrix $M^2_S(m_{S(\ell)}^2)$ can be even  
non-diagonalizable. We also assume that the derivative 
$M^2{}^{\prime}_{\!\!\!\!S\, } (m_{S(\ell)}^2)$ is infrared-finite, so that the 
singularities of the propagator are poles rather than branch points (see e.g. 
\cite{Kibble}).
}
\eq{\label{Eq:D-as-GENERAL-scalars-indi} 
\widetilde{G}^{kj}(p,-p)
=
\sum_\ell \sum_r
\, 
\ze^k_{S[\ell_r]} 
\,
\frac{i}{p^2-m^2_{S(\ell)}}
\,
\ze^j_{S[\ell_r]}
+\text{[non-pole part]}
\,.
}
Moreover, if Feynman integrals contributing to $M^2_S(p^2)$ do not acquire 
imaginary parts in a left neighborhood $\sU_\ell\subset\bR$ of 
$p^2=(m^{\rm tree}_{S(\ell)})^2$, so that the following reality condition
\eq{\label{Eq:RealityCond-scalar}
M^2_S(s)=M^2_S(s)^\star
\,,\qquad \ 
\forall_{s\in\sU_\ell}\,,
}
 is satisfied, then all terms of a formal power series 
\eq{\nn
m_{S(\ell)}^2=(m^{\rm tree}_{S(\ell)})^2+\cO(\hb)\,,
}
are real and there exist vectors $\zeta_{S[\ell_r]}$ obeying Eqs. 
\refer{Eq:xi-Eig-GENERAL-scalars}-\refer{Eq:norm-cond-GENERAL-scalars} 
and such that 
$\zeta_{S[\ell_r]}=\zeta_{S[\ell_r]}^{\ \star}$ for all $r$.

Some comments are in order. The normalization conditions 
\refer{Eq:norm-cond-GENERAL-scalars} have here a different form than 
the ones given in \cite{OnFeRu} but are, nevertheless, equivalent to
them; their form \refer{Eq:norm-cond-GENERAL-scalars}
will be more convenient in what follows. 
The left hand side of Eq. \refer{Eq:norm-cond-GENERAL-scalars}
is symmetric in the indices $r$ and $q$. Therefore
starting with an arbitrary basis of 
the eigenspace, say a set of vectors $\{\xi_{[\ell_r]}\}$, one can 
construct vectors obeying Eq. 
\refer{Eq:norm-cond-GENERAL-scalars} provided  $\xi_{[\ell_r]}$ are in a
one-to-one correspondence with the eigenvectors of the tree-level mass 
matrix.\footnote{\label{Foot:Patho}
One could worry that the condition \refer{Eq:norm-cond-GENERAL-scalars} 
cannot be imposed since e.g.
$[1,\ \ i]\,[1,\ \ i]^{\rmt}=0$, 
however such a pathology is impossible at the tree-level, and thus it is 
impossible 
for 
the formal power series.
}
\kD

The form \refer{Eq:D-as-GENERAL-scalars-indi} of the propagator
uniquely determines the form 
\eq{\label{Eq:AsymField}
\vphi^j= 
{\sum_{  \ell}}^\prime\sum_r
{\zeta}^j_{S[\ell_r]}  
\Phi^{\ell_r}\,,
}
of the asymptotic (in the LSZ sense) field
corresponding to $\phi^j$. The prime on the first sum in (\ref{Eq:AsymField}) indicates that it runs 
only over indices $\ell$ labeling poles of the propagator 
\refer{Eq:D-as-GENERAL-scalars-indi} located at real values of $p^2$ 
(we assume the corresponding vectors ${\zeta}_{S[\ell_r]}$ are choosen real).
The operators $\Phi^{\ell_r}$ in (\ref{Eq:AsymField}) are Hermitian scalar
free field operators built out of the creation and annihilation 
operators of spin 0 particles of mass $m_{S(\ell)}$ acting in the standard way 
in the Fock space of the in (or out) 
states, and are such that one-particle
states created by $\Phi^{\ell_r}$ and by $\Phi^{\ell'_{r'}}\neq \Phi^{\ell_r} $
from the  
vacuum are  orthogonal.\footnote{
For completeness, the explicit form of $\Phi^{\ell_r}$ in our conventions is 
given is Sec. \ref{Sec:Sub:AsStGauge} below, cf. Eq. \refer{Eq:Phi}.
}
This form of (\ref{Eq:AsymField}) guarantees that (the Fourier transform of)
$\langle0|T\vphi^j(x)\vphi^j(y)|0\rangle$ reproduces the behavior of 
\refer{Eq:G2-scalar} in the vicinity of all poles located on the real 
axis. 
The asymptotic field $\vphi^j$ allows us to 
write the LSZ formula for the $S$-operator in a compact form \cite{Bec} 
\eq{\label{Eq:S}
S=\,\,:\!\exp\!\left\{
-\int\,\volel{x}\,\vphi^{j}(x)\!
\int\,\volel{y}\,\Gamma_{jk}(x,y)\,
\derf{}{J_k(y)}
\right\}\!:\, \exp(i\, W[J])\bigg|_{J=0}\,,
}
which when inserted between states of the asymptotic in (or out) Fock space 
yields $S$-matrix elements corresponding to transitions between stable 
particles. $\Gamma_{jk}(x,y)$ in (\ref{Eq:S}) is the Fourier transform of 
\refer{Eq:Gamma2-scalar} and the normal ordering refers to the free quantum 
fields $\vphi^j$. 
\kG 
The functional $W[J]$  generating connected Greens functions is related 
through the Legendre transform to the (renormalized) 1PI effective action 
$\Gamma[\phi]$ 
\eq{\label{Eq:Leg}
\Gamma[\phi]=W[\cJ^\phi] - \int\!d^4x~\!\cJ^\phi_j(x)\!\cdot\!\phi^j(x)\,,
\qquad \ 
\derf{W[J]}{J_{j}(x)}\Bigg|_{J=\cJ^\phi}
=\phi^{j}(x)\,.
}
\kH 
In practical terms 
the formula \refer{Eq:S} means that to obtain the correctly normalized 
(i.e. consistent with unitarity) amplitude of a process involving a particle 
corresponding to the field operator $\Phi^{\ell_r}$, the eigenvector  
$\zeta^j_{S[\ell_r]}$ has to be contracted with the appropriate amputated 
correlation function $\cA_{j\ldots}(p,\ldots)$ of the scalar field 
$\phi^j$ evaluated at $p^2=m^2_{S(\ell)}$.

As we already said, vectors 
$\ze_{S[\ell_r]}$ corresponding to complex poles $m_{S(\ell)}^2$ in 
Eq. \refer{Eq:D-as-GENERAL-scalars-indi}, even though they are not associated 
with asymptotic fields, are useful in the 
study of properties  
of unstable particles, as they govern the behavior of amplitudes for 
$s\approx \re(m_{S(\ell)}^2)$ \cite{PilafRL1,PlumOld} (see also 
\cite{PilafUnd,PlumNew}). In particular, the imaginary part of $\ze_{S[\ell_r]}$ 
is one of the sources of CP-asymmetry in decays of unstable states 
\cite{PilafRL1}.

We also note that the formuale \refer{Eq:AsymField} and 
\refer{Eq:xi-Eig-GENERAL-scalars}-\refer{Eq:norm-cond-GENERAL-scalars} are 
obvious at the tree-level. In particular, $M_S^{2\,\prime}(p^2) = \cO(\hb)$ and 
therefore \refer{Eq:AsymField} is nothing but an expansion of the scalar 
field in an orthonormal basis of eigenvectors of the mass-squared matrix. From 
this point of view, Eq. \refer{Eq:norm-cond-GENERAL-scalars}  defines a
``quantum-corrected metric" which  fixes correct normalization of the 
eigenvectors in higher 
orders in $\hb$.  

\subsection{Vector and scalar fields} \label{Sec:Sub:Prescr-Vectors}

We consider now a set $\{A^\al_\mu\}$ of (renormalized) Hermitian vector 
fields, together with a set $\{\phi^j\}$ of Hermitian scalar fields (having 
vanishing vacuum expectation values). In order to fix the conventions, we 
give here an expression for a covariant derivative of scalars 
\eq{\label{Eq:CovDerSc}
(D_\mu\phi)^j =\partial_\mu \phi^j
+ A^\alpha_\mu[{\TS}_\alpha]^{j}_{\ k}(\phi^k+v^k)\,,
}
where ${\TS}_\alpha$ are real antisymmetric generators of the
gauge group in the representation formed by the scalars; they contain gauge 
couplings and satisfy the commutation relations
$[{\TS}_\alpha,~\!{\TS}_\beta]={\TS}_\gamma~\! 
e^\gamma_{\phantom{a}\alpha\beta}$ 
with real structure constants $e^\gamma_{\phantom{a}\alpha\beta}$. 
As said, $\phi^j$ have vanishing VEVs; 
$v^j$ are the VEVs of ``fields in the symmetric phase" 
$
\phi^j_{\rm sym}
\equiv
\phi^j + v^j\,.
$ 
Thus, $v^j$ are determined by the condition that the complete tadpole of 
$\phi^j$ vanishes (cf. Eq. \refer{Eq:TadCondFormal}), which gives $v^j$ as 
a formal power series in $\hb$ 
\footnote{\label{Foot:Stue}
In order to simplify the notation, we have  assumed in \refer{Eq:CovDerSc}
that none of the components $\phi^j$ is a Stueckelberg field (see e.g. 
\cite{Stue} and references therein). 
Nonetheless, everything what we say here works \kUUUU   
also in the presence of Stueckelberg scalars, provided one makes the 
replacement
\eq{\nn 
{\TS}_\alpha v \mapsto {\TS}_\alpha v +\bar{P}_\alpha\,,
}
where coefficients $\bar P_\alpha$ obey $\TS_\beta \bar{P}_\alpha=0$ and (in a 
natural basis of the gauge Lie algebra) can be nonzero only for indices 
$\alpha=\al_A$ associated with the Abelian ideal. 
} 
\eq{\label{Eq:VEV-expansion}
v^j=v^j_{(0)}+\hb\, v^j_{(1)}+\cO(\hb^2)\,.
}

The most general form of the renormalized
\footnote{   
Recall that we allow for completely arbitrary renormalization 
conditions that are consistent with Slavnov-Taylor identities, see e.g. \cite{PiguetSorella}. 
 }
1PI two-point function of vector fields is
\eqs{\label{Eq:Gamma2-vector} 
\widetilde{\Ga}^{\mu\nu}_{\al\be}(-q,q)
&\equiv&
-\eta^{\mu\nu}\Big[
q^2\id-M^2_V(q^2)
\Big]_{\al\be}
+q^\mu q^\nu\sL_{\al\be}(q^2)
\,.
}
The general form of the 1PI two-point function of scalar fields, which 
must be considered in parallel to that of vector fields, if some of gauge
symmetries are broken by VEVs of scalars, is still given by 
\refer{Eq:Gamma2-scalar}.
Even though the mixed vector-scalar two point function 
$\widetilde{\Ga}^{\mu}_{\al j}(-q,q)$ 
is, in general, non-vanishing,  
the Landau gauge condition ensures that the mixed propagator vanishes 
\eq{\label{Eq:G2-scalar-vector}
\widetilde{G}^{\,j \be}_{\ \,\nu}(q,-q)=0.
}
Thus, in addition to 
the propagator \refer{Eq:G2-scalar} of scalar fields, 
for practical purposes  
it suffices to consider the propagator 
of vector fields which takes (see Section \ref{Sec:Proof}) the form:
\eq{\label{Eq:G2-vector} 
\widetilde{G}^{\be\de}_{\nu\rho}(q,-q)
=-i
\left[\eta_{\nu\rho}-\frac{q_\nu q_\rho}{q^2}\right]
\Big[ \big(q^2\id-M^2_V(q^2)\big)^{-1} \Big]^{\be\de}
\,.\
}
Since the ``denominator'' of \refer{Eq:G2-vector} has the same structure
as that of \refer{Eq:Gamma2-scalar}, one can immediately write
\eq{\label{Eq:As-vect-part} 
\Big[ \big(q^2\id-M^2_V(q^2)\big)^{-1} \Big]^{\be\de}
=
\sum_\la \sum_r
\, 
\ze^\be_{V[\la_r]} 
\,
\frac{1}{q^2-m^2_{V(\la)}}
\,
\ze^\de_{V[\la_r]}
+\text{[non-pole part]}
\,.
}
The complete pole 
part of the full propagator \refer{Eq:G2-vector} 
will be given in Section \ref{Sec:Proof} (the formula 
\refer{Eq:G2-vector:pole-part}).
The formula \refer{Eq:As-vect-part} is however all 
one needs to write 
down those terms
of the asymptotic vector field $\vA^\al_{\mu}$ which are relevant for
computing $S$-matrix amplitudes of processes with stable spin 1 
paricles in the initial and/or final states:
\eq{\label{Eq:AsymField-vec}
\vA^\al_{\mu}
 = 
{\sum_{  \la}}^\prime\sum_r
{\zeta}^\al_{V[\la_r]}  
\bA^{\la_r}_{\mu} 
+ \ldots\,.
} 
As in \refer{Eq:AsymField}
the prime over the first sum indicates that it runs only over the indices
$\lambda$ labeling poles at real values $m^2_{V(\la)}$ of $q^2$. With each
idependent eigenvector $\ze_{V[\la_r]}$ corresponding to such a pole associated 
is in \refer{Eq:AsymField-vec} a free Hermitian vector field 
$\bA^{\la_r}_{\mu}$ built out of the spin 1, mass $m_{V(\la)}$ particle
annilation and creation operators acting 
in the Fock space of the in (or out) states. 
The operator\footnote{The explicit form of the operator $\bA^{\la_r}_{\mu}$ 
in our conventions is given is Sec. \ref{Sec:Sub:AsStGauge} 
(the formula \refer{Eq:bA}).} 
$\bA^{\la_r}_{\mu}$ has the unitarity gauge structure 
(if $m_{V(\la)}\neq0$), or the Coulomb gauge structure (if $m_{V(\la)}=0$). 
As in the case of the asymptotic field \refer{Eq:AsymField}, 
the one-particle states 
created/annihilated by $\bA^{\la_r}_{\mu}$ and by
$\bA^{\la'_{r'}}_{\mu}\neq \bA^{\la_r}_{\mu} $ are orthogonal 
to each other.
The ellipsis in \refer{Eq:AsymField-vec} stand for free operators 
creating/annihilating in the in (or out) Fock space states belonging 
to Kugo-Ojima quartet representations \cite{KugoOjima};  
the explicit formulae for these operators are given in Sec. 
\ref{Sec:Sub:AsStGauge}. 
 With all these operators taken into account
the (Fourier transform of the) two point function 
$\langle0|T(\vA^\al_{\mu}(x)\vA^\beta_{\nu}(y))|0\rangle$ 
reproduces the behavior of 
\refer{Eq:G2-vector} near all poles located on the real axis. 
Using the asymptotic field \refer{Eq:AsymField-vec} in the formula 
\refer{Eq:S} for the $S$-operator 
\footnote{
In our conventions, Eqs. \refer{Eq:S}-\refer{Eq:Leg} 
are valid in the generic case, provided that 
indices $j$ and $k$ run over all components of all fields, 
including vectors, fermions, (anti)ghosts and Nakanishi-Lautrup multipliers, 
see e.g. \cite{Bec}. 
} 
then shows that 
the amplitude of a process
with a stable spin 1 particle corresponding to $\bA^{\la_r}_{\mu}$
in the initial or final state is obtained by 
contracting the amputated correlation functions 
$\cA^\mu_{\al\ldots}(p,\ldots)$  of fields $A^\al_\mu$ with the 
eigenvector $\ze^\al_{V[\la_r]}$ and the appropriate (canonically normalized) 
polarization vector $e_\mu(\vvp,m_{V(\la)})$ or $e_\mu(\vvp,m_{V(\la)})^\star$.

In the presence of spontaneous breaking of some gauge symmetries, 
it is also necessary  to identify those terms in the decomposition 
\refer{Eq:AsymField} of the asymptotic scalar field which create/annihilate 
physical states. 
This is particularly easy if 
there are no Goldstone bosons of spontaneously broken global symmetries\footnote{
Of physical spin 0 particles, only Goldstone bosons can naturally be massless. }
as then all fields $\Phi^{\ell_r}$ corresponding to $m_{S(\ell)}=0$ create  
would-be Goldstone bosons 
while all remaining fields are associated with 
physical particles. 
If the true Goldstone bosons are present (e.g. in the singlet Majoron model 
\cite{Majo}, see also Section \ref{Sec:SubSec:Examples:Maj}), 
we need a prescription for identifying massless eigenvectors $\ze_{S[\ell_r]}$  
associated with them. 
The gauge symmetry 
implies that the eigenvectors $\zeta_{S[\ell_r]}$ corresponding to the would-be 
Goldstone bosons are linear combinations of vectors\footnote{More precisely, 
this fact follows from the ``non-renormalization theorem" expressed by the 
relation \refer{Eq:Antigh-Conclusion}, which is a manifestation of an 
additional symmetry of the action specific for the Landau gauge 
\cite{BPS:anti-ghost-eq}.} 
${\TS}_\alpha v$. 
The orthogonality condition \refer{Eq:norm-cond-GENERAL-scalars} then 
suggests 
that of all vectors $\zeta_{S[\ell_r]}$  associated with poles at $p^2=0$, 
to physical massless states 
should 
correspond vectors $\zeta_{S[\ell_r]}$ 
such that
\eq{\label{Eq:norm-cond-GENERAL-scalars-phys}
\ze_{S[\ell_r]}^{\,\, \rmt} 
\left[
\mathds{1} - M^2{}^{\prime}_{\!\!\!\!S\, } (0)
\right]
\TS_\al v
=0\,,
}
for all indices $\alpha$. 
\footnote{ At the tree-level this reduces to a well-known condition 
(see e.g. (1.1) in \cite{Weinberg}). 
}
In Sec. \ref{Sec:Sub:AsStGauge} we will show that the states of the asymptotic
Fock spaces associated with eigenvectors $\ze_{S[\ell_r]}$ obeying this condition 
do indeed belong to the physical subspace of the kernel of BRST charge.  
\vspace*{4 pt}

It should be also stressed that the normalization condition, which 
for vectors $\zeta^j_{S[\ell]}$ takes 
the form \refer{Eq:norm-cond-GENERAL-scalars},
has to be  slightly modified 
in order to avoid  \emph{spurious} infrared divergences. 
Take, for instance, the $Z$--photon 
block of the Standard Model (SM, see e.g. \cite{WeinT2}); the 2-by-2 matrix 
$M^2_V{}^{\prime}(0)$ (more 
precisely, its $ZZ$ entry) 
is IR divergent at one-loop order, 
however the photonic singularity is still a pole. 
The IR-finiteness of the whole matrix $M^2_V{}^{\prime}(0)$ is therefore too 
strong a requirement. 
In Sec. \ref{Sec:Sub:AsymProp} we will show that Eq. 
\refer{Eq:As-vect-part} holds provided $M^2_V(s)$ is continuous at each 
$m^2_{V(\la)}$ and that the limit \koment{str.PSMW:4}
\eq{\label{Eq:Lim1Der}
\lim_{\ \  q^2\to m^2_{V(\la)}} 
\left\{
M^2_V{}^{\prime}(q^2)\, \xi 
\right\}\,,
}
exists for each $\xi$ belonging to the eigenspace \footnote{As before, we 
have to assume that each generalized eigenvector  \kE  of $M^2_V(m_{V(\la)}^2)$ 
associated with the eigenvalue $m_{V(\la)}^2$ is an ordinary eigenvector.
} 
$M^2_V{}(m^2_{V(\la)})$ associated with $m^2_{V(\la)}$. 

The vectors $\ze_{V[\la_r]}$ appearing in \refer{Eq:As-vect-part} are then 
elements of a basis of the eigenspace 
\eq{\label{Eq:xi-Eig-GENERAL-vectors}
M^2_V(m_{V(\la)}^2)\,  \ze_{V[\la_r]} 
= 
m_{V(\la)}^2\,
\ze_{V[\la_r]}\,,
}
obeying the normalization conditions \koment{str.PSMW:1} \kM 
\eq{\label{Eq:norm-cond-GENERAL-vectors}
\lim_{\ \  q^2\to m^2_{V(\la)}} 
\left\{\ze_{V[\la_r]}^{\,\, \rmt} \,  
\left[
\mathds{1} - M^2{}^{\,\prime}_{\!\!\!\!V\, } (q^2)
\right]\,
\ze_{V[\la_t]}\right\}
=\delta_{r t}\,. 
}
Furthermore, if {\it massless spin 1} particles are present,  
an additional assumption is necessary to ensure that the singularity of the 
full propagator \refer{Eq:G2-vector} at $q^2=0$ is a (second order) pole: 
the limit \koment{str.PSMW:12}  
\eq{\label{Eq:Lim2Der}
\lim_{q^2\to 0} 
\left\{
M^2_V{}^{\prime\prime}(q^2)\, \xi 
\right\}\,,
}
has to exist for each $\xi$ belonging to a basis of the null eigenspace of 
$M^2_V{}(0)$. 
In Sec. 
\ref{Sec:SubSec:Examples:Z-A-mixing} we will show that the limits 
\refer{Eq:Lim1Der} and \refer{Eq:Lim2Der} 
are indeed finite for the photonic eigenvector $\xi$ in the SM 
at one-loop order.

The discussion of physically meaningful infrared divergences 
(i.e. the ones that lead to divergent residues)
is beyond 
the scope of this paper, as they change the structure of asymptotic states 
\cite{Kibble}.  \kL
In what follows, it will be assumed that an IR regulator has been introduced, 
if necessary, so that the limits \refer{Eq:Lim1Der} and \refer{Eq:Lim2Der} are 
finite. \kN  \kDDDDD \kCCCCC\\

We end this section with an alternative prescription for  
finding the directions of eigenvectors $\ze_{V[\la_r]}$ 
corresponding to massless spin 1 particles. 
At the tree level, when 
\eq{\label{Eq:m_V}
M^2_V(q^2)_{\al\be}=\matri{m_V^2}_{\alpha\beta}\equiv
(\TS_\al\, v_{(0)})^{\rmt}(\TS_\be\, v_{(0)})
\,,
}
the eigenvectors of the matrix
$M^2_V(0)_{\al\be}$ corresponding to its zero eigenvalues are directly related 
to the unbroken generators of the gauge group
\cite{Weinberg} 
\eq{\nn
m^2_{V\al\be}\,\,\th^\be=0 
\qquad \Leftrightarrow \qquad
\th^\be\,\TS_\be\, v_{(0)}=0\,.
}
This immediately determines the vectors $\ze_{V[\la_r]}$ corresponding to 
massless gauge bosons 
at the zeroth order. 
In the  Landau gauge this 
prescription   
generalizes to hihger orders owing to the antighost identity 
\cite{BPS:anti-ghost-eq}, specific for this gauge, 
which guarantees that 
quantum corrections to 1PI correlation functions of ghosts vanish 
at zero momentum. This fact is particularly useful when applied to 
the functions representing  the corrections to 
BRST transformations. 
Because  in the Landau gauge the (anti)ghosts are massless to all orders,
the function ${\Om}(q^2)^{\al}_{\ \be}$ which parametrizes the 
(renormalized) 1PI two-point ghost-antighost function 
\eqs{\label{Eq:Gh-AntiGh}
\left.\derf{}{\hat{\om}^\be(q)}
\derf{}{\hat{\anti{\om}}_\al(p)}\Gamma\right|_{0}
&=&
(2\pi)^4 \delta^{(4)}(q+p)
\left\{-q^2 {\Om}(q^2)^{\al}_{\ \be}\right\}\,,
}
must (in our conventions) have the form
\koment{str.MPW:28}
\eq{\label{Eq:Om-Expansion}
{\Om}(q^2)^{\al}_{\ \be}=-\de^\al_{\ \be}+\cO(\hb)\,.
}
Existence of unbroken gauge symmetries means that there are vectors
$\Th^\be$ such that 
\footnote{
 Recall that $v$ is the complete (and renormalized) VEV, as 
in Eq. \refer{Eq:VEV-expansion}.} 
\eq{\label{Eq:Th-cond}
\Th^\be\,\TS_\be\, v=0\,.
}
From the antighost identity combined with a Slavnov-Taylor identity
it then follows (see the discussion below the formula \refer{Eq:STid1prime} 
in Sec. \ref{Sec:Sub:STids}) that
\koment{str.MPW:105} \kK 
\eq{\label{Eq:Massless-Eigenvectors}
\lim_{q^2\to 0} 
\left\{
M^2_V(q^2)_{\be\al}\,{\Om}(q^2)^{\al}_{\ \ga}\,\Th^\ga
\right\}
=0\,,
}
which means that the vectors $\ze_{V[\la_r]}$ corresponding to 
massless gauge bosons are up to normalization given by 
$\ze^{\al}_{V}\propto{\Om}(0)^{\al}_{\ \ga}\,\Th^\ga$. 

The identity \refer{Eq:Massless-Eigenvectors} is interesting in its own right, 
as it immediately shows, for instance,  that the photon in the SM remains 
massless to all orders. It will also play an important role   
in the analysis of the unphysical asymptotic states in Sec. 5, in particular in    
showing that they form Kugo-Ojima quartets.

\section{Results in a general renormalizable model}\label{Sec:Res} 

In this section we give one-loop expressions for matrices 
$M_S^2(p^2)$ and $M_V^2(p^2)$, cf. Eqs. \refer{Eq:Gamma2-scalar} and 
\refer{Eq:Gamma2-vector}, in a general renormalizable 
gauge field theory model.

\subsection{Parametrization of the action } \label{Sec:SubSec:Action}

We assume that
the gauge group is a direct product of an arbitrary number of compact simple 
Lie groups and $U(1)$ groups and  that the gauge fields are 
coupled to scalar and fermionic fields  forming arbitrary representations
of the gauge group (we assume the representation formed by fermions is 
nonanomalous). 
We work with real scalars $\phi^j$, real vectors $A^\al_\mu$ 
and Weyl fermions $\chi^a_A$ (together with their complex conjugates 
$\anti{\chi}^a_{\dot{A}}$). 
Recall that the fields $\phi^j$ are assumed to have all vanishing VEV, and 
are related to
``the symmetric phase'' field by
$\phi^j_{\rm sym}=\phi^j+v^j$. The classical gauge-invariant action $I_0^{GI}$ 
is 
the integral of the Lagrangian density 
(we follow the conventions of \cite{ChLM}) 
\begin{eqnarray}\label{Eq:LagrTreeGI}
\mathcal{L}^{GI}_0=-\frac{1}{4}
\delta_{\alpha \beta}F^\alpha_{\phantom{a}\mu\nu}F^{\beta~\!\!\mu\nu}+
\frac{1}{2}\delta_{ij}(D_\mu\phi)^i(D^\mu\phi)^j\!
-\mathcal{V}(\phi+v)
+\mathcal{L}^{F}_0\,.
\end{eqnarray}
Lorentz indices are lowered/raised  with the aid of the Minkowski metric 
$\eta_{\mu\nu}$. 
The potential $\mathcal{V}(\phi_{\rm sym})$ is a fourth order polynomial 
parametrized below by the following coupling constants and mass parameters:
\begin{eqnarray}\label{Eq:RegFey:LamRhoMs}
\lambda_{ijkl} = \mathcal{V}^{(4)}_{ijkl}(v_{(0)}),\phantom{aaa}~
\rho_{ijk} = \mathcal{V}^{\prime\prime\prime}_{ijk}(v_{(0)}),\phantom{aaa}~
m_{Sij}^{2}
=\mathcal{V}^{\prime\prime}_{ij}(v_{(0)}),
\phantom{a}
\end{eqnarray}
where $v_{(0)}$, determined by the condition 
$\mathcal{V}^\prime_{i}(v_{(0)})=0$,  is the first term of the expansion 
\refer{Eq:VEV-expansion} of the complete VEV. 
The covariant derivative of scalars is given by \refer{Eq:CovDerSc}, and
the explicit form of $F^\alpha_{\phantom{a}\mu\nu}$ is
\begin{eqnarray}\nn
F^\alpha_{\text{ }\,\mu\nu}=\partial_\mu A^\alpha_\nu\!
-\partial_\nu A^\alpha_\mu\!+e^\alpha_{\phantom{a}\beta\gamma}A^\beta_\mu A^\gamma_\nu\,,
\end{eqnarray} 
with real structure constants $e^\gamma_{\phantom{a}\alpha\beta}\,$ 
which include the gauge couplings
and are
defined by the relation
$
[{\TS}_\alpha,~\!{\TS}_\beta]={\TS}_\gamma~\! e^\gamma_{\phantom{a}\alpha\beta}
$).

The fermionic part of the Lagrangian density reads 
\eqs{\label{Eq:LagrWeylOg}
\mathcal{L}^{F}_0
&=&
i\, \de_{ab}\,\anti{\chi}^a\anti{\sigma}^\mu\partial_\mu\chi^b
+i\, \TF_{\alpha ab}\,\anti{\chi}^a\anti{\sigma}^\mu\chi^b
A^\alpha_\mu
+{}\nn\\
{}&{}&
-
\frac{1}{2}\left(
\,\,\anti{\!\!M\!}_{F\,ab}\,\chi^{a}\chi^{b}+
\,\,\anti{\!\!M\!}^{\,\,\star}_{F\,ab}\,\anti{\chi}^{a}\anti{\chi}^{b}
\right)
-
\frac{1}{2}\phi^j\left(
{Y}_{j ab}\,\chi^{a}\chi^{b}+
{Y}^{\star}_{j ab}\,\anti{\chi}^{a}\anti{\chi}^{b} 
\right),
}
where $SL(2,\bC)$ indices have been suppressed 
\eq{\nn
\chi^{a}\chi^{b}\,\equiv\, \chi^{a\,A}\,\chi^{b}_A\,,
\qquad \quad
\anti{\chi}^a\ \! \anti{\sigma}^\mu {\chi^b}
\,\equiv\, 
\anti{\chi}^a_{\dot{B}}\,{\anti{\sigma}^\mu}^{\dot{B}A}\, \chi^b_A\,,
}
etc. Here $\TF_{\alpha a b} = - \TF_{\alpha b a}^\star$ are matrix elements 
of anti-Hermitian gauge-group generators 
$([{\TF}_\alpha,~\!{\TF}_\beta]={\TF}_\gamma~\! 
e^\gamma_{\phantom{a}\alpha\beta})$, 
while $\YF_{j a b}=\YF_{j b a}$ are elements of symmetric Yukawa 
matrices $Y_j$. The fermionic matrix $\,\anti{\!\!M\!}_{F\,}$ depends on $v$
\eq{\label{Eq:M_F}
\,\anti{\!\!M\!}_{F\,ab}\,={\sM}_{ab}+{Y}_{j ab}\,v^j\,.
}
The coefficients ${\sM}_{ab}$, ${Y}_{j ab}$, etc. are, of course,
constrained by the gauge (and global) symmetries.\footnote{These constraints 
imply, in particular, that ${\sM}_{ab}\equiv 0$ in the SM.}

In calculating diagrams we find it more convenient to work with 
four-component Majorana spinors  $\psi^a$  
\eq{\label{Eq:Majo-spinor}
\psi^{a}
=
\left[
\begin{array}{l}
\chi^{a}_A\\[4 pt]
\anti{\chi}^{a\dot{A}}\\
\end{array}
\right]
\,.
}
For this reason solid lines in diagrams displayed below represent 
Majorana fields (and are, consequently, non-oriented). We therefore rewrite 
the fermionic part of the Lagrangian in the following form (discarding 
total derivatives) 
\koment{por.\cite{OnFeRu}} 
\eqs{\label{Eq:LagrTreeGI-Majo}
\mathcal{L}^{F}_0&=&
+\frac{1}{2}~\!\bar{\psi}^a\!\left\{\delta_{ab}\,
i\,\gamma^\mu\, \pa_\mu\psi^b
-\,
\left(  \,\,\anti{\!\!M\!}_{F\,ab}\, P_L
+\,\,\anti{\!\!M\!}^{\,\,\star}_{F\,ab}\, P_R \right)
\psi^b 
\right\}
+\nn\\[4pt]&{}&
+\frac{1}{2!}\,
i\,A^\alpha_\mu\,\,
\bar{\psi}^a\,\gamma^\mu 
\left(  \TF_{\alpha a b}\, P_L+\TF^\star_{\alpha a b}\, P_R \right)
\psi^b+\nn\\[4pt]
&{}& 
-
\frac{1}{2!}\,\phi^j\,
\bar{\psi}^a 
\left(  \YF_{j a b}\, P_L+\YF^\star_{j a b}\, P_R \right)
\psi^b \,,
}
where $P_{L,R}$ are chiral projections and 
$\bar\psi\equiv\psi^\dagger\,\ga^0 = \psi^\rmt\,\cC$ 
is the Dirac-conjugate field.

To generate Green's functions of the quantum theory, the classical action
$I_0^{GI}$ is supplemented with a gauge fixing term  and with the ghost
fields action, what leads to the BRST invariant tree-level action 
\begin{eqnarray}
I_0=I_0^{GI}+I_0^{Rest}=\int\!{\rm d}^4 x~\! (\mathcal{L}^{GI}_0
+\mathcal{L}^{Rest}_0)~\!,\label{Eq:I_0=I_0^GI+I_0^Rest}
\end{eqnarray}
where ${\cal L}_0^{Rest}$ depends on the Nakanishi-Lautrup fields $h_\beta$ and 
the ghost and antighost fields $\omega^\alpha$ and $\overline{\omega}_\alpha$; 
in order to control quantum corrections to the gauge transformations one also 
introduces terms with the external sources (antifields) 
\cite{BRS1,BRS2,PiguetSorella} $K_i$, $\bar{K}_a$, $K^\mu_\al$ and $L_\al$:
\begin{eqnarray}\label{Eq:LagrTreeRest}
\mathcal{L}^{Rest}_0=
s\!\left(-\overline{\omega}_\alpha \,
\partial_\mu A^{\alpha\mu} 
\right)
+ L_\alpha\, s({\omega}^\alpha)+K_i\, s(\phi^i)
+\bar{K}_a\, s(\psi^a)+K^\mu_\alpha\, s(A^\alpha_\mu)~\!,\phantom{aa}
\end{eqnarray}
where the action on fields of the BRST differential $s$ is
given by \cite{BRS1,BRS2,PiguetSorella}
\begin{eqnarray}\label{Eq:BRS}
&s(\phi^i)=\omega^{\alpha}
\left[\TS_\alpha(\phi\!+\!v)   
\right]^i,\qquad\qquad\qquad
&s(\psi^a) = \omega^{\alpha}([\TF_{\al}]^a_{\ b}P_L
                            +[\TF_{\al}^\star]^a_{\ b} P_R)\psi^b,
\nonumber\\[5pt]
&s(A^\gamma_\mu)=-\partial_\mu\omega^{\gamma}\!+\!
e^{\gamma}_{\phantom{a}\alpha \beta}~\!\omega^{\alpha}A^\beta_\mu,
\qquad\qquad
&s(\omega^\alpha)=\frac{1}{2}~\!
e^{\alpha}_{\phantom{a}\beta \gamma}~\!\omega^\beta\omega^\gamma,
\nonumber\\[5pt]
&s(\overline{\omega}_\alpha) =h_\alpha,
\qquad\qquad\qquad\qquad\qquad\ \ \ 
&s(h_\alpha)=0.
\end{eqnarray}
The ``flavor" indices on constant tensors parameterizing the action are 
raised$\,$/ lowered with the aid of standard (Kronecker delta) metrics that 
appear in Eq. \refer{Eq:LagrTreeGI}. In particular,  
$[\TF_{\al}]^a_{\ b}\equiv \TF_{\al a b}$, etc.

The first term in ${\cL}^{Rest}_0$ represents the gauge-fixing and ghosts 
Lagrangian in the Landau gauge. 
Setting $s(K_i)=s(\bar{K}_a)=s(K^\mu_\alpha)=s(L_\alpha)=0$ makes the action
$I_0^{Rest}$ a BRST-exact functional: $I_0^{Rest}=s\,W $. The complete
action  \refer{Eq:I_0=I_0^GI+I_0^Rest} is then BRST-invariant, $sI_0=0$,  
due to the nilpotency $s^2=0$.

All fields and parameters introduced above are understood as renormalized 
quantities. In other words, the one-loop action has the form 
\eq{\nn
I_1 = I_0 - \hb\, \de\!I_1\,, 
}
and contains counterterms $\de\!I_1$; in the $\anti{{\rm MS}}$ scheme of 
dimensional regularization each term in $\de\!I_1$ is a singular part of an
appropriately chosen 1PI one-loop effective vertex.  
We also note that $I_0$ 
itself contains terms with all powers of $\hb$, as it depends on the 
complete (but renormalized) VEV $v^i$ (cf. the formula 
\refer{Eq:VEV-expansion}). 
\kCC

\vskip0.2cm

\subsection{One-loop self-energies}

The formulae  collected in this section are valid in 
the Landau gauge, and are renormalized in the $\anti{{\rm MS}}$ scheme \cite{MSbar}
of dimensional regularization
with the anticommuting $\gamma^5$ matrix
which in non-anomalous theories is consistent at the one-loop order and 
preserves chiral gauge symmetries. \footnote{Since we use the dimensional regularization 
(rather than dimensional reduction), 
additional finite counterterms 
have to be adjusted in supersymmetric models to restore supersymmetry. 
We do not give explicit expressions for them in what follows. 
}
All loop integrals associated with the diagrams listed in this 
section were checked against the FeynCalc \cite{FC} results.

Without loss of generality we assume that the components $\phi^j$, $A^\al_{\mu}$ 
and $\chi^a_A$ are chosen in such a way that the tree-level mass-squared 
matrices are diagonal (and nonnegative) 
\eq{\nn
\big[m^2_{Sij}\big] =  \diag(m^2_{S1},\ldots)\,,
\qquad 
\big[m_{V\al\be}^2\big]   =  \diag(m^2_{V1},\ldots)\,,
}
(cf. Eqs. \refer{Eq:RegFey:LamRhoMs} and \refer{Eq:m_V}) 
and \kO 
\footnote{In realistic model there are Dirac particles and it is more 
convenient to keep $M_{F}$ non-diagonal, diagonalizing only the product 
$M_{F}M_{F}^\star$.}
\eq{\label{Eq:MF-kw-tree}
M_{F}M_{F}^\star = \diag(m_{F1}^2\,,\ldots)\,,
}
where (cf. Eqs. \refer{Eq:M_F} and \refer{Eq:VEV-expansion})
\eq{\nn
M_{F\,ab}\,={\sM}_{ab}+{Y}_{j ab}\,v^j_{(0)}\,.
}
In particular, the pole masses $m^2_{S(\ell)}$ and $m^2_{V(\la)}$ are $\cO(\hb)$ 
perturbations of the appropriate tree-level masses $m^2_{Sj}=m^2_{Sjj}$ and
$m^2_{V\al}=m^2_{V\al\al}$. 

\begin{figure}[pt]
\centering
\includegraphics[width=0.7\textwidth]{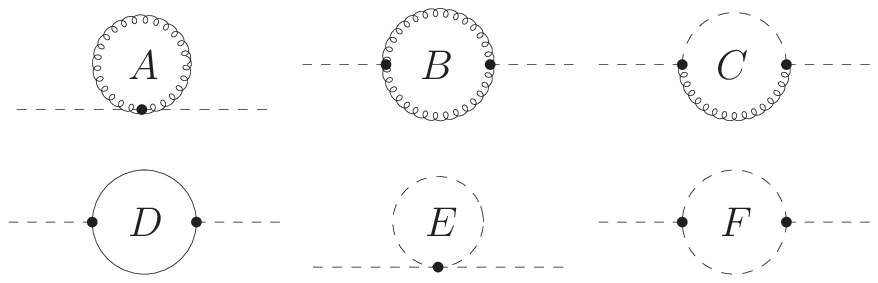}
\caption{One-loop contributions to the two-point functions 
$\tig_{i_1 i_2}(p,-p)$ of scalar fields. 
Solid lines represent Majorana fermions \refer{Eq:Majo-spinor} 
(see the Lagrangian \refer{Eq:LagrTreeGI-Majo}). 
At order $\cO(\hb)$ to $\tig_{i_1 i_2}(p,-p)$ contributes also 
the correction to the VEV, cf. Eqs. \refer{Eq:MV-and-MS}.
}
\label{Rys:PhiPhi}
\end{figure}

\begin{figure}[pt]
\centering
\includegraphics[width=0.7\textwidth]{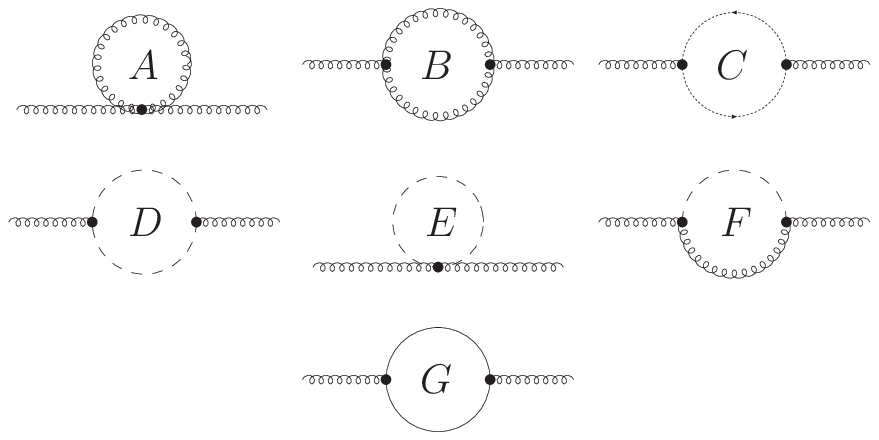}
\caption{
One-loop contributions to the self-energy
$\tig^{\mu\nu}_{\alpha\beta}(p,-p)$ of vector fields.
Diagram $C$ represents the ghost-antighost loop.  At order $\cO(\hb)$ to 
$\tig^{\mu\nu}_{\alpha\beta}(p,-p)$ contributes also the 
correction to the VEV, cf. Eqs. \refer{Eq:MV-and-MS}. 
}
\label{Rys:AA}
\end{figure}

One-loop 1PI diagrams contributing to $M_S^2(p^2)$ and $M_V^2(p^2)$ in the 
Landau gauge are shown in Figs. \ref{Rys:PhiPhi} and \ref{Rys:AA}, respectively. 
Finite (minimally subtracted) parts of these contributions are denoted by 
$-(4\pi)^{-2}\De^{\!S}(p^2)$ and $+(4\pi)^{-2}\Si^V(p^2)$. In addition, the 
quantum correction  $v_{(1)}$ to the VEV in Eq. \refer{Eq:VEV-expansion} 
contributes to both matrices $M_S^2(p^2)$ and $M_V^2(p^2)$; thus $(s\equiv p^2)$ 
\koment{str.OC75$_A$ i OC46}
\eqs{\label{Eq:MV-and-MS}
M_V^2(s)_{\al\be} &=& 
(\TS_\al\, v_{(0)})^{\rmt}\!(\TS_\be\, v_{(0)})
\!-\!
\hb\,
v_{(0)}^{\,\rmt}\{\TS_\al\,,\ \TS_\be\,\} v_{(1)}
\!+\! \frac{\hb}{(4\pi)^2}\,\Si^V\!(s)_{\al\be} 
\!+\!\cO{(\hb^2)}\,,\nn
\\[5pt]
M_S^2(s)_{ij} &=& \cV^{\prime \prime}_{ij}(v_{(0)}+\hb\,v_{(1)}) 
- \frac{\hb}{(4\pi)^2}\,\De^{\!S}(s)_{ij} 
+\cO{(\hb^2)}\,.
}
Matrices $\De^{\!S}(p^2)$ and $\Si^{V}\!(p^2)$ can be expressed in terms 
of the (minimally subtracted) one-loop functions $a^R$ and $b^R_0$ 
in the dimensional regularization (see e.g. \cite{Chank})
\eqs{\label{Eq:Fun-aR}
&{}&
a^R(m)
=
m^2\left\{\ln\frac{m^2}{\bar\mu^2}-1\right\}
\,, \nn\\[6pt]
&{}&
 b_0^R(p^2,m_1,m_2)
=
\int_0^1{\rm d}x\ln\frac{x(x-1)p^2+(1-x)m_1^2+x\,m_2^2-i\,0}{\bar\mu^2}
\,,\  
}
where $\bar{\mu}$ is the renormalization scale of the $\anti{\rm MS}$ scheme, related to the usual \mbox{'t Hoot} 
mass unit $\mu_H$ via  
$\bar{\mu}\equiv\mu_H \sqrt{4\pi}\, e^{-{\gamma_E}/{2}}$. \\

\begin{figure}[pt]
\centering
\includegraphics[width=0.7\textwidth]{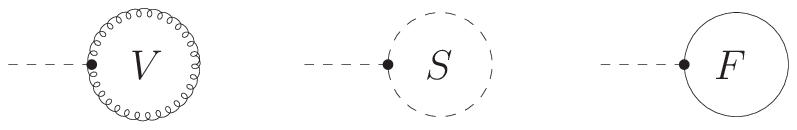}
\caption{One-loop contributions to the scalar one-point functions 
$\tig^{}_{i}(0)$ necessary to determine quantum-corrected VEV
(cf. Eq.~\refer{Eq:TadCond}). 
}
\label{Rys:Phi}
\end{figure}

We begin with one-loop corrections to scalar tadpoles which are shown in Fig. 
\ref{Rys:Phi}. They yield the following equation for $v_{(1)}$ 
\koment{str.OC46,OC44}  \kPP 
\eqs{\label{Eq:TadCond}
0=
-\cV^{\prime }_{i}(v_{(0)}+\hb\,v_{(1)}) 
+
\frac{\hb}{(4\pi)^2}
\Big\{&{}&
3\sum_{\al j} [\TS_\al^2]_{ij}\,v_{(0)}^j
\left[a^R(m_{V\al})+\frac{2}{3}m_{V\al}^2\right]
+\nn\\[4mm]
&{}&\hspace*{-155 pt}
-\frac{1}{2}\sum_j \rho_{ijj}\,a^R(m_{Sj})
+\sum_{bc} ( M_{Fbc}^{\phantom{\star}} Y_{icb}^\star
             \!+\!M_{Fbc}^\star Y_{icb}^{\phantom{\star}}) 
            a^R(m_{Fb})
\Big\}
+\cO(\hb^2)\,.
}
The contribution $\De^{\!S}(p^2)_{ij}$ to the scalar two-point function reads ($s\equiv p^2$)
\kP 
\koment{str.OC32,OC41,OC39,OC14,OC31; por OC45}
\eqs{\label{Eq:De_S}
\De^{\!S}(s)_{ij}
&=&
3\sum_{\al} [\TS_\al^2]_{ij}
       \left[a^R(m_{V\al})+\frac{2}{3}m_{V\al}^2\right]
-4\sum_{\al k}\TS_{\al i k}\,\TS_{\al k j}\,S_C(s,m_{V\al},m_{Sk})
+\nn\\[4 pt]
&{}&
-\frac{1}{2}\sum_{\al \be}
   [\{\TS_\al\,,\ \TS_\be\}v_{(0)}]_i\,
   [\{\TS_\al\,,\ \TS_\be\}v_{(0)}]_j\,
   S_B(s,m_{V\al},m_{V\be})
+\nn\\[4 pt]
&{}&
+\sum_{abcd}
\Big\{
     (Y_{iab}^{\phantom{\star}}\,\de_{bc}^{\phantom{\star}}
      \,Y_{jcd}^{\star}\,\de_{da}^{\phantom{\star}}
      +{\rm cc}.)\,
     S_D(s,m_{Fb},m_{Fd})+
\nn\\[4 pt]
&{}& \phantom{ +\sum_{abcd}\Big\{ }
+
  (Y_{iab}^{\phantom{\star}}\,M_{Fbc}^{\star}\,
   Y_{jcd}^{\phantom{\star}}\,M_{Fda}^{\star}+{\rm cc}.)\,
   b_0^R(s,m_{Fb},m_{Fd})
\Big\}
+\nn\\[4 pt]
&{}&
-\frac{1}{2}\sum_k \la_{ijkk}\,a^R(m_{Sk})
-\frac{1}{2}\sum_{kn} \rho_{kin}\,\rho_{njk}\,b^R_0(s,m_{Sn},m_{Sk})
\,,
} 
where ${\rm cc}.$ indicates the complex conjugation of the preceding term.  
The following combinations of basic one-loop functions have been introduced \kQ
\eqs{\label{Eq:S_C}
S_C(s,m_{V},m_{S})
&=&
\frac{1}{4} 
\Bigg\{ \ 
a^R\!\left(m_V\right)-a^R\!\left(m_S\right)+\left(s-m_S^2\right) \frac{a^R\!\left(m_V\right)}{m_V^2}
+\nn\\[4pt]
&{}&\qquad
+\left(2 s+2 m_S^2-m_V^2\right)
   b_0^R\!\left(s,m_V,m_S\right)
+\nn\\[4pt]
&{}&\qquad
- 
\frac{\left(s-m_S^2\right)^2}{{m_V^2}}
\left[
{b_0^R\!\left(s,m_V,m_S\right)-b_0^R\!\left(s,0,m_S\right)}
\right]
 \Bigg\}
\,,\ \ 
}
\kR
\eqs{\label{Eq:S_B}
S_B(s,m_1,m_2)&=&
2
+\frac{1}{4\, m_1^2\, m_2^2}
 \bigg\{m_2^2\, a^R\!\left(m_1\right)
       +m_1^2\, a^R\!\left(m_2\right)
       +s^2\, b_0^R(s,0,0) +
\nn\\[4pt]
&{}&
       -(m_2^2-s)^2\, b_0^R\!\left(s,0,m_2\right)
       -(m_1^2-s)^2\, b_0^R\!\left(s,m_1,0\right)
+
\nn\\[4pt]
&{}&
 +\big[\, (m_2^2-s) (m_1^2-s)
        +m_1^2 (m_1^2-s)
        +m_2^2 (m_2^2-s)+ 
\nn\\[4pt]
&{}&\quad\ 
        +9\, m_1^2\, m_2^2\,
  \big]\, 
   b_0^R\!\left(s,m_1,m_2\right)
\bigg\}\,,
}
and \kT \kS 
\eqs{\label{Eq:S_D}
S_D(s,m_1,m_2)&=&
a^R(m_2)+\left\{m_1^2-\frac{s}{2}\right\}\,b_0^R(s,m_1,m_2)\,.
}
The reality conditions \refer{Eq:RealityCond-scalar} are violated whenever 
$b_0^R$ has a non-vanishing imaginary part.  

Contributions of massless vectors is obtained by taking in the formulae
\refer{Eq:S_C}-\refer{Eq:S_B} the limits $m_V\to 0$. 
We also note that $S_C(0,m_{V},m_{S})=0$.

\vspace*{20 pt}

\begin{figure}[pt]
\centering
\includegraphics[width=0.3\textwidth]{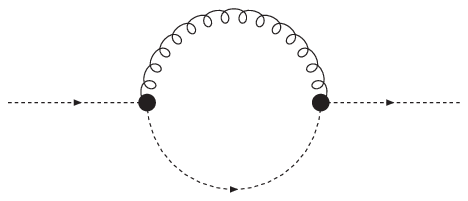}
\caption{One-loop contribution to the ghost-antighost self-energy.
\hspace*{30pt}
}
\label{Rys:KmuOm} 
\end{figure}

\begin{figure}[pt]
\centering
\includegraphics[width=0.9\textwidth]{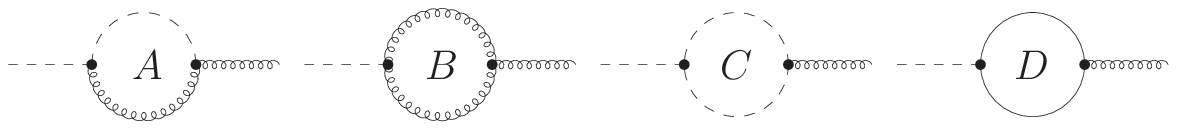}
\caption{One-loop contributions to $\tig^{\ \!\nu}_{i\beta}(p,-p)$.
At order $\cO(\hb)$ to $\tig^{\ \!\nu}_{i\beta}(p,-p)$ contributes also
the correction to the VEV (cf. Eq.~\refer{Eq:MixSV}).}
\label{Rys:PhiA}
\end{figure}


The contribution $\Si^{V}\!(s)_{\al\be}$ to the two-point function of vector 
fields reads \kV 
\kU 
\koment{str.OC64,OC68,OC70,OC75}
\eqs{\label{Eq:Si_V}
\Si^{V}\!(s)_{\al\be}
&=&
\sum_{\ep\ga} e^{\ep}_{\ \al\ga}\, e^{\ga}_{\ \be\ep}\,
V_{\!{\scriptscriptstyle ABC }}(s,m_{V\ep},m_{V\ga})
+\sum_{ij}\TS_{\al i j}\,\TS_{\be j i}\,
V_{\!{\scriptscriptstyle DE }}(s,m_{Si},m_{Sj})
+\nn\\[4 pt]
&{}&
+\sum_{\ga i}
   [\{\TS_\ga\,,\ \TS_\al\}v_{(0)}]_i\,
   [\{\TS_\ga\,,\ \TS_\be\}v_{(0)}]_i\,
   V_F(s,m_{V\ga},m_{Si})
+\nn\\[4 pt]
&{}&
+\sum_{abcd}
\Big\{\,
     (\TF_{\al ab}\,\de_{bc} \,\TF_{\be cd}\,\de_{da}
      +{\rm cc}.)\,
     V_G(s,m_{Fb},m_{Fd})+
\nn\\[4 pt]
&{}& \phantom{ +\sum_{abcd}\Big\{ }
-
  (\TF_{\al ab}^{\phantom{\star}}\,M_{Fbc}^{\star}\,
   \TF_{\be cd}^{{\star}}\,M_{Fda}^{\phantom{\star}}+{\rm cc}.)\,
   b_0^R(s,m_{Fb},m_{Fd})
\Big\}
\,.
} 
The functions $V$ are defined 
in terms of the auxiliary function \kX 
\koment{OC69 i OC66$_A$}
\eqs{
A(s,m_1,m_2)&=&
\frac{m_1^2-m_2^2}{12 s} 
  \left[a^R(m_1)-a^R(m_2)-\left(m_1^2-m_2^2\right) b_0^R(s,m_1,m_2)  \right]
+\nn\\[4pt]
&{}&
+\frac{1}{12} \left[2 m_1^2+2 m_2^2-s\right] b_0^R(s,m_1,m_2)
+\nn\\[4pt]
&{}&
+\frac{1}{12} \left[a^R(m_1)+a^R(m_2)\right]
+\frac{s}{18}-\frac{1}{6} \left(m_1^2+m_2^2\right)
\,,
} 
and read \kZ 
\koment{str.OC65 i OC66$_B$ -- $V_{\!{\scriptscriptstyle ABC }}$ to Gtot}
\eqs{
V_{\!{\scriptscriptstyle ABC }}(s,m_{1},m_{2})
&=&
\frac{5 s}{3}-\left\{a^R(m_1)+a^R(m_2)\right\}
+\nn\\[4pt]
&{}&\hspace*{-75 pt}
+\frac{1}{2 m_1^2 m_2^2}\Big\{
 \left[m_1^4+10\, m_1^2\, m_2^2+m_2^4+10 \left(m_1^2+m_2^2\right) s+s^2\right]
   A(s,m_1,m_2)
+\nn\\[4pt]
&{}&\hspace*{-75 pt} \phantom{   +\frac{1}{2 m_1^2 m_2^2}\Big\{   }
 -\left[m_1^4+10\, m_1^2\, s+s^2\right] A(s,m_1,0)
+\nn\\[4pt]
&{}&\hspace*{-75 pt} \phantom{   +\frac{1}{2 m_1^2 m_2^2}\Big\{   }
 -\left[ m_2^4+10\, m_2^2\, s+s^2\right] A(s,0,m_2)
+\nn\\[4pt]
&{}&\hspace*{-75 pt} \phantom{   +\frac{1}{2 m_1^2 m_2^2}\Big\{   }
 +\left[s^2-2\, m_1^2\, m_2^2\right] A(s,0,0)
\,\Big\}\,,
}
\kW
\eq{
V_{\!{\scriptscriptstyle DE }}(s,m_{1},m_{2})
=2\, A(s,m_1,m_2)-\frac{1}{2}a^R(m_1)-\frac{1}{2}a^R(m_2)\,,
}
\koment{str.OC70:} 
\eq{
V_{F}(s,m_{V},m_{S})
=
b_0^R(s,m_V,m_S)
-\frac{1}{m_V^2}
\big\{A(s,m_V,m_S) - A(s,0,m_S)  \big\}
\,,
}
and 
\koment{str.OC75, tam to jest $F$} 
\eqs{
V_{G}(s,m_{1},m_{2})
&=&
\frac{1}{2}
\bigg\{
a^R(m_1)+a^R(m_2)
+\left(m_1^2+m_2^2-s\right)b_0^R(s,m_1,m_2)
+\nn\\[4 pt]
&{}&
\quad\ 
-4 A(s,m_1,m_2)
\bigg\}
\,.
}
Again, in contributions of massless gauge bosons the limit
$m_V\to 0$ is understood 
\kY 
and, again, the imaginary part of 
$b_0^R$ violates the reality of $M_V^2(s)$.\\

It will be useful to have also the one-loop correction to the ghost-antighost 
self-energy which at one-loop is given by the single
diagram of Fig. \ref{Rys:KmuOm}. It gives the following factor 
${\Om}(q^2)^{\al}_{\ \ga}$ 
in the two-point function \refer{Eq:Gh-AntiGh} 
\eq{\label{Eq:Omega}
{\Om}(q^2)^{\al}_{\ \ga}=-\de^\al_{\ \ga}
+
\frac{\hb}{(4\pi)^2}
\sum_{\be\ep}e^\al_{\ \be \ep}\,e^{\ep}_{\ \be \ga}\,
\cH(q^2,m_{V\be})+
\cO(\hb^2)\,,
}
where 
\eqs{
\cH(s,m)&=&
\frac{1}{2}\, b_0^R(s,m,0)
-\frac{1}{4 s} \left\{m^2\, b_0^R(s,m,0)-a^R(m)\right\}
+\nn\\[4pt]
&{}&
-\frac{1}{4 m^2} \left\{ s\left[b_0^R(s,m,0)-b_0^R(s,0,0)\right]-a^R(m)\right\}
\,.
}
Notice that because $a^R(m)= m^2\, b_0^R(0,m,0)$, the function
$\cH(s,m)$ does not have a pole at $s=0$.
\koment{str.OC45 } 
\kFF 

Finally, 1PI diagrams contributing to the scalar-vector two-point function 
at one-loop are shown in Fig. \ref{Rys:PhiA}. There is also an additional 
contribution originating from a correction to the VEV; schematically we 
can write \koment{str. OC88} 
\eq{\label{Eq:MixSV}
\widetilde{\Gamma}_{j\be}^{\ \nu}(p,-p)
= 
-i\, p^\nu\,\TS_{\be j k} (v_{(0)}^k+\hb\,v_{(1)}^k) 
+[{\rm Fig.\, \ref{Rys:PhiA} }]
+\cO(\hb^2)
\,.
}
We do not need the expression for $\widetilde{\Gamma}_{j\be}^{\ \nu}(p,-p)$ 
(just as we do not need the expression for the matrix $\sL(q^2)$ in Eq. 
\refer{Eq:Gamma2-vector}). Nevertheless, the fermionic contribution to this 
function (i.e. the finite part of diagram $D$ in Fig. \ref{Rys:PhiA}) 
will turn out to be useful in Sec. \ref{Sec:SubSec:Examples:Maj} 
\kBB 
\eq{\label{Eq:MixSV-FermLoop}
\widetilde{\Gamma}_{j\be}^{\ \, \nu}(p,-p)_{[\ref{Rys:PhiA}.D]}
=
\hb\,\frac{2\,i\,p^\nu}{(4\pi)^2}\,
\sum_{abcd}
\Big\{
  (Y_{jab}^{\phantom{\star}}\,M_{Fbc}^{\star}\,
   \TF_{\be cd}^{{\star}}\,\de_{da}^{\phantom{\star}}+{\rm cc}.)\,
   J(p^2,m_{Fd},m_{Fb})
\Big\}\,,
}
where 
\eqs{\nn
J(s,m_1,m_2)&=&
\frac{1}{2s}
\left\{
a^R(m_1)-a^R(m_2)+
\left[m_2^2-m_1^2-s\right] b_0^R(s,m_1,m_2)
\right\}\,.
}
For future reference, we note that contributions of fermionic loops of Figs. 
\ref{Rys:PhiPhi}.D, \ref{Rys:PhiA}.D and \ref{Rys:Phi}.F are related 
as follows 
\koment{str.OC23}
\eq{\label{Eq:nie-WT-fermion}
0=-i\, p_\mu\, \widetilde{\Gamma}_{j\al}^{\  \mu}(p,-p)_{[\ref{Rys:PhiA}.D]}
+(\TS_\al\,v_{(0)})^k\, 
  \widetilde{\Gamma}_{j k}(p,-p)_{[\ref{Rys:PhiPhi}.D]}
+[\TS_\al]^{k}_{\ j} \widetilde{\Gamma}_{k}(0)_{[\ref{Rys:Phi}.F]}
\,,
}
where $(4\pi)^2\,\widetilde{\Gamma}_{i j}(p,-p)_{[\ref{Rys:PhiPhi}.D]}$ is given 
by the fourth term in Eq. \refer{Eq:De_S} (with $s\equiv p^2$), while 
$\widetilde{\Gamma}_{i}(0)_{[\ref{Rys:Phi}.F]}$ represents the term with $a(m_{F})$ 
on the right-hand-side of Eq. \refer{Eq:TadCond}. \kDD 

It is perhaps worth stressing, for completeness, that the matrix $\sL(q^2)$ 
in Eq. \refer{Eq:Gamma2-vector} as well as the two-point function 
\refer{Eq:MixSV} are (at one-loop order) entirely fixed in terms of the 
matrices $M_V^2(p^2)$,   $M_S^2(p^2)$ and \refer{Eq:Omega} by the 
gauge-symmetry (see a discussion below Eq. \refer{Eq:would-be-Gold-Th} 
in Sec. \ref{Sec:Sub:STids}). Thus, the above results give the complete 
set of bosonic  two-point functions in the Landau gauge.

\section{Examples}\label{Sec:Examples} 

 \subsection{Corrections to electroweak mixing} 
   \label{Sec:SubSec:Examples:Z-A-mixing} 
\kTTTT
The matrix $M^2_V(p^2)$ 
parameterizing the two-point function \refer{Eq:Gamma2-vector} of the SM
vector fields is block-diagonal. It has two $2\times2$ blocks corresponding to 
the pairs $(Z_\mu,\, A_\mu)$, $(W^1_\mu,\, W^2_\mu)$ and one $8\times8$ block 
corresponding to gluons; in the last two blocks the matrix $M^2_V(p^2)$ is 
proportional to the identity matrix (see e.g. \cite{WeinT2}). 
Here $Z_\mu$ and $A_\mu$ denote, as usually, the eigenfields of the tree-level 
mass-squared matrix with eigenvalues $m^2_Z$ and 0. 
The generic formulae of Sec. \ref{Sec:Res} yield the following one-loop 
expression for the $(Z,\, A)$ block 
of $M^2_V(0)$ in the $\anti{{\rm MS}}$ scheme \koment{str.OC89}
\eq{\label{Eq:M_V--ZA}
M^2_V(0)
=
\left[
\begin{array}{cc}
m_Z^2+\hb\,\ca & \hb\, \cb \\
\hb\,\cb  & 0
\end{array}
\right]+\cO(\hb^2)\,,
}
with  
\eqs{\nn
\ca
&=&
\frac{1}{(4\pi)^2 v_{H(0)}^2}
\bigg\{
\frac{6 m_H^2 m_Z^4}{m_H^2\!-\!m_Z^2}
  \left[\ln\! \left(\frac{m_H}{\bar{\mu}}\right)\!-\!\frac{5}{12}\right]
\!-\!\frac{6 m_Z^6}{m_H^2\!-\!m_Z^2} 
  \left[\ln\!\left(\frac{m_Z}{\bar{\mu}}\right)\!-\!\frac{5}{12}\right]
\!+\!\nn\\[4pt]
&{}&
+\frac{1}{2} m_H^2 m_Z^2
\!-\!\left(24 m_W^4\!-\!12 m_W^2 m_Z^2\right) 
\left[\ln\! \left(\frac{m_W}{\bar{\mu}}\right)-\frac{5}{12}\right]
\!+\!\nn\\[4pt]
&{}&
-12m_Z^2\sum_{quarks} m_q^2 \ln\! \left(\frac{m_q}{\bar{\mu}}\right)
-4m_Z^2\sum_{\ell=e\mu\tau} m_\ell^2 \ln\! \left(\frac{m_\ell}{\bar{\mu}}\right)
\bigg\}
+2m_Z^2 \frac{v_{H(1)} }{v_{H(0)}}\,,
}
where $m_X$ is the tree-level mass of the particle $X$
and $v_{H(1)}$ in the last term denotes the corrections to the tree-level VEV 
$v_{H(0)}$ of the (symmetric phase) Higgs doublet field 
\eq{\label{Eq:Higgs-doublet}
\sH\equiv \sH_{\rm sym}=
\frac{1}{\sqrt{2}}
\left(
\begin{array}{c}
G_1+i\,G_2\\
H+i\,G_Z
\end{array}
\right)
+\frac{1}{\sqrt{2}}
\left(
\begin{array}{c}
0\\
v_{H(0)}+\hb\,v_{H(1)}+\cO(\hb^2)
\end{array}
\right)\,.
}
The  formula \refer{Eq:TadCond} for the one-loop correction to 
the VEV 
yields here 
\eqs{\nn 
 v_{H(1)}
&=&
\frac{2}{(4 \pi )^2 m_H^2\, v_{H(0)}} 
\bigg\{
6\sum_{quarks} m_q^2\, a^R({m_q})
\!+\!
2\sum_{\ell=e\mu\tau} m_\ell^2\, a^R(m_\ell)
\!-\!\frac{3}{4} m_H^2\,a^R(m_H)
\!+\!
\nn\\[4pt]
&{}&\quad 
-3 m_W^2 \left[a^R(m_W)+\frac{2}{3}m_W^2\right]
-\frac{3}{2} m_Z^2 \left[a^R(m_Z)+\frac{2}{3}m_Z^2\right]
\bigg\}
\,.
}
At one-loop the factor $\cb$ in the
off-diagonal element of \refer{Eq:M_V--ZA} reads
\eq{\nn
\cb
=-\frac{3\,e}{8\pi^2} \frac{m_Z}{ v_{H(0)}}
  \,m_W^2\!\left[\ln\! \left(\frac{m_W}{\bar{\mu}}\right)-\frac{5}{12}\right]\,,
}
where
\eq{\nn
e=2 \frac{m_W}{v_{H(0)}} \sqrt{1-\frac{m_W^2}{m_Z^2}}\,\,,
}
is the renormalized charge coupling constant. 

We see that the vector
\eq{\nn
\ze_{[\rm photon]}
=
\cN 
\left[
\begin{array}{c}
-\hb\, \cb / m_Z^2
\\[4 pt]
1
\end{array}
\right]+\cO(\hb^2)\,,
}
corresponds to the zero eigenvalue of the matrix \refer{Eq:M_V--ZA}:
\eq{\label{Eq:eigen-eq}
M_V^2(0)\,  \ze_{[\rm photon]}
=\cO(\hb^2)
\,, 
}
which means that
the photon is massless to one-loop accuracy. The normalization factor 
$\cN$ can be obtained from Eq. \refer{Eq:norm-cond-GENERAL-vectors}. 
To this end one needs the derivative of of the matrix $M_V^2(q^2)$. 
Its $ZZ$ element in the limit $q^2\to 0$ is singular: 
\eq{\label{Eq:M_VprimeZZ}
M_V^{2\,{\prime}}(q^2)_{ZZ}
=\frac{\hb\,m_Z^2}{48\pi^2 v_{H(0)}^2}
\big\{
 (6-1) \ln(-q^2/\bar{\mu}^2)+\cO((q^2)\!\!{\phantom{|}}^0)
\big\}
+\cO(\hb^2)\,.
} 
The factor of 6 in the bracket originates from contributions of 
neutrinos, while $-1$ is the contribution of unphysical massless gauge 
degrees of freedom. The $AA$ and $AZ$ elements of this derivative are 
regular and read
\koment{str.OC90}
\eqs{\nn
M_V^{2\,{\prime}}(0)_{AA}
&=&
\frac{\hb\,e^2}{6\pi^2}
\Big\{
\sum_{\ell=e\mu\tau}\ln(m_\ell/\bar{\mu})
\!+\!3\sum_{quarks} Q_q^2\ln(m_q/\bar{\mu})
\!-\!3\ln(m_W/\bar{\mu})-\frac{11}{16}
\Big\}
\!+\!\nn\\[4pt]
&{}&
+\cO(\hb^2)\,, 
}
with $Q_q=+2/3, -1/3$ denoting the electric charge of quark $q$, 
and 
\eqs{\nn
M_V^{2\,{\prime}}(0)_{ZA}
&=&\frac{\hb\,e}{(4 \pi )^2} \frac{1}{18 m_Z v_{H(0)}} 
\Big\{
24 \left(4 m_W^2-3 m_Z^2\right) 
     \sum_{\ell=e\mu\tau}  \ln\! \left(m_{\ell }/\bar{\mu}\right)
+\nn\\[4pt]
&{}&\hspace*{-50 pt}
+16 \left(8 m_W^2-5 m_Z^2\right) \!\!\!\!
            \sum_{up-quarks}  \ln\! \left(m_q/\bar{\mu}\right) 
+8 \left(4 m_W^2-m_Z^2\right)  \!\!\!\!\!\!
         \sum_{down-quarks} \ln\! \left(m_{q}/\bar{\mu}\right)
+\nn\\[4pt]
&{}&\hspace*{-50 pt}
-12 \left(24 m_W^2+m_Z^2\right) 
\ln\! \left(m_W/\bar{\mu}\right)-66 m_W^2+41 m_Z^2
\Big\}
+\cO(\hb^2)\,.
} 
We thus see that, despite the singular behavior of $M^{2\,\prime}_V(q^2)_{ZZ}$, 
the product $M^2_V{}^{\prime}(q^2)\, \ze_{[\rm photon]}$ is 
finite in the limit $q^2\to 0$ to one-loop accuracy, in agreement with 
conditions formulated 
in Sec. \ref{Sec:Sub:Prescr-Vectors}. \kKK 
We have 
checked that $M^2_V{}^{\prime\prime}(q^2)\, \ze_{[\rm photon]}$ is also finite for 
$q^2\to 0$,  
\kGG  
and therefore the propagator of vector fields has 
in the $(Z,\,A)$ block a pole at $q^2=0$. In particular, 
the correctly normalized eigenvector $\ze_{[\rm photon]}$ has the form 
\koment{str.OC89}
\eq{\label{Eq:zeta-photon}
\ze_{[\rm photon]}
=
(1+\frac{1}{2}\,M_V^{2\,{\prime}}(0)_{AA}) 
\left[\!\!
\begin{array}{c}
-\hb\, \cb / m_Z^2
\\[4 pt]
1
\end{array}
\!\! \right]+\cO(\hb^2)
=
\left[\!\!
\begin{array}{c}
-\hb\, \cb / m_Z^2
\\[4 pt]
1+\frac{1}{2}\,M_V^{2\,{\prime}}(0)_{AA}
\end{array}
\!\!\right]
+\cO(\hb^2).
}
Using Eq. \refer{Eq:AsymField-vec} we get the decomposition of the
asymptotic fields $\vZ_\mu$ and $\vA_\mu$ corresponding to ${Z_\mu}$ and $A_\mu$ 
\eqs{ \label{Eq:ZA-asymp} 
\vZ_\mu&=&
-\frac{\hb\, \cb}{m_Z^2}\, {\bA_\mu}+\cO(\hb^2)+\ldots\,,
\nn\\[5pt]
\vA_{\mu}&=& 
\left\{1+\frac{1}{2}\,M_V^{2\,{\prime}}(0)_{AA}\right\}
{\bA_\mu}
+\cO(\hb^2)+\ldots\,, 
}
where $\bA_{\mu}$ is a canonically normalized free massless vector field in the 
Coulomb gauge. The ellipsis indicates the contributions of unphysical 
modes  discussed in Section \ref{Sec:Proof}. \footnote{We stress that, 
while rigorously only eigenvectors $\ze_{V[\la_r]}$ corresponding to stable 
particles enter the decomposition \refer{Eq:AsymField-vec}, the 
factorization \refer{Eq:As-vect-part} of pole residues 
is correct for complex poles as well. 
In particular, a (complex) eigenvector $\ze_{[Z]}$  associated with the $Z$ 
boson can be useful in the study of properties of the resonance 
\cite{PilafRL1,PlumOld,PilafUnd,PlumNew}. 
If, however,  the $Z$ boson is treated as a stable particle, then 
the corresponding free vector field $\bZ_\mu$ (in the unitarity gauge) 
should be also included in Eqs. \refer{Eq:ZA-asymp}. Its 
``content''
in the asymptotic fields $\vZ_\mu$ and $\vA_\mu$ is then determined by the 
eigenvector $\ze_{[Z]}\approx \re(\ze_{[Z]})$ associated with the $Z$ pole.   
}

\begin{figure}[pt]
\centering
\includegraphics[width=0.25\textwidth]{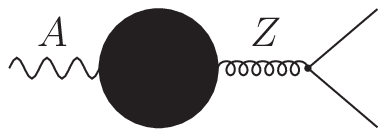}
\caption{
Diagrams with the external line corrections that reproduce  
the operator in Eq. \refer{Eq:S-mix-ZA}.  
}
\label{Rys:AZ-Mix}
\end{figure}

Because $\cb\neq 0$, the amputated correlation functions of 
the $Z_\mu$ field  
contribute to transition amplitudes with photons. 
Taking, for instance, 
the coupling between $Z_\mu$ and fermions (cf. Eq. \refer{Eq:LagrTreeGI-Majo})
\eqs{\nn
\mathcal{L}^{F}_0&\supset&
\frac{1}{2}\,
i\,Z_\mu\,\,
\bar{\psi}^a\,\gamma^\mu 
\left(  \TF_{Z a b}\, P_L+\TF^\star_{Z a b}\, P_R \right)
\psi^b\,,
}
we see that $\cb$ gives the following contribution to the $S$-operator
\eqs{\label{Eq:S-mix-ZA}
S_{\rm mix}&=&
\frac{\hb\, \cb}{2 m_Z^2}\,
\int\ \volel{x}\, 
\bA_\mu\,
\bar{\vpsi}^{a_1}\,\gamma^\mu 
\left(  \TF_{Z a_1 a_2}\, P_L+\TF^\star_{Z a_1 a_2}\, P_R \right)
\vpsi^{a_2}\,.
}
$\vpsi^a$  are here the asymptotic fields corresponding to $\psi^a$. 
In certain extensions of the SM this term contributes to e.g. decays of 
heavy neutrinos into light ones and photons. \kSS 
The $S_{\rm mix}$ term is by no means surprising; it can be recovered by ignoring
the LSZ formalism and including instead the terms of the 
Dyson series corresponding to 
diagrams  shown  in 
Fig.~\ref{Rys:AZ-Mix}. By contrast, 
in the proper LSZ approach which we have extended here to the case
of fields subject to mixing, 
the amplitudes are inferred directly from the amputated correlation functions.
With our prescription, one can find 
the external line factors $\ze^\al_{V[\la_r]}$ 
which are correctly normalized also at higher orders, what is essential 
for unitarity.\footnote{We also note that while the term 
$q_\mu q_\nu/q^2$ in the numerator of the propagator of the $Z$ field, 
makes the diagram of Fig. \ref{Rys:AZ-Mix} somewhat singular in the Landau 
gauge, the determination of $\ze^\al_{\rm [photon]}$ factor
is completely free from singularities. \kTT
}

 We can also use the example of the $Z$-photon mixing to demonstrate how 
the relation \refer{Eq:Massless-Eigenvectors} determines the direction (but 
not the normalization) of the eigenvector $\ze_{[\rm photon]}$. 
The advantage of this prescription lies in the small number 
of diagrams contributing to the ghost-antighost self-energy;  at one-loop 
in a general renormalizable gauge theory there is only one diagram 
(shown in Fig. \ref{Rys:KmuOm}) contributing to the ghost-antighost 
self-energy, while seven diagrams (those of Fig. \ref{Rys:AA}) can contribute 
to the self-energy of vector bosons. 
In the SM, the matrix $\Om(q^2)$ appearing in  the 1PI two-point function 
\refer{Eq:Gh-AntiGh} has the same block structure as the matrix $M_V^2(q^2)$ 
discussed above. We are interested in  its 
$(Z,\, A)$ block,  which has the form (cf. Eq. \refer{Eq:Omega}): 
\eq{\label{Eq:Om-SM-ZA-Jawne}
\Om(0)=
-\mathds{1}
- 
\frac{\hb}{8\pi^2}
\cH(0,m_W)
\left[
\begin{array}{cc}
\frac{4 m_{W}^4}{m_Z^2\,v_{H(0)}^2}       &   \frac{2\,e\, m_W^2}{m_Z\,v_{H(0)}}
\\[8pt]
\frac{2\,e\, m_W^2}{m_Z\,v_{H(0)}}     &   e^2 
\end{array}
\right]
+\cO(\hb^2)\,,
}
with
\eq{\nn
\cH(0,m_W)=\frac{1}{8}\big\{  12 \ln(m_W/\bar{\mu})-5 \big\}\,, 
}
(the correction to $\Om(0)$ has vanishing determinant, which reflects 
the fact that ghost of the Abelian ideal $U(1)_Y$ are \kJJ   noninteracting). 
In the SM case, the quantum-corrected VEV $v$ has the same direction 
as  does the tree-level one $v_{(0)}$, and therefore the generator 
$\TS_\al=\TS_A$, to which the $A_\mu$ field couples at the tree-level,   
remains unbroken also at one-loop order. Thus, the vector $\Th$ that 
fulfills the condition \refer{Eq:Th-cond} can be 
chosen as (in the $(Z,\,A)$ subspace)  \koment{OC91}
\eq{\nn
\Th
=
\left[
\begin{array}{c}
0
\\[4 pt]
-1
\end{array}
\right]\,,
}
so that  
\eq{\label{Eq:OmTh-SM}
\Om(0)\Th
=
\big\{1 + e^2 \frac{\hb}{8\pi^2} \cH(0,m_W)\big\}
\left[
\begin{array}{cc}
 \frac{\hb}{4\pi^2} \frac{e\, m_W^2}{m_Z\,v_{H(0)}} \cH(0,m_W)
 \\[8pt]
 1 
\end{array}
\right]
+\cO(\hb^2)\,.
}
This is, up to a proportionality factor, the photon eigenvector  
\refer{Eq:zeta-photon}, as expected.

In order to illustrate the role of the limit in Eq. 
\refer{Eq:Massless-Eigenvectors},
we give here results in the gluonic block, where 
$M_V^2(q^2)\propto \mathds{1}$ and   $\Om(q^2)\propto \mathds{1}$  
with the following proportionality factors 
\kHH
\eq{\nn
M_V^2(p^2)
=
\frac{\hb\, g_s^2\, p^2}{12 \pi^2}
\Big\{
\frac{1}{16} \left[ 97\! -\! 78\, \ln(-p^2/\bar{\mu}^2) \right]
+\sum_{quarks}\ln(m_q/\bar{\mu})
+\cO(p^2)
\Big\}
+\cO(\hb^2),
}   
and 
\eq{\nn
\Om(p^2)
=
-1-\frac{3\,\hb\,g_s^2}{64\pi^2}
\big\{
3\,\ln(-p^2/\bar{\mu}^2) -4\big\}
+\cO(\hb^2)\,.
}   
Hence, Eq. \refer{Eq:Massless-Eigenvectors} holds also for vectors $\Th$ 
pointing in the directions of $SU(3)_C$ generators, 
at least in the perturbative regime.

\subsection{Singlet Majoron Model}\label{Sec:SubSec:Examples:Maj} 

In order to illustrate the usefulness of the condition 
\refer{Eq:norm-cond-GENERAL-scalars-phys}, which determines the 
vectors $\zeta_{S[\ell_r]}$ associated with physical massless spin 0
particles, we study in this section the singlet Majoron model \cite{Majo}. 
Its additional (with respect to the SM - see e.g. \cite{WeinT2}) 
fermionic fields are made up of three gauge-sterile Weyl fields (``neutrino 
singlets")  $N^i_A$, $i=1,2,3$, and their complex conjugates 
$\anti{N}^i_{\dot{A}}$. The scalar sector of the model consists of the usual 
electroweak scalar doublet  \refer{Eq:Higgs-doublet},  
and a new gauge-sterile complex scalar $\cphi\equiv\cphi_{\rm sym}$
which carries two units of the lepton number. This field couples 
only to the sterile neutrinos and to the electroweak doublet $\sH$; the 
scalar potential consistent with gauge 
symmetries and the lepton number symmetry reads
\eq{\nn 
\cV(\sH,\,\cphi)
=- \, m_1^2\, \sH^\dagger\! \sH
-m_2^2\, \cphi^\star \cphi \,
+\lambda_1\,(\sH^\dagger\! \sH)^2
+2\lambda_3\, \sH^\dagger\! \sH\,
            \cphi^\star \cphi
+\lambda_2\,( \cphi^\star \cphi )^2 .
}
The  Yukawa couplings of the model are given by
\eqs{\nn
\cL_{\rm Y} =\cL^{{\rm SM}}_{\rm Y}
+
\big\{
  Y_{ji}^\nu N^{jA} {\sH}^\rmt\!\ep{L}^{i}_A \, 
 - \frac{1}{2}Y^M_{ji} \cphi N^{j A} N^i_A
\big\} \, +\, {\rm H.c.}
\,.
\qquad
}
$\cL^{{\rm SM}}_{\rm Y}$ represents here the Yukawa couplings of the SM 
\cite{WeinT2}, $L^i_A$ are lepton $SU(2)_L$ doublets
\eq{\nn
 L^i_A  \equiv \left(\begin{array}{c} \nu^i_A\\[4pt]
                                   e^i_A \end{array}\right)\,,
}
with the index $i=1,2,3$ labeling the three families 
and $\ep$ is the antisymmetric  $SU(2)_L$ metric. 

We are interested here
in the phase in which both symmetries: the electroweak one 
and lepton number one, are spontaneously broken. Exploiting the 
symmetries of the action we can assume that $\VEV{\cphi}$ is real 
\eq{\nn
\cphi\equiv\cphi_{\rm sym}=
\frac{1}{\sqrt{2}}(S+i\,G_\cphi)
+
\frac{1}{\sqrt{2}}(v_{\cphi(0)}+\hb\, v_{\cphi(1)}+\cO(\hb^2))\,.
}
The parametrization of $\sH$ is given by \refer{Eq:Higgs-doublet}. The 
tree-level VEVs are related to the mass parameters of the potential  by
\eq{\nn 
m_1^2=\lambda_{3}\,v_{\cphi(0)}^{2}+\lambda_{1}\,v_{H(0)}^{2}\,,\qquad\qquad
m_2^2=\lambda_{3}\,v_{H(0)}^{2}+\lambda_{2}\,v_{\cphi(0)}^{2}\,.
}
Linear combinations $(h,\, \underline{h})$ of the fields $(S,\,H)$ are then 
eigenstates of the tree-level mass-squared matrix, with the eigenvalues 
$m_{\rm I}^2$ and $m_{\rm II}^2$; all other scalars are massless 
(would-be) Goldstone bosons.

By a unitary rotation of the three sterile neutrinos $N^i$ 
the Yukawa matrix $Y^M$  can be brought into a diagonal
and non-negative form. The matrix $Y^\nu$ is then, in general, non-diagonal. 
However, as the sole purpose of this section is to illustrate the use 
of the condition \refer{Eq:norm-cond-GENERAL-scalars-phys}, we  will 
simply assume that also $Y^\nu$ is positive and diagonal so that 
both matrices,  $Y^M$ and $Y^\nu$ can be unambigously expressed in 
terms of the masses of the physical light and heavy 
neutrinos, denoted (with a little abuse of notation) by  
$m_{\nu_i}$ and $m_{N_i}$. At the one-loop order, the matrix $M^2_S(p^2)$ 
obtained using the formulae
\refer{Eq:MV-and-MS} and \refer{Eq:De_S} is then 
block diagonal  \kSSSS  with the blocks corresponding to pairs 
\footnote{In the first block $M^2_S(p^2)$ is proportional to the identity 
matrix.}    
$(G_{1},\,G_{2})$, 
$(G_{Z},\,G_\cphi)$, 
and
$(h,\underline{h})$. 
For vanishing $p^2$, the first two blocks of the matrix $M^2_S(p^2)$ 
vanish in agreement with the Goldstone theorem; since this 
results from a  nontrivial cancellations between the 
contributions to the one-loop 1PI 
self-energies and  the 
one-loop corrections to the VEVs, the explicit expressions for the  VEVs
(obtained from  the general formula \refer{Eq:TadCond}) 
are given in Appendix \ref{Sec-App:ResMajoron} 
(Eqs. \refer{Eq:App:vH_1}-\refer{Eq:App:vcphi_1}) 
for completeness.  

We are interested in the block of $M^2_S(p^2)$ corresponding to the 
neutral (would-be) Goldstone bosons $(G_{Z},\ G_\cphi)$. The matrix  
that appears in the normalization condition 
\refer{Eq:norm-cond-GENERAL-scalars} 
has, after reduction to this  block, 
the following form  \koment{str.OC86} \kQQ 
\eq{\label{Eq:ScProd}
\mathds{1}-M^{2\,\prime}_S(0)
=
\left[
\begin{array}{cc}
1+\hb\,\va & \hb\, \vb \\
\hb\, \vb  & 1+\hb\,\vc
\end{array}
\right]+\cO(\hb^2)\,.
}
The formulae for $\va$, $\vb$ and $\vc$ follow from 
Eqs. \refer{Eq:MV-and-MS} and \refer{Eq:De_S}, and are given in Appendix 
\ref{Sec-App:ResMajoron}, Eqs. \refer{Eq:App:ScProd-a}-\refer{Eq:App:ScProd-b}.

The null eigenvector $\ze_{S[\rm Maj]}$ corresponding to the physical Goldstone 
boson (Majoron) has to obey the condition 
\refer{Eq:norm-cond-GENERAL-scalars-phys}. The  gauge symmetry
generators relevant to our problem are  $\TS_\al=\TS_Z,\,\TS_A$ to which the 
$Z_\mu$ and $A_\mu$ fields couple at the tree-level.  \kOO  The latter is 
unbroken, $\TS_Av=0$, while $\TS_Z v$, in the 
$(G_1,\,G_2,\,G_Z,\,G_\cphi,\,h,\,\underline{h})$ coordinates, 
reads 
\eq{\nn
\TS_Z\, v
=
(m_Z\,
{v_H}/{v_{H(0)}})\,
\left[
0\,,\  0\,,\ 
1\,,\  0\,,\
  0\,,\  0
\right]^{\rmt}\,,
}
($m_Z$ is the tree-level mass of the $Z$ boson). 
Thus, taking into account the normalization condition 
\refer{Eq:norm-cond-GENERAL-scalars} we get \koment{str.OC87}
\eq{\nn
\ze_{S[\rm Maj]}
=
\left[
0,\ 0,\ 
-\hb\,\vb,\ 
1-\hb\,{\vc}/{2}, \    
0,\ 0
\right]^\rmt+\cO(\hb^2)\,.
}
The correctly normalized eigenvector associated with 
the unphysical neutral would-be Goldstone boson has the form \kTTTT 
\eq{\nn
\ze_{S[\rm unph]}
=\left[
0,\ 0,\ 
1-\hb\,\va/2,\ 
0,\  
0,\ 0
\right]^\rmt+\cO(\hb^2)\,.
}
The formula \refer{Eq:AsymField} gives, therefore,
the following decomposition of 
asymptotic fields $\vG_{Z}$ and $\vG_\cphi$ corresponding to $G_{Z}$ and $G_\cphi$ 
\eqs{\nn
\vG_{Z}&=&(1-\hb\,\va/2) {\bG}_{Z} - \hb\,\vb\, {\bG}_{\cphi} +\cO(\hb^2)\,,
\nn\\[5pt]
\vG_{\cphi}&=&(1-\hb\,\vc/2) {\bG}_{\cphi} +\cO(\hb^2)\,,
}
where ${\bG}_{\cphi}$ (${\bG}_{Z}$) is  the canonically normalized free 
scalar field constructed out of the operators creating/annihilating states
of the physical (unphysical) massless spin 0 particles. In 
particular, the amputated correlation functions of $G_Z$ contribute to 
transition amplitudes of the Majoron. 
By contrast, the amputated correlation functions of  the 
$Z_\mu$ field  
in the Landau gauge cannot contribute to transition amplitudes of (physical) 
scalar particles.  

\begin{figure}[pt]
\centering
\includegraphics[width=0.7\textwidth]{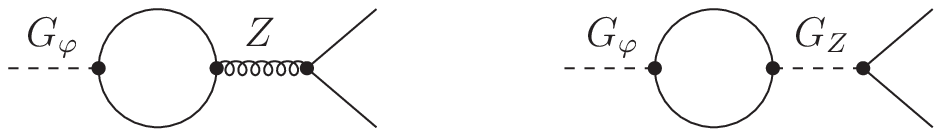}
\caption{ External line corrections reproducing the effects of the
operator in Eq. \refer{Eq:S-mix-Majo}. Only fermions (more specifically, 
neutrinos) contribute to  the mixed self-energies. 
}
\label{Rys:MajMix}
\end{figure}

From the Lagrangian \refer{Eq:LagrTreeGI-Majo} 
one can read off that $\vb\neq0$ gives 
rise to the following term in the $S$-operator  
\eqs{\label{Eq:S-mix-Majo}
S_{\rm mix}&=&
\hb\,\vb\,
\frac{i}{2!}\!
\int\,\volel{x}\,
\bG_\cphi\,
\bar{\vpsi}^{a_1} \!
\left\{  \YF_{z a_1 a_2}\, P_L+\YF^\star_{z a_1 a_2}\, P_R \right\}
\vpsi^{a_2} \,,
}
where the index $z$ on the Yukawa matrices corresponds to the 
$\phi^z\equiv G_Z$ component of the scalar field, and $\vpsi^a$ 
are the asymptotic fields associated with the interpolating fields 
$\psi^a$. This one-loop result is consistent with the one obtained by ignoring the LSZ 
formalism and including the terms of the Dyson series shown in Fig.~\ref{Rys:MajMix} \cite{Majo,Majo:Pi}.~\footnote{The external line corrections are the sole source of the $\cO(\hb)$ 
couplings between the Majoron and quarks. By contrast,  
proper vertex corrections contribute at $\cO(\hb)$ to the couplings between the Majoron and charged leptons \cite{Majo,Majo:Pi}.  
}
The sum of these diagrams (in an arbitrary $R_\xi$ gauge)  can be written as 
\koment{str.OC25,OC88}
\eqs{\label{Eq:S-mix-Majo-Inne}
S_{[{\rm Fig.}\ref{Rys:MajMix}]}
&=&
\frac{i}{2}\,
\int\,\volel{x}\,
\vG_\cphi\,
\bar{\vpsi}^{a_1} 
\left\{  \YF_{z a_1 a_2}\, P_L+\YF^\star_{z a_1 a_2}\, P_R \right\}
\vpsi^{a_2}\,\times
\nn\\[4pt]
&{}&
\qquad\qquad \times
\frac{1}{0-\xi\,m_Z^2}
\left\{\widetilde{\Ga}_{m z}(0,0)_{[\ref{Rys:PhiPhi}.D]}
+\xi\, m_Z\, P_{Zm}(0)_{[\ref{Rys:PhiA}.D]}
\right\}\,, \ \ 
}
where the $m$ index corresponds to the $\phi^m\equiv G_\cphi$ field, while 
$P_{Zm}(0)_{[\ref{Rys:PhiA}.D]}$ parametrizes the mixed $Z_\nu$-$G_\cphi$ two-point function \koment{str.OC22}
\eq{\nn
\widetilde{\Gamma}_{m\al}^{\ \  \nu}(q,-q)_{[\ref{Rys:PhiA}.D]}
=i\, q^\nu\, P_{\al m}(q^2)_{[\ref{Rys:PhiA}.D]}\,,
}
with $A^\al_\nu\equiv Z_\nu$. 
The subscripts ${}_{[\ref{Rys:PhiPhi}.D]}$ and ${}_{[\ref{Rys:PhiA}.D]}$ 
indicate that in the present model the mixed self-energies are produced by  
the fermionic loops (Figs. \ref{Rys:PhiPhi}$.D$ and \ref{Rys:PhiA}$.D$). 
The explicit expression for 
$\widetilde{\Gamma}_{m\al}^{\ \  \nu}(q,-q)_{[\ref{Rys:PhiA}.D]}$ is given in Eq. 
\refer{Eq:MixSV-FermLoop}; see also  the remarks below Eq. 
\refer{Eq:nie-WT-fermion}. To obtain Eq. \refer{Eq:S-mix-Majo-Inne} we have 
used the fact that $\vpsi^{a}(x)$  satisfies the free equations of 
motions with (up to negligible corrections) the tree-level mass matrices 
$M_F^{\phantom{\star}}$ and $M_F^{{\star}}$, which are related to the Yukawa 
matrices by the gauge invariance \koment{(D:C.38)}
\eq{\nn
\TF_\al^\rmt\, M_F+M_F\,\TF_\al
=-(\TS_\al v_{(0)})^j\, Y_j\,,
} 
were $(\TS_\al v_{(0)})^j=m_Z\, \de^{j}_{\ z}$ for $\al=Z$. \kLL 

We have $\widetilde{\Ga}_{m z}(0,0)_{[\ref{Rys:PhiPhi}.D]}=0$, while the 
``Ward identity" \refer{Eq:nie-WT-fermion} gives (cf. Eq. \refer{Eq:ScProd})
\koment{OC88} \kNN
\eq{\nn
P_{Zm}(0)_{[\ref{Rys:PhiA}.D]}
=-\hb\,m_Z\, \vb\,,
}
(the Majoron $G_\cphi=\phi^m$ is gauge-sterile and therefore 
$[\TS_\al]^j_{\ m}\equiv 0$). Thus, Eq. \refer{Eq:S-mix-Majo-Inne} 
agrees with the result of  the properly generalized LSZ 
prescription in which
the $S$-matrix elements are always extracted from (completely) amputated 
correlation functions.

\section{Derivation of the prescription for vector fields}\label{Sec:Proof}

In this section we carefuly investigate the structure of propagators 
of a system of mixed vector (gauge) and scalar fields using the 
relevant Slavnow-Taylor identities, thereby justifying the practical
prescriptions given in Section \ref{Sec:Sub:Prescr-Vectors}. For completeness
we construct the corresponding asymptotic fields (which enter the 
formula \refer{Eq:S} for the $S$-operator) including also those terms which 
create/annihilate particle states which are not physical (in the sense of the
BRST clasification). 

\subsection{Slavnov-Taylor identities} \label{Sec:Sub:STids}

We begin by recalling the identities satisfied by the renormalized 1PI 
effective action $\Gamma$ of a non-anomalous gauge theory. Firstly,
it must obey the Zinn-Justin identity \cite{Z-J:74,BRS1,BRS2} 
(see also \cite{PiguetSorella})
\begin{eqnarray}
{\cal S}(\Gamma)=0~\!,
\label{eqn:ZJidBasic}
\end{eqnarray}
in which $\mathcal{S}(F)$ for an arbitrary functional $F$ of fields and 
antifields is given by (cf. Eq. \refer{Eq:LagrTreeRest})
\begin{eqnarray}\nn
\mathcal{S}(F)\equiv
\derf{F}{K^\mu_\alpha}\!\cdot\!\derf{F}{A^\alpha_\mu}+
\derf{F}{K_i}\!\cdot\!\derf{F}{\phi^i}+
\derf{F}{\bar{K}{}^{\phantom{b}}_a}\!\cdot\!\derf{F}{\psi^a}+
\derf{F}{L_\alpha}\!\cdot\!\derf{F}{\omega^\alpha}+
h_\alpha\!\cdot\!\derf{F}{\anti{\omega}_{\alpha}}\,.
\end{eqnarray}
(We use here the abbreviated notation 
${k}\cdotp\!{g}\equiv\intt{4}{x}{k(x)~\!g(x)}$). In the lowest order, 
$\Ga=I_0+\cO(\hb)$ and $\cS(I_0)=0$ is nothing but the condition of 
BRST-invariance of the tree-level action \refer{Eq:I_0=I_0^GI+I_0^Rest}.
Secondly, $\Gamma$ satisfies also
the auxiliary identities: the gauge condition identity and the ghost 
identity \cite{PiguetSorella} which in the Landau gauge take the 
forms\footnote{These two identities generalize (in different forms) also 
to other gauge conditions.} 
\kUU
\eq{\label{Eq:Gauge}
\derf{\Ga}{h_\beta(x)}=-\partial^\nu\!\!A_\nu^\beta(x) \,,
} 
\eq{\label{Eq:Ghost}
\left\{
\derf{}{\anti{\omega}_{\alpha}\!\bracket{x}}
-\derp{}{x_\mu}\derf{}{{K}^\mu_\alpha(x)} 
\right\} \Ga=0\,.
}
Finally, $\Gamma$ satisfies also
the antighost identity 
\eqs{\label{Eq:AntiGhost}
\int\!{{\rm d}^4}{x}
\bigg\{\derf{}{{\omega}^\alpha\!\bracket{x}}
-\anti{\omega}_\gamma(x)\mind{e}{\gamma}{\alpha\beta}
\derf{}{h_\beta\!\bracket{x}}\bigg\}
\Ga
&=&
\int\!{{\rm d}^4}{x}\bigg\{L_\beta\mind{e}{\beta}{\!\alpha\gamma}\omega^\gamma
-K^\mu_\beta\mind{e}{\beta}{\!\alpha\gamma}A^\gamma_\mu
+\nn\\[4pt]
&{}& \hspace*{-100 pt}
-K_i\matri{\TS_\alpha(\phi+v)}^i
+
\bar{K}_a ([\TF_{\al}]^{a}_{\ b}P_L+[\TF_{\al}^\star]^{a}_{\ b} P_R)\psi^b
\bigg\}.
}
which 
is specific only for the Landau gauge~\cite{BPS:anti-ghost-eq}. 
\footnote{ 
In 
theories with Abelian ideals, additional auxiliary identities are 
satisfied \cite{BBBC2,Grassi}.
They encode the lack of certain quantum 
corrections (e.g. they enforce the vanishing of a determinant of  
the matrix in Eq. \refer{Eq:Om-SM-ZA-Jawne}), 
and therefore play an important role in the proof 
of renormalizability of non-semisimple gauge-models \cite{BBBC2,Grassi}. 
Nonetheless, we do not use them in what follows: we treat 
Abelian gauge fields on an equal footing with non-Abelian ones.  
}

The identities \refer{eqn:ZJidBasic}-\refer{Eq:AntiGhost} differentiated with 
respect to fields give, after restriction to vanishing configurations of 
fields (cf. Eq. \refer{Eq:TadCondFormal}), relations between  two-point 
functions which will be useful in investigation of propagators 
(in general, relations originating from the Zinn-Justin identity are 
usually called the Slavnov-Taylor identities (STids) \cite{BRS1,BRS2}). 
Before we write down the relevant identities, we need to parameterize 
the 1PI two-point functions. Those of scalar and vector fields are 
parametrized as in \refer{Eq:Gamma2-scalar} and 
\refer{Eq:Gamma2-vector} (cf. Eq. \refer{Eq:Notacja1}). 
The mixed, vector-scalar correlation function 
$\VEV{\hat{A}^{\al}_{\mu}(-q)\hat{\phi}^j(q)}$  is written as
\koment{str.MWP10,MPW11}
\eq{\label{Eq:Gamma2-vec-sc}
\widetilde{\Ga}^{\mu}_{\al j}(-q,q)
\equiv
i\, q^\mu\, P_{\al j}(q^2)
=
-\widetilde{\Ga}^{\ \mu}_{j \al}(-q,q)
\,.
}
The correlation functions of Nakanishi-Lautrup multipliers: 
$\VEV{\hat{h}_\al(-q) \hat{A}^{\be}_{\nu}(q)}$, \\
$\VEV{\hat{h}_\al(-q) \hat{\phi}^{j}(q)}$ 
and 
$\VEV{\hat{h}_\al(-q) \hat{h}_\be(q)}$, 
are  
uniquely fixed to all orders by the the gauge condition \refer{Eq:Gauge}
\koment{str. MPW10,MPW13}
\eqs{\nn
\widetilde{\Ga}^{\al\nu}_{\ \be}(-q,q)
&\equiv& i\,\de^{\al}_{\ \be}\, q^\nu
=-\widetilde{\Ga}^{\nu\al}_{\be}(-q,q)\,,
\nn\\[8pt]
\widetilde{\Ga}^{\al}_{\ j}(-q,q)
&\equiv&0 = \widetilde{\Ga}^{\ \al}_{j}(-q,q)\,,
\nn\\[8pt]
\widetilde{\Ga}^{\al\be}(-q,q)
&\equiv&0\,. 
}
The correlation functions of antifields are parametrized as 
\koment{str.MPW11} \kWW
\eqs{\label{Eq:Antifields-Gh:Scalar}
\left.\derf{}{\hat{\om}^\ga(p)}
\derf{}{\hat{K}_i(q)}\Gamma\right|_{0}
&=&
(2\pi)^4 \delta^{(4)}(q+p)\, 
B(q^2)^i_{\ \ga} \,\,,
 \\[8pt]
\label{Eq:Antifields-Gh:Vector}
\left.\derf{}{\hat{\om}^\ga(p)}
\derf{}{\hat{K}^\mu_\al(q)}\Gamma\right|_{0}
&=&
(2\pi)^4 \delta^{(4)}(q+p)
\left\{i\,q_\mu\, {\Om}(q^2)^{\al}_{\ \ga}\right\}\,.
}
Notice, that the ghost identity \refer{Eq:Ghost} ensures that the same 
matrix ${\Om}(q^2)$ appears in the above function and in \refer{Eq:Gh-AntiGh}. 

We are now ready to write the required STids for the two-point functions 
\refer{Eq:Gamma2-vector} and \refer{Eq:Gamma2-vec-sc}.  These are 
\koment{str.MPW12, por.MPW26,MPW71} \kRRRR 
\eq{
\label{Eq:STid1prime}
P_{\be j}(q^2)\,B(q^2)^j_{\ \ga}
=  
\left\{ q^2\, \sL_{\al\be}(q^2) + [M_V^2(q^2)-q^2\mathds{1}]_{\al\be} \right\}
\Om(q^2)^\al_{\ \ga}\,,
}
and
\eq{
\label{Eq:STid7prime}
q^2\,  P_{\al j}(q^2)\,\Om(q^2)^\al_{\ \ga} 
=  
[q^2\mathds{1}-M_S^2(q^2)]_{ij}\, B(q^2)^i_{\ \ga}
\,.
} 
Moreover, from the antighost identity \refer{Eq:AntiGhost} 
it follows that 
\eq{\label{Eq:Antigh-Conclusion}
B(0)^i_{\ \ga}=(\TS_\ga v)^i\,,
}
where $v$ is the  (exact)  vacuum expectation value 
\refer{Eq:VEV-expansion}. \kXX \kZZ

These relations allow us to prove the prescription \refer{Eq:Massless-Eigenvectors} 
for massless eigenvectors $\ze_{V[\la_r]}$: 
since the 1PI functions in four dimensions do 
not have poles,\footnote{
This statement is correct in finite orders of perturbation theory. 
} 
we see that Eq. \refer{Eq:Massless-Eigenvectors} follows immediately from 
\refer{Eq:STid1prime} after contracting both sides with a vector 
$\Th=(\Th^\ga)$ obeying Eq. \refer{Eq:Th-cond}. \kYY  \kPPPP 
Similarly, combining \refer{Eq:STid7prime} with \refer{Eq:Antigh-Conclusion},  
one immediately obtains the 
Goldstone theorem:
\eq{\label{Eq:would-be-Gold-Th}
M_S^2(0)_{ij}(\TS_\ga v)^j = 0
\,.
}

It is also worth noticing, that in the Landau gauge there are no 
one-loop diagrams contributing to the two-point function 
\refer{Eq:Antifields-Gh:Scalar}; therefore 
\koment{por. (D:I.57)}
\eq{\label{Eq:B(q^2)}
B(q^2)^i_{\ \ga}=(\TS_\ga v_{(0)})^i+\hb(\TS_\ga v_{(1)})^i+\cO(\hb^2)\,,
}
in agreement with \refer{Eq:Antigh-Conclusion}. Thus, the 
STids \refer{Eq:STid1prime}-\refer{Eq:STid7prime}, together with the 
invertibility of 
$\Om(q^2)$ (cf. Eq. \refer{Eq:Omega})  allow us to express the form-factors 
$\sL_{\al\be}(q^2)$ and $P_{\al j}(q^2)$ at one-loop order in terms of 
quantities which  at 
one-loop have been explicitly calculated in 
Sec. \ref{Sec:Res}.

\subsection{Propagators} \label{Sec:Sub:Propagators}

Inverting the complete matrix of  the 1PI two-point functions 
$\widetilde{\Ga}$  whose different blocks have been paramterized
in the previous section,  that is solving the algebraic equation
\eq{\label{Eq:Prop-Gen}
\widetilde{\Ga}_{IJ}(-p,p)\, \widetilde{G}^{JK}(p,-p)
=i\, \de_{I}^{\ K}\,,
} 
we find the  matrix $\widetilde{G}$ of propagators with 
(resummed) quantum corrections \koment{str.MPW14}
The indices $I$, $J$ and $K$ run {here} over components of bosonic 
fields $\phi^n$, $A^\al_\mu$ and $h_\be$. \kAAA 
The resulting expressions for {the} $\phi \phi$ and $AA$ propagators are 
given by {the formulae} \refer{Eq:G2-scalar} and \refer{Eq:G2-vector}, 
respectively. 
\kBBB 
The mixed scalar-vector propagators vanish, as has been already 
said (see \refer{Eq:G2-scalar-vector}). \kCCC 
The propagators  which mix the Nakanishi-Lautrup fields 
$h_\beta$ with vectors, scalars and themselves have the form 
\koment{str.MPW27,por.MPW26,por.MPW18}
\eqs{
\label{Eq:G2:hA}
\tG_{\!\be \mu}^{\!\ \,\al}(q,-q)
&=&-\de^{\al}_{\ \be}\,\frac{q_\mu}{q^2}
=-\tG_{\, \mu\be}^{\,\al}(q,-q)\,,
\\[5pt]\nn
\tG_{\be }^{\ \,n}(q,-q)
&=&i\,P_{\be j}(q^2)\left[\left(q^2\mathds{1}-M_S^2(q^2)\right)^{-1}\right]^{jn}
=\tG_{\, \ \be }^{\,n}(q,-q)
\,,\\[5pt]\nn 
\tG_{\be\ga }(q,-q)
&=&i
\bigg\{
P_{\be n}(q^2)\left[\left(q^2\mathds{1}-M_S^2(q^2)\right)^{-1}\right]^{nj}
              P_{\ga j}(q^2)
+\nn\\[4pt]
&{}&\quad\ 
+\de_{\be\ga}-\frac{1}{q^2}M_V^2(q^2)_{\be\ga}-\sL_{\be\ga}(q^2)
\bigg\}
\,.\nn
}
The last two propagators can be simplified {by exploiting} the STids 
\refer{Eq:STid1prime}-\refer{Eq:STid7prime} {which lead to}
\koment{str.MPW30-31} \kFFF    \kEEE
\eqs{
\label{Eq:G2:h-phi}
\tG_{\be }^{\ \,n}(q,-q)
&=&\frac{i}{q^2}
\,B(q^2)^{n}_{\ \ga}
   \left[\Om(q^2)^{-1}\right]^{\ga}_{\ \be}
=\tG_{\, \ \be }^{\,n}(q,-q)
\,,\\[5pt]
\label{Eq:G2:hh}
\tG_{\be\ga }(q,-q)
&=&0 
\,.
} 
Finally, the ghost-antighost propagator has the form
\koment{str.MPW32,por.MPW30} \kDDD 
\eq{\label{Eq:G2:gh-antigh}
\widetilde{\cG}^{\be}_{\ \al}(q,-q)
=-\frac{i}{q^2}\, \left[\Om(q^2)^{-1}\right]^{\be}_{\ \al}
\,,}
where the matrix $\Om(q^2)$ is defined by \refer{Eq:Gh-AntiGh} or, 
equivalently, by \refer{Eq:Antifields-Gh:Vector}.  \\

\subsection{Pole structure of the propagators} \label{Sec:Sub:AsymProp}

The first step in finding the asymptotic fields that appear in the LSZ formula 
\refer{Eq:S} 
for the $S$-operator is to determine the behavior of {all} propagators 
of the theory in 
the vicinity of their singularities located on the real axis \cite{Bec}. 
As we have already said in Sec. \ref{Sec:Sub:Prescr-Vectors}, the discussion of 
infrared divergences is beyond the scope of the present paper. Therefore we 
assume that an IR regulator has been introduced, if 
necessary,  
so that the limits \refer{Eq:Lim1Der} and \refer{Eq:Lim2Der} (as well as 
the $\Om(0)$ matrix in Eq. \refer{Eq:Gh-AntiGh}) are finite. 
\kOOOO 

With this proviso,\footnote{
As we have seen at the end of Sec. 
\ref{Sec:SubSec:Examples:Z-A-mixing}, the 
$\Om(0)$ matrix is IR-divergent in QCD; therefore Eq. 
\refer{Eq:G2:gh-antigh:pole-part} illustrates the need for an IR regulator. 
} 
{from} \refer{Eq:G2:gh-antigh} {we} immediately 
obtain the pole part of the (anti)ghosts propagator
\eq{\label{Eq:G2:gh-antigh:pole-part}
\widetilde{\cG}^{\be}_{\ \al}(q,-q)_{\rm pole}
=-\frac{i}{q^2}\, \left[\Om(0)^{-1}\right]^{\be}_{\ \al}
\,.
}
Similarly, {the formulae} \refer{Eq:G2:hA}-\refer{Eq:G2:hh} give 
the near-pole behaviour of the nontrivial propagators of the  
Nakanishi-Lautrup fields:
\eqs{
\label{Eq:G2:hA:pole-part}
\tG_{\be \mu}^{\ \,\al}(q,-q)_{\rm pole}
&=&-\tG_{ \mu\be}^{\al}(q,-q)_{\rm pole}
=-\de^{\al}_{\ \be}\,\frac{q_\mu}{q^2}
\,,\\[5pt]
\label{Eq:G2:h-phi:pole-part}
\tG_{\be }^{\ \,n}(q,-q)_{\rm pole}
&=&
\tG_{\, \ \be }^{\,n}(q,-q)_{\rm pole}
=
\frac{i}{q^2}
\,B(0)^{n}_{\ \ga}
   \left[\Om(0)^{-1}\right]^{\ga}_{\ \be}
=
\nn\\[4pt]
&{}&\qquad\qquad\qquad\qquad \!\!\!\!\!\!
=\frac{i}{q^2}
\,(\TS_\ga\,v)^{n}
 \left[\Om(0)^{-1}\right]^{\ga}_{\ \be}
\,.\phantom{a}
}
Obviously, the pole part of the $hh$ propagator $\tG_{\be\ga }\equiv0$, 
as well as of the mixed scalar-vector propagator 
$\tG^{\,j \be}_{\ \,\nu}=\tG^{\, \be j}_{\,\nu}=0$, vanish. 
The relevant behavior of the scalar fields propagator can be obtained 
directly from its form \refer{Eq:D-as-GENERAL-scalars-indi}:
\eq{\label{Eq:G2-scalar:pole-part}
\widetilde{G}^{kj}(p,-p)_{\rm pole}
=
{\sum_{\ell}}^\prime  \sum_r
\, 
\ze^k_{S[\ell_r]} 
\,
\frac{i}{p^2-m^2_{S(\ell)}}
\,
\ze^j_{S[\ell_r]}
\,,
}
(recall that the prime indicates restriction of the summation 
to the poles at real values of $p^2=m^2_{S(\ell)}$). As explained in Sec. 
\ref{Sec:Sub:Scalars}, the corresponding coefficients $\ze^k_{S[\ell_r]}$ 
can be chosen to be real; we assume that this choice has been made here. 

It remains to investigate the propagators \refer{Eq:G2-vector}
of the vector fields.  
Clearly, all poles of \refer{Eq:As-vect-part} located at real values of 
$q^2$ should be taken into account. 
Just as in \refer{Eq:G2-scalar:pole-part}
we assume that {vectors} $\ze^\be_{V[\la_r]}$ 
corresponding to these poles in \refer{Eq:As-vect-part} 
have been chosen to be real. 
Moreover, it will be convenient to single out the pole located at 
$q^2=0$ and to label it by $\lambda=\mathbf{0}$. 
The behavior of the propagator \refer{Eq:G2-vector} near its real poles
\koment{str.MPW55}  
can be then written in the form
\eqs{\label{Eq:G2-vector:pole-part} 
\widetilde{G}^{\be\de}_{\nu\rho}(q,-q)_{\rm pole}
&=&
-i
{\sum_{\la\neq\mathbf{0} }}^\prime 
\bigg[\eta_{\nu\rho}-\frac{q_\nu q_\rho}{m_{V(\la)}^2}\bigg]
\frac{1}{q^2-m^2_{V(\la)}}
\sum_r
\ze^\be_{V[\la_r]} 
\ze^\de_{V[\la_r]}
+\nn\\[4pt]
&{}&
-\frac{i}{q^2}\,\eta_{\nu\rho}\,\cZ^{\be\de}
+
i\,\frac{q_\nu\, q_\rho}{q^2}\,\cR^{\be\de}
+
i\,\frac{q_\nu\, q_\rho}{(q^2)^2}\,\cZ^{\be\de}
\,,
}
{in which} 
\koment{MPW54}
\eq{\label{Eq:cZ}
\cZ^{\be\de}=
\sum_r  \ze^\be_{V[{\bf 0}_r]} 
        \ze^\de_{V[{\bf 0}_r]}\,,
}
{and} $\cR$ is given by  the formulae 
\refer{Eq:cR-fin-form}-\refer{Eq:M_V^2Z} below. 
The remainder of this section is devoted to {the} derivation of 
\refer{Eq:G2-vector:pole-part}. Construction of {the} casymptotic states 
corresponding to the propagators in Eqs. 
\refer{Eq:G2:gh-antigh:pole-part}-\refer{Eq:G2-vector:pole-part} is given 
in Sections \ref{Sec:Sub:AsStGen} and \ref{Sec:Sub:AsStGauge}.

Let us start with {the equality} \refer{Eq:As-vect-part}. 
If the limit $q^2\to \mpo$  of  $M_V^{2\,\prime}{(q^2)}$ is finite,  
the form of the  right hand side of \refer{Eq:As-vect-part} follows 
immediately from the analysis 
of the scalar fields propagator carried out in \cite{OnFeRu}. 
However in Sec. \ref{Sec:SubSec:Examples:Z-A-mixing}, we have encountered a 
physically important example in which some of the matrix elements of 
$M_V^{2\,\prime}{(q^2)}$ were IR divergent. Therefore, as proposed
in Sec. 
\ref{Sec:Sub:Prescr-Vectors}, we will only assume that there exists a finite 
limit \refer{Eq:Lim1Der} for each eigenvector $\xi$ of the matrix 
$M_V^{2}(m^2_{V(\la)})$ associated with its eigenvalue $m^2_{V(\la)}$. 
This requires a slight modification of the reasoning presented in 
\cite{OnFeRu}. 

We first need some facts proved in \cite{OnFeRu}. 
Let \koment{str.PSMW2} 
\eq{\nn
R_\la(s)=\left(s\mathds{1}-M^2_V(\mpo)\right)^{-1}\,,
}
($s\equiv q^2$) be a resolvent of $M^2_V(\mpo)$. 
Assuming that each generalized eigenvector (see e.g. \cite{Axler})  of 
$M^2_V(\mpo)$ associated with the eigenvalue $\mpo$ is an (ordinary) 
eigenvector, and using the explicit form \cite{OnFeRu} of $R_\la(s)$ 
written in the 
Jordan basis of $M^2_V(\mpo)$ we can write \kAAAA \koment{str.PSMW3}
\eq{\label{Eq:Res-Form}
(s-\mpo)\,R_\la(s)=\bP(\la)+(s-\mpo)F_\la(s)\,,
}
where $F_\la(s)$ has for $s\to \mpo$ a finite limit $F_\la(\mpo)$, 
while $\bP(\la)$ is the projection onto the eigenspace of $M^2_V(\mpo)$ 
corresponding to its eigenvalue $\mpo$ along the direct sum of 
remaining generalized eigenspaces of $M^2_V(\mpo)$. 
\footnote{The decomposition \refer{Eq:Res-Form} is obvious if 
$M^2_V(\mpo)$ is a diagonalizable matrix.
}
As was shown in \cite{OnFeRu}, the projection $\bP(\la)$ can be written 
as the sum of products 
\eq{\label{Eq:bPla}
\bP(\la)=\sum_r \xi_{[\la_r]}^{\phantom{{\ \rmt}}}\xi_{[\la_r]}^{\ \rmt}\,,
}
where the vectors $\{\xi_{[\la_r]}\}$  form a basis of the 
eigenspace corresponding to $\mpo$ 
and fulfill the following normalization conditions 
\eq{\nn
\xi_{[\la_r]}^{\ \rmt}\xi_{[\la_s]}^{\phantom{{\ \rmt}}}=\de_{rs}\,.
}

Now, let $\sA_\la(s)$ be a matrix such that \koment{str.PSMW2} 
\eq{\label{Eq:LagrTh}
M^2_V(s)=M^2_V(\mpo)+(s-\mpo)\sA_\la(s)\,.
}
Applying the Lagrange's mean value theorem to the matrix elements 
of $M^2_V(s)\proj$, we see that $\sA_\la(s)$ has the following property 
\koment{str.PSMW2,PSMW4,por.PSMW9}    
\eq{\label{Eq:Lim1Der-Gdef}
\lim_{\ \  s\to m^2_{V(\la)}} 
\big\{
\sA_\la(s)\proj\, 
\big\}\,
=
\lim_{\ \  s\to m^2_{V(\la)}} 
\big\{
M^2_V{}^{\prime}(s)\proj
\big\}\,
\equiv \sG_\la\,,
}
because the second limit 
is finite by our assumptions. \kYYY

It is convenient to denote
\eq{
\left(s\mathds{1}-M^2_V(s)\right)^{-1}\equiv R_{\rm tot}(s)\,.
}
We have the obvious equality
\koment{str.PSMW4,por.PSMW2}
\eq{\label{Eq:Rtot-Simp}
R_{\rm tot}(s)=
R_{\la}(s)\!
 \left\{\mathds{1}-(s-\mpo)\sA_\la(s)\,R_{\la}(s)\right\}^{-1}\,,
}
from which it follows that 
\eqs{\label{Eq:Delta-Def-scalars}
\lim_{\ \, s\to \mpo}
\Big\{
(s-\mpo)R_{\rm tot}(s)
\Big\}
&=&\ 
\bP(\la)\big\{1-\sG_\la\big\}^{-1}\,
\,.
}
It is also easy to check that \kBBBB \koment{PSMW5} 
\eq{\label{Eq:zeta-V-sum}
\bP(\la)\big\{\mathds{1}-\sG_\la\big\}^{-1}
=
\sum_r \ze_{V[\la_r]}^{\phantom{{\ \rmt}}}\ze_{V[\la_r]}^{\ \rmt}\,,
}
where the vectors 
\eq{\label{Eq:zeta-V}
\zeta_{V[\la_r]}=\sum_s\cN(\la)^{s}_{\ r}\,\xi_{[\la_s]}\,,
}
form a basis of the eigenspace and obey the normalization condition 
\refer{Eq:norm-cond-GENERAL-vectors}. 
This completes the derivation of the general decomposition
\refer{Eq:As-vect-part}. 

The decomposition \refer{Eq:As-vect-part} is all we need to obtain 
the behavior of propagator \refer{Eq:G2-vector} near its poles at 
$q^2\neq0$. Poles located at $q^2=0$ require, however,  
a refined treatment because of the factor $q_{\mu}q_{\nu}/q^2$. 
Namely, we have to show that \koment{str.MPW53,por.PSMW10-11}
\eq{\label{Eq:Rtot-exp}
s\,R_{\rm tot}(s)
= 
\sum_r \ze_{V[{\bf 0}_r]}^{\phantom{{\ \rmt}}}\ze_{V[{\bf 0}_r]}^{\ \rmt}
+s\,\cR
+s\,\sR(s)\,,
}
where $\sR(s)\to0$ for $s\to0$. If  
\refer{Eq:Rtot-exp} holds, it will directly lead to the decomposition
\refer{Eq:G2-vector:pole-part}. \kCCCC
To ensure that Eq. \refer{Eq:Rtot-exp} does indeed hold, 
we need to assume that the limit
\eq{\label{Eq:cB0}
\cB_0=\frac{1}{2} \lim_{s\to 0} 
\left\{
M^2_V{}^{\prime\prime}\!(s)\,\,\bP({\bf 0}) 
\right\}\,,
} 
is finite. 
The Taylor's theorem then implies that 
\eq{\label{Eq:Taylor}
M^2_V(s)\,\bP({\bf 0})
=
M^2_V(0)\,\bP({\bf 0})
+
s\,\sG_{{\bf 0}}
+s^2\, 
\cB(s)
\,,
}
where $\sG_{{\bf 0}}$ is the limit defined (for $\la={\bf 0}$) 
by  \refer{Eq:Lim1Der-Gdef}, while $\cB(s)\to \cB_0$ when $s\to 0$. 
Moreover, for $s=0$ the imaginary parts of 
all Feynman diagrams contributing to the two-point 1PI function
vanish which implies that the symmetric matrix $M_V^2(0)$ is 
always 
real and, consequently, diagonalizable. \kDDDD 
In particular, the equality \refer{Eq:Res-Form} takes then the form 
\koment{PSMW7-8,por.MPW59}
\eq{\nn 
R_{{\bf 0}}(s)=\frac{1}{s}~\!\bP({{\bf 0}})+F_{{\bf 0}}(s)\,,
}
with 
\eq{\nn
F_{{\bf 0}}(s)=\sum_{M\neq0}(s-M)^{-1}P_M\,,
}
where $M$ runs over (different) nonzero eigenvalues of $M_V^2(0)$ and 
$P_{M}$ is {the} projection onto the eigenspace associated with $M$ 
along the direct sum of remaining eigenspaces of $M_V^2(0)$. 
Defining  {now} 
\kEEEE \koment{str.PSMW8,por.7}
\eq{\nn
{\cX}(s)
=-(\id-\sG_{{\bf 0}}^{\,\rmt})^{-1}
\big\{
s\,\cB(s)^{\rmt}+F_{{\bf 0}}(s)\big[M_V^2(s)-M_V^2(0)\big]
\big\}\,,
}
(clearly, ${\cX}(s) \to 0$ for $s\to0$), and using the relation
\eq{\nn
\bP({\bf 0})(\id-\sG_{{\bf 0}})^{-1}=
(\id-\sG_{{\bf 0}}^{\,\rmt})^{-1}\,\bP({\bf 0})\,,
}
(which is an immediate consequence of the relation 
\refer{Eq:zeta-V-sum}), 
one can prove the following identity \kHHHH \koment{str.PSMW10}
\eqs{
s\,R_{\rm tot}(s)
&=&\ 
\bP({\bf 0})(\id-\sG_{{\bf 0}})^{-1}
+
s\,(\id-\sG_{{\bf 0}}^{\,\rmt})^{-1}\,F_{{\bf 0}}(s)
+\nn
\\[4pt]
&{}&
+s\,(\id+\cX(s))^{-1}
   (\id-\sG_{{\bf 0}}^{\,\rmt})^{-1}
   \big\{\cB(s)^{\rmt} \bP({\bf 0})+F_{{\bf 0}}(s)\sG_{{\bf 0}}\big\}
   (\id-\sG_{{\bf 0}})^{-1}
+\nn
\\[4pt]
&{}&
+s^2\,(\id+\cX(s))^{-1}
   (\id-\sG_{{\bf 0}}^{\,\rmt})^{-1}
    F_{{\bf 0}}(s)\cB(s)
   (\id-\sG_{{\bf 0}})^{-1}
+\nn
\\[4pt]
&{}&
-s\,(\id+\cX(s))^{-1}\cX(s)
   (\id-\sG_{{\bf 0}}^{\,\rmt})^{-1}
    F_{{\bf 0}}(s)
\,.\nn
}
The last two terms tend to zero faster than $s$, 
what gives us the decomposition \refer{Eq:Rtot-exp}; 
looking at the $\cO(s)$ terms we obtain the following formula for $\cR$
\eqs{\nn
\cR
&=&
(\id-\sG_{{\bf 0}}^{\,\rmt})^{-1}\,F_{{\bf 0}}(0)\,\,(\id-\sG_{{\bf 0}})^{-1}
+  (\id-\sG_{{\bf 0}}^{\,\rmt})^{-1}
   \, \cB_0^{\rmt} \bP({\bf 0})\,
   (\id-\sG_{{\bf 0}})^{-1}
\,.
}
Notice that the matrix $\cR$ is symmetric (cf. Eq.~\refer{Eq:cB0}), 
as it should  be. \kFFFF 
For future reference, we rewrite this {formula} in a simpler form. 
To this end, we note that \refer{Eq:Lim1Der-Gdef} and 
\refer{Eq:zeta-V-sum}, together with the definition \refer{Eq:cZ} of 
the $\cZ$ matrix, give 
\koment{por.MPW95.por.62}
\eqs{\label{Eq:cR-fin-form}
\cR
&=&
-\big[(\id-\sG_{{\bf 0}})^{-1}\big]^{\rmt}
\Big\{ 
\sum_{M\neq0}M^{-1}P_M
\Big\}
 (\id-\sG_{{\bf 0}})^{-1}
+\frac{1}{2} \lim_{s\to 0} 
\left[
\cZ M^2_V{}^{\prime\prime}\!(s) \cZ 
\right],\qquad
}
where $(\id-\sG_{{\bf 0}})^{-1}$ can be represented as \kGGGG 
\eq{\label{Eq:M_V^2Z}
(\id-\sG_{{\bf 0}})^{-1}=\id+\lim_{s\to0}[M_V^{2\,\prime}(s)\,\cZ]\,.
}

\kNNNN

\subsection{Propagators with non-simple poles}\label{Sec:Sub:AsStGen} 

As follows from the formulae
\refer{Eq:G2-vector:pole-part}-\refer{Eq:cZ},  in the Landau gauge the
propagators of vector fields have second order poles if in the particle
spectrum of the considered theory massless spin 1 particles are present. 
Therefore, as the first step, we explicitly construct in this section
the generic free field operator whose time-ordered propagator has 
second order poles. In the second step, 
the asymptotic states of a general renormalizable model are reconstructed in Sec.~\ref{Sec:Sub:AsStGauge} 
on the basis of the structure of real poles of the theory propagators 
\refer{Eq:G2:gh-antigh:pole-part}-\refer{Eq:G2-vector:pole-part}.

Consider a set of annihilation and creation operators satisfying 
the following (anti)commutation relations  \koment{str.MPW33}
\eqs{\label{Eq:RelKom-a-a^dagger}
\left[a_{A}(\ve{p}),\ a_{B}(\ve{p'})^\dagger\right]_{\mp}
&=&
g_{A B}\, 2 E_A(\ve{p}) \,(2\pi)^3 \delta^{(3)}(\ve{p}-\ve{p}')\,,
\nn\\[5pt]
\big[a_{A}(\ve{p'}),\ a_{B}(\ve{p})\big]_{\mp}&=&0\,,
}
with upper and lower signs for bosons and fermions, respectively. 
Here the labels $A$, $B$, etc. distinguish different states 
$a^\dagger_{A}(\ve{p})|{\rm 0}\rangle$ with the same momentum $\ve{p}$;  
$E_A(\ve{p})=\sqrt{m^2_A+\ve{p}^2}$ is the energy of the state, while  
$g_{A B}=g_{BA}^{\ \star}$ is a matrix that determines the scalar product 
in the pseudo-Fock space (see e.g. \cite{NAKANISHI}). We assume that
\eq{\nn
g_{AB}=0\,, \qquad {\rm if} \qquad   m_A\neq m_B\,. 
}
and that $g_{AB}\neq0$ only if both states $A$ and $B$ are bosonic 
or both are fermionic.

Out of the operators $a_{A}(\ve{p})$, $a^\dagger_{A}(\ve{p})$
one can construct free fields 
\eq{\label{Eq:PsiTot}
\Psi^I(x)\equiv
\Psi^I_{(-)}(x)+\Psi^I_{(+)}(x)\,,
}
where \koment{str.MPW33} \kJJJ 
\eq{\label{Eq:PsiMin}
\Psi^I_{(-)}(x)
=
\sum_{A}\!\int\!\frac{{\rm d}^3\ve{k}}{(2\pi)^3}\,
\left\{
\frac{{U}^I_{A}(\ve{k})}{2E_A(\ve{k})}
+i\,x_0\,\frac{{R}^I_{A}(\ve{k})}{4E_A(\ve{k})^2}\right\}
 \exp(-i\, \bar{k} x) a_A(\ve{k})
\,,
}
and \koment{str.MPW35}
\eq{\label{Eq:PsiPlus}
\Psi^I_{(+)}(x)
=
\sum_{A}\!\int\!\frac{{\rm d}^3\ve{k}}{(2\pi)^3}\,
\left\{
\frac{{V}^I_{A}(\ve{k})}{2E_A(\ve{k})}
-i\,x_0\,\frac{{S}^I_{A}(\ve{k})}{4E_A(\ve{k})^2}\right\}
 \exp(+i\, \bar{k} x) a_A(\ve{k})^\dagger
\,.
}
{$\bar{k}$ denotes here the on-shell 
four-momentum, $\bar{k}=(\bar{k}^\mu)=(E_A(\ve{k}),\ \ve{k})$, and 
$U$, $R$, $S$ and $V$ are certain functions. The  
non-exponential dependence on time $x_0$ implies \cite{Bec} that the Fourier 
transform $\hat{\Psi}^I(q)$ contains, in addition to the delta function 
$\de(q^2-m_A^2)$, also its derivative $\de '(q^2-m_A^2)$. Such a 
time-dependence is characteristic of 
a non-diagonalizable (\emph{pseudo}Hermitian) Hamiltonian \cite{NAKANISHI}.

The time-ordered propagator \koment{str.MPW34}
\eqs{\label{Eq:PropDef}
\vG^{IJ}\!(x,y)
=
\langle T(\Psi^I(x)\Psi^J(y))\rangle
&=&\!\!\!\!
\phantom{\pm}
\Theta(x^0-y^0)
\left<{\rm 0}\right|\!\Psi^I(x)\Psi^J(y)\!\left|{\rm 0}\right>
+
\nn\\[4pt]
&{}&\!\!\!\!
\pm
\Theta(y^0-x^0)
\left<{\rm 0}\right|\!\Psi^J(y)\Psi^I(x)\!\left|{\rm 0}\right>
\!,\ \ \ \ \ \ \ 
}
(with  the upper and lower signs corresponding to bosons and 
fermions, respectively) can be easily found by  applying the standard 
textbook procedure \cite{WeinT1}. \kHHH
In particular, the $\Th$ functions can be traded for an integral over an 
independent time component $k^0$ of the momentum. 
In order that the explicit time factors do not spoil the 
translational invariance of the propagator,  
the functions $U$, $R$, $S$ and $V$ have to satisfy, 
(for each value of mass $m$) 
the following 
consistency conditions  
\eq{\label{Eq:TrInvCond1}
{\sum_A}^{(m)} {\sum_B}^{(m)}
\left\{
  {R}^I_{A}(\ve{k})\,g_{AB}\,{V}^J_{B}(\ve{k})
   -
  {U}^I_{A}(\ve{k})\,g_{AB}\,{S}^J_{B}(\ve{k})
\right\}
=0\,,
}  
\eq{\label{Eq:TrInvCond2}
{\sum_A}^{(m)} {\sum_B}^{(m)}
  {R}^I_{A}(\ve{k})\,g_{AB}\,{S}^J_{B}(\ve{k})
=0\,,
}
in which the sums run over the indices $A$ and $B$ labeling the states of mass 
$m$.
If  these conditions are satisfied, the propagator $\vG^{IJ}\!(x,y)$ 
still contains
explicit factors of time, but only in the combination  $(x^0-y^0)$ 
which can be eliminated by integrating by parts.
It is this operation which
gives rise to second order poles in the momentum space propagator
$\widetilde{\vG}^{IJ}\!(k,-k)$ in 
\koment{str.MPW15,MPW44,MPW42} \kIII
\eq{\label{Eq:Prop-Full-Fourier-Conv}
\vG^{IJ}\!(x,y)
=\int\volfour{k}\,\, e^{-i\,k(x-y)}\,\,\widetilde{\vG}^{IJ}\!(k,-k)\,, 
}
which takes then the form 
\eqs{\label{Eq:PropJawGeneral}
\widetilde{\vG}^{IJ}\!(k,-k)
&=&i\,\sum_m
\bigg\{
\frac{k^0\,\sA_m^{IJ}\!(\ve{k})+\sB_m^{IJ}\!(\ve{k})-\sC_m^{IJ}\!(\ve{k})}
     {k^2-m^2+i\,\vep}
+\nn\\[4pt] 
&{}&\phantom{i\,\sum_m\bigg\{}
-2\,
\frac{k^0\,\sD_m^{IJ}\!(\ve{k})+E_m(\ve{k})^2\,\sC_m^{IJ}(\ve{k})}
     {\left[k^2-m^2+i\,\vep\right]^2}
\bigg\},
}
with 
$E_m(\ve{k})=\sqrt{m^2+\ve{k}^2}$, and
\eqs{\nn
\sA_m^{IJ}\!(\ve{k})&=& \frac{\sQ_m^{(+)IJ}\!(\ve{k})-\sQ_m^{(-)IJ}\!(\ve{k})}
                             {2\,E_m(\ve{k})}\,,
\nn\\[5pt]
\sB_m^{IJ}\!(\ve{k})&=& \frac{\sQ_m^{(+)IJ}\!(\ve{k})+\sQ_m^{(-)IJ}\!(\ve{k})}
                             {2}\,,
\nn\\[5pt]
\sC_m^{IJ}\!(\ve{k})&=& \frac{\sT_m^{(+)IJ}\!(\ve{k})+\sT_m^{(-)IJ}\!(\ve{k})}
                             {4\,E_m(\ve{k})^2}\,,
\nn\\[5pt]
\sD_m^{IJ}\!(\ve{k})&=& \frac{\sT_m^{(+)IJ}\!(\ve{k})-\sT_m^{(-)IJ}\!(\ve{k})}
                             {4\,E_m(\ve{k})}\,,
}
where \koment{str.MPW40,str.MPW44}
\eqs{\nn
\sQ_m^{(+)IJ}\!(\ve{k})
 &=&
{\sum_A}^{(m)} {\sum_B}^{(m)} {U}^I_{A}(\ve{k})\,g_{AB}\,{V}^J_{B}(\ve{k})\,,
\nn\\[5pt] 
\sQ_m^{(-)IJ}\!(\ve{k})
 &=&
 \pm{\sum_A}^{(m)} {\sum_B}^{(m)}
 {U}^J_{A}(-\ve{k})
                                 \,g_{AB}\,{V}^I_{B}(-\ve{k})
 =\pm\sQ_m^{(+)JI}\!(-\ve{k})\,,
\nn\\[5pt] 
\sT_m^{(+)IJ}\!(\ve{k})
 &=&
 {\sum_A}^{(m)} {\sum_B}^{(m)}{R}^I_{A}(\ve{k})\,g_{AB}\,{V}^J_{B}(\ve{k})\,,
\nn\\[5pt] 
\sT_m^{(-)IJ}\!(\ve{k})
 &=&
 \pm{\sum_A}^{(m)} {\sum_B}^{(m)}{R}^J_{A}(-\ve{k})
                                 \,g_{AB}\,{V}^I_{B}(-\ve{k})
 =\pm\sT_m^{(+)JI}\!(-\ve{k})\,,
}  
with upper and lower signs for bosons and fermions, respectively. 

\subsection{Asymptotic states} \label{Sec:Sub:AsStGauge}  

We are now in a position to reconstruct the asymptotic states 
on the basis of the stucture 
\refer{Eq:G2:gh-antigh:pole-part}-\refer{Eq:G2-vector:pole-part}
of real poles of the theory propagators, which is the essence of the
LSZ asymptotic formalism.
Similar analysis was carried out in Ref.~\cite{KugoOjima},
where it was applied to some specific 
gauge theory models.
Here we generalize it (in the Landau gauge) to the general case allowing
for an arbitrary mixing of fields.
The first step is to 
choose the basis of the subspace of unphysical states. 
It will be convenient to work with the  state vectors 
$b_\be(\ve{k})^\dagger|{\rm 0}\rangle$  representing the 
Nakanishi-Lautrup (NL) modes, and the 
states $d^\be(\ve{k})^\dagger|{\rm 0}\rangle$ of ``scalar gauge bosons".  
\kWWWW  
Writing now the  asymptotic field 
\eq{\label{Eq:h_be-asymp}
\vh_\be(x)
=
\int\!\frac{{\rm d}^3\ve{k}}{(2\pi)^3\,2|\ve{k}|}\,
\left\{
 \exp(-i\, \bar{k} x)\, b_\be(\ve{k})
 +
 \exp(+i\, \bar{k} x)\, b_\be(\ve{k})^\dagger
\right\}
\,,
}
associated with the Nakanishi-Lautrup multiplier $h_\be(x)$,
in which $\bar{k}=(\bar{k}^\mu)\equiv(|\ve{k}|,\ \ve{k})$, and taking into 
account that the propagator of the (interpolating) NL fields 
has no real poles (in fact it vanishes identically, cf. \refer{Eq:G2:hh})
we conclude that 
\eqs{\label{Eq:RelKom-bb}
\left[b_{\al}(\ve{p'}),\ b_{\be}(\ve{p})^\dagger\right]_{-}
=\left[b_{\al}(\ve{p'}),\ b_{\be}(\ve{p})^{\phantom{\dagger}}\!\!\right]_{-}=0
\,.
}
From this it follows that $b_\be(\ve{k})^\dagger|{\rm 0}\rangle$ 
is a zero-norm state. 

Next, we assume the following decomposition of the 
asymptotic vector field \kXXX
\eq{\label{Eq:AsymField-vec-TOTAL}
\vA^\al_{\mu}
 = 
\bV^\al_{\mu}
+
\pa_\mu \bS^\al + \bL^\al_{\mu} 
\,,
}
in which the scalar field $\bS^\al$ is built out of the annihilation and 
creation operators of the ``scalar gauge bosons'' 
\eq{\label{Eq:S^al-asymp}
\bS^{\al}(x)
=
\int\!\frac{{\rm d}^3\ve{k}}{(2\pi)^3\,2|\ve{k}|}\,
\left\{
 \exp(-i\, \bar{k} x)\, d^\al(\ve{k})
 +
 \exp(+i\, \bar{k} x)\, d^\al(\ve{k})^\dagger
\right\}
\,,
}
and the ``longitudinal" massless
vector field $\bL^\al_{\mu}$ involves the operators of the NL zero norm states 
\eq{\label{Eq:L_al-asymp}
\bL^\al_{\mu}(x)
=
\cZ^{\al\be}\!
\int\!\frac{{\rm d}^3\ve{k}}{(2\pi)^3\,2|\ve{k}|}\,
\left\{
  e^{-i\, \bar{k} x}\, 
  \left[i\frac{(\cP\bar{k})_\mu}{4|\ve{k}|^2}
        -\frac{\bar{k}_\mu}{2|\ve{k}|}   x_0
 \right]
  b_\be(\ve{k})
 +{\rm H.c.}
 \right\}
\,,
}
(here $\cP\bar{k}=((\cP\bar{k})^\mu)\equiv(|\ve{k}|,\ -\ve{k})$ denotes 
the parity transformed momentum $\bar{k}$). 
The field $\bV^\al_{\mu}$ creates and annihilates only the physical states. 
(Clearly, the creation and annihilation operators of 
physical states commute with the ones associated with the unphysical states). 

As can be easily checked, the asymptotic fields $\vh_\be$ and 
$\vA^\al_{\mu}$ correctly reproduce the behavior of the 
mixed propagator \refer{Eq:G2:hA:pole-part} 
in the vicinity of its pole  if 
\koment{str.MPW65}
\eqs{\label{Eq:RelKom-b-d^dagger}
\left[b_{\be}(\ve{k}),\ d^\al(\ve{q})^\dagger\right]_{-}
&=&
\left[d^\al(\ve{q}),\ b_{\be}(\ve{k})^\dagger\right]_{-}
=
\de^{\al}_{\ \be}\, 
2 |\ve{k}| \,(2\pi)^3 \delta^{(3)}(\ve{k}-\ve{q})\,.\phantom{aa}
}
Moreover, the pole structure
\refer{Eq:G2-vector:pole-part} is 
reproduced by the time-ordered propagator of  the  asymptotic fields 
$\vA^\al_\mu$ provided \koment{str.MPW65} \kLLL
\eq{\label{Eq:RelKom-d-d^dagger}
\left[d^\al(\ve{k}),\ d^{\be}(\ve{q})^\dagger\right]_{-}
=\cR^{\al\be}\, 2 |\ve{k}| \,(2\pi)^3 \delta^{(3)}(\ve{k}-\ve{q})\,,
}
and  
\eq{\label{Eq:bV}
\bV^\al_{\mu}
=
{\sum_{  \la}}^\prime\sum_r
{\zeta}^\al_{V[\la_r]}  
\bA^{\la_r}_{\mu}\,.
}
(As before, the prime over the first sum indicates that the summation 
is restricted to indices $\la$ corresponding to poles on the real axis). 
Here $\bA^{\la_r}_{\mu}$ is  the free vector field  (in the unitary 
gauge  if 
$m_{V(\la)}\neq0$,  
or 
in the Coulomb gauge, if $m_{V(\la)}=0$)
of spin 1 particles of mass $m_{V(\la)}$. 
The $\bA^{\la_r}_{\mu}$ field is  canonically normalized. 
For completeness we give here its explicit form\footnote{See e.g. 
\cite{WeinT1}; we use slightly  more common 
normalization conventions, however.}  
\eq{\label{Eq:bA}
\bA^{\la_r}_{\mu}(x)
=
\sum_{\ch}\int\!\frac{{\rm d}^3\ve{k}}{(2\pi)^3\,2\sqrt{m_{V(\la)}^2+\ve{k}^2}}\,
\left\{
 \exp(-i\, \bar{k} x)\, e_\mu^\ch(\ve{k},m_{V(\la)})\, a^{\la_r}_\ch(\ve{k})
 +
{\rm H.c.} 
\right\} ,
}
with $\bar{k}=(\bar{k}^\mu)\equiv(\sqrt{m_{V(\la)}^2+\ve{k}^2},\ \ve{k})$, and
\eqs{\nn
\left[a^{\la\phantom{'}_{\!\!r\phantom{'}}}_{\ch \!\!\phantom{'}}\!\!(\ve{k}),\ a^{\la'_{r'}}_{\ch'}\!(\ve{q})^\dagger\right]_{-}
&=&
\de_{\ch \!\!\phantom{'} \ch'} \, 
\de_{{\la\!\!\phantom{'}} \la'} \, 
\de_{{r\!\!\phantom{'}} r'} \, 
2 \sqrt{m_{V(\la)}^2+\ve{k}^2} \,(2\pi)^3 \delta^{(3)}(\ve{k}-\ve{q})\,,
}
where $\ch$ and $\ch'$ run over  the helicity values
$\pm1,0$ (if $m_{V(\la)}\neq0$) or $\pm1$ (if $m_{V(\la)}=0$).
The explicit form of the polarization vectors is 
\eq{\nn 
e_\mu^{-}(\ve{p},m)
=
-e_\mu^{+}(\ve{p},m)^\star
=-\frac{1}{\sqrt{2}\,|\ve{p}|\,\sqrt{(p^1)^2+(p^2)^2}}
\left[\begin{array}{c}
0\\[4pt]
p^1 p^3 + i\, p^2 |\ve{p}|
\\[4pt]
p^2 p^3 - i\, p^1 |\ve{p}|
\\[4pt]
-(p^1)^2-(p^2)^2
\end{array}\right]_{\!{}^\mu}
\,,
}
and 
\eq{\nn
e^0_\mu(\vk,m)
=
-\,\frac{\sqrt{\vk^2+m^2}}{m\,|\vk|} 
\left[
\begin{array}{c}
\frac{\vk^2}{\sqrt{\vk^2+m^2}}\\[7pt]
-\vk
\end{array}
\right]_{\!{}^\mu}\,.
}

It is worth stressing that it is precisely the 
second 
line of \refer{Eq:G2-vector:pole-part} which
fixes the form \refer{Eq:L_al-asymp} of the longitudinal field 
$\bL^\al_{\mu}$. 
In particular, $\bL^\al_{\mu}$  is nonzero only if in the theory spectrum there 
are massless spin 1 particles (cf. the definition \refer{Eq:cZ} of the 
$\cZ$ matrix).  
For this reason we have called $\bL^\al_{\mu}$ the ``longitudinal" field: 
it creates massless gauge bosons of zero helicity (clearly, they form a subspace 
of the Nakanishi-Lautrup modes).  

It remains to construct 
the asymptotic fields $\vphi^i$ associated with the interpolating
scalar fields 
$\phi^i$. The decomposition of $\vphi^i$ which reproduces the 
structure \refer{Eq:G2-scalar:pole-part} of the poles on the real axis
of the scalar fields propagator 
follows immediately from the prescription  formulated 
in Sec.~\ref{Sec:Sub:Scalars} and is given by  \refer{Eq:AsymField}. 
However,  it is still necessary to split this asymptotic field
into its parts creating/annihilating 
physical and unphysical states. 
To this end it is better to forget Eq. \refer{Eq:AsymField} altogether and 
write down the decomposition of $\vphi^i$  in terms of fields creating 
(yet unknown) physical and unphysical states \koment{str.MPW65}
\eqs{\label{Eq:AsymField-NOWE}
\vphi^j
&=&
\vphi^j_{\rm ph}+\vphi^j_{\rm unph}
\,.
}
Let us introduce  the matrix (cf. Eqs. 
\refer{Eq:Antifields-Gh:Scalar}-\refer{Eq:Antifields-Gh:Vector}) 
\koment{str.MPW84}
\eq{\label{Eq:Def-tC}
{\tC}^{j}_{\ \ga}(q^2)=B(q^2)^{j}_{\ \be}\,[\Om(q^2)^{-1}]^{\be}_{\ \ga}
\,,
}
together with its limit
\eq{\label{Eq:Def-C}
{C}^{j}_{\ \ga}={\tC}^{j}_{\ \ga}(0)\,.
} 
The structure \refer{Eq:G2:h-phi:pole-part} of the 
pole
of the mixed $\phi h$ propagator 
and the lack of poles of the (vanishing identically in the Landau gauge)
mixed scalar-vector propagator (cf.
\refer{Eq:G2-scalar-vector})
are correctly reproduced by  $\vphi^j_{\rm unph}$ of 
\refer{Eq:AsymField-NOWE}, if  
\koment{str.MPW65,MPW66,MPW70}
\eqs{\label{Eq:AsymField-NOWE-unph}
\vphi^j_{\rm unph}(x)
&=&
C^j_{\ \ga}\, \bS^\ga(x)
-C^j_{\ \be}\,\cR^{\be\ga}\, \vh_\ga(x)
\,.
}
Vanishing of the mixed propagator of the asymptotic fields
$\vphi^j$ 
and $\vA^\al_{\mu}$ (given by
\refer{Eq:AsymField-vec-TOTAL}), necessary to reproduce
\refer{Eq:G2-scalar-vector}, hinges on 
the following relation \koment{str.MPW70,por.MPW66} \kKKK 
\eq{\label{Eq:ConstCondImp}
C^j_{\ \ga}\, \cZ^{\ga\be} = 0
\,,
}
whose validity can be seen as follows. 
Let us rewrite the STids \refer{Eq:STid1prime}-\refer{Eq:STid7prime} as 
\koment{str.MPW71}
\eq{
\label{Eq:STid1prime-mod}
P_{\be j}(q^2)\,\tC^j_{\ \al}(q^2)
=  
\left\{ q^2\, \sL_{\al\be}(q^2) + [M_V^2(q^2)-q^2\mathds{1}]_{\be\al} \right\}
\,,
}
and \koment{MPW84}
\eq{
\label{Eq:STid7prime-mod}
q^2\,  P_{\al j}(q^2)
=  
[q^2\mathds{1}-M_S^2(q^2)]_{ij}\, \tC^i_{\ \al}(q^2)
\,.
}
For $q^2\to0$ these relations reduce to 
\koment{str.MPW71}
\eq{
\label{Eq:STid1prime-lim}
M_V^2(0)_{\be\al}
=
P_{\be j}(0)\,C^j_{\ \al}
\,,
}
and \koment{MPW84}
\eq{
\label{Eq:STid7prime-lim}
M_S^2(0)_{ji}\, C^i_{\ \al}=0
\,.
}
Now Eq. \refer{Eq:STid7prime-mod} gives \koment{str.MPW84}
\eq{
\label{Eq:STid7prime-mod2}
P_{\al j}(q^2)\,C^j_{\ \be}
=  
\frac{1}{q^2}\, \tC^i_{\ \al}(q^2)\,
   [q^2\mathds{1}-M_S^2(q^2)]_{ij}\, 
C^j_{\ \be}
\,,
}
and using \refer{Eq:STid1prime-lim}-\refer{Eq:STid7prime-lim} we get for 
$q^2\to 0$  \koment{str.MPW84}
\eq{\label{Eq:M_V^2(0)}
M_V^2(0)_{\al\be}
=  
\lim_{q^2\to 0} 
\left\{
C^i_{\ \al}\,
   [\mathds{1}-M_S^{2\,\prime}(q^2)]_{ij}\, 
C^j_{\ \be}
\right\},
}
provided the limit 
\eq{\nn
\lim_{q^2\to 0} 
\left\{M_S^{2\,\prime}(q^2)_{ij}\, C^j_{\ \be}\right\},
}
exists. \kMMM \kOOO \kUUU 
Since for $q^2=0$ the reality of $M_V^2(q^2)$ cannot be violated, 
the matrix $M_V^2(0)$ has an orthonormal 
basis of real eigenvectors $\th_{(M,n)}=(\th_{(M,n)}^\be)$, where $n$ 
distinguishes different eigenvectors $\th_{(M,n)}$ corresponding to the 
eigenvalue $M$. 
Let us introduce the following set of vectors 
\koment{str.MPW92,MPW88,MPW106}
\eq{\label{Eq:zeta-Gold}
\ze_{(M,n)}^j
=\frac{1}{\sqrt{M}}\,C^{j}_{\ \ga}\,\th_{(M,n)}^{\ga}\,, 
\qquad {\rm for} \ \quad M\neq 0\,,
}  
and \koment{str.MPW106}
\eq{\nn 
\xi_{(n)}^j
=C^{j}_{\ \ga}\,\th_{(0,n)}^{\ga}\,.
}  
Eq. \refer{Eq:M_V^2(0)} now yields 
\eq{\label{Eq:norm-cond-Gold}
\lim_{q^2\to 0} 
\left\{
\ze_{(M,n)}^{\,\, \rmt} \,  
\left[
\mathds{1} - M^2{}^{\prime}_{\!\!\!\!S\, } (q^2)
\right]\,
\ze_{(M',n')}
\right\}
=\delta_{MM'}\, \delta_{nn'}\,, 
}
as well as 
\eq{\label{Eq:norm-cond-nieGold2}
\lim_{q^2\to 0} 
\left\{
\ze_{(M,n)}^{\,\, \rmt} \,  
\left[
\mathds{1} - M^2{}^{\prime}_{\!\!\!\!S\, } (q^2)
\right]\,
\xi_{(n')}
\right\}
=0\,,
}
and 
\eq{\label{Eq:norm-cond-nieGold}
\lim_{q^2\to 0} 
\left\{
\xi_{(n)}^{\,\, \rmt} \,  
\left[
\mathds{1} - M^2{}^{\prime}_{\!\!\!\!S\, } (q^2)
\right]\,
\xi_{(n')}
\right\}
=0\,.
}
Eq. \refer{Eq:norm-cond-Gold} shows that the vectors
$\ze_{(M,n)}$  are linearly independent. 
Then \refer{Eq:norm-cond-nieGold2} shows that none of 
$\xi_{(n')}$ is a linear combination of $\ze_{(M,n)}$. In fact, 
Eq. \refer{Eq:norm-cond-nieGold} implies that all $\xi_{(n)}$ vanish; this would 
be obvious if the limit $M_S^{2\,\prime}(0)$ was finite.  
Indeed, $M_S^{2\,\prime}(q^2)=\cO(\hb)$ and $M_S^{2\,\prime}(q^2)$ is  for $q^2<0$
a real symmetric matrix; thus $\id-M_S^{2\,\prime}(0)$ 
is positive definite (for perturbative values of coupling constants)  provided 
the limit exists. 
We do not assume finiteness of the whole matrix $M_S^{2\,\prime}(0)$.   
Nonetheless, finiteness of 
the limit in Eq. \refer{Eq:norm-cond-nieGold} simply means the 
cancellation of certain $\ln(q^2)$ divergences; 
therefore Eq. \refer{Eq:norm-cond-nieGold} cannot be satisfied for a nonzero 
vector $\xi_{(n')}=\xi_{(n)}$, at least in the perturbative regime.  \kXXXX
Hence, we have the following equivalence  \kNNN 
\eq{\label{Eq:M_V-equiv}
M_V^2(0)_{\al\be}\La^\be=0
\qquad\quad 
\Leftrightarrow
\qquad\quad 
C^i_{\ \be}\,\La^\be=0\,.
} 
Recall now that coefficients $\ze^\be_{V[\mathbf{0}_r]}$ in the formula 
\refer{Eq:cZ} for the $\cZ^{\be\de}$ matrix are null eigenvectors of 
$M_V^2(0)$; thus we have proved Eq. \refer{Eq:ConstCondImp}. 

Finally we have to consider the scalar-scalar propagators. 
The time-ordered propagator 
\eq{\label{Eq:PropUnph-Formal}
\langle T(\vphi^l_{\rm unph}(x)\vphi^j_{\rm unph}(y))\rangle
=\int\volfour{k}\,\, e^{-i\,k(x-y)}\,\,\widetilde{\vG}^{lj}\!(k,-k)_{\rm unph}
\,,
}
of unphysical fields \refer{Eq:AsymField-NOWE-unph} is easy to find: 
\eq{\nn
\widetilde{\vG}^{lj}\!(k,-k)_{\rm unph}
=-\frac{i}{k^2}\,C^{l}_{\ \be}\,C^{j}_{\ \ga}\, \cR^{\be\ga}\,.
}
The explicit form \refer{Eq:cR-fin-form}-\refer{Eq:M_V^2Z} of the matrix 
$\cR^{\be\ga}$ together with the identity \refer{Eq:ConstCondImp} allow us 
to simplify this expression to
\eq{\label{Eq:PropUnph}
\widetilde{\vG}^{lj}\!(k,-k)_{\rm unph}
=\frac{i}{k^2}\,C^{l}_{\ \be}\,C^{j}_{\ \ga}\, 
\sum_{M \neq 0} \frac{1}{M}\,(P_{M})^{\be\ga}
\,.
}
Recall that $M$ runs over (different) nonzero eigenvalues of $M_V^2(0)$ 
while $P_{M}$ is a  
projection onto the eigenspace associated with $M$ along the direct sum of 
the remaining eigenspaces of $M_V^2(0)$; in 
particular \koment{str.MPW88}
\eq{\nn
(P_{M})^{\be\ga}=\sum_n \th_{(M,n)}^{\be}\th_{(M,n)}^{\ga}\,.
}
Comparing  \refer{Eq:PropUnph} rewritten in terms of the vectors 
\refer{Eq:zeta-Gold},
\eq{\nn
\widetilde{\vG}^{lj}\!(k,-k)_{\rm unph}
=\frac{i}{k^2} 
\sum_{M \neq 0}\sum_n \ze^{l}_{(M,n)}\,\ze^{j}_{(M,n)}
\,,
}
with Eq. \refer{Eq:G2-scalar:pole-part}, one 
can identify $\ze_{(M,n)}$ as the 
eigenvectors $\ze_{S[\ell_r]}$ associated with the would-be Goldstone bosons. 
Indeed, the equality \refer{Eq:STid7prime-lim} shows that $\ze_{(M,n)}$ 
are null eigenvectors 
of $M_S^2(0)$, while \refer{Eq:norm-cond-Gold} implies that they  satisfy 
the (refined version of the) normalization conditions 
\refer{Eq:norm-cond-GENERAL-scalars}, 
in complete agreement with the general prescription for finding 
$\ze_{S[\ell_r]}$ described in Sec. \ref{Sec:Sub:Scalars}.
The number of vectors \refer{Eq:zeta-Gold} 
equals to the dimension of the gauge group minus the number of massless gauge 
bosons, as required by the counting of degrees of freedom based
on the Goldstone theorem. \kQQQ \kVVV

Of course, all the eigenvectors ${\zeta}_{S[\ell_r]}$ corresponding to the 
same pole at a value $m_{S(\ell)}^2$ of $p^2$} have to obey the 
orthogonality conditions  
\refer{Eq:norm-cond-GENERAL-scalars} in order to ensure the expansion 
\refer{Eq:D-as-GENERAL-scalars-indi} of the propagator.\footnote{More 
precisely, as we have shown in Sec. \ref{Sec:Sub:AsymProp}, only 
the ``refined version" \refer{Eq:norm-cond-GENERAL-vectors} of 
the normalization 
conditions is needed for Eq. \refer{Eq:As-vect-part} to hold true. 
Clearly, the same is true for its scalar counterpart 
\refer{Eq:D-as-GENERAL-scalars-indi}. } 
Therefore the physical massless eigenvectors ${\zeta}_{S[{\ell}_r]}$ 
must be orthogonal (with respect to the scalar product 
\refer{Eq:norm-cond-GENERAL-scalars}) to the unphysical ones  
\refer{Eq:zeta-Gold}. This is equivalent to the condition 
\refer{Eq:norm-cond-GENERAL-scalars-phys}  owing to the 
``non-renormalization theorem" \refer{Eq:Antigh-Conclusion}. 
In particular, the physical part $\vphi_{\rm ph}$ of the asymptotic scalar field 
\refer{Eq:AsymField-NOWE} can be written as 
\eq{\label{Eq:phys-scalar-field}
\vphi^j_{\rm ph} = 
\sum_{ {{\rm phys.}\,r}}
{\zeta}^j_{S[{\bf 0}_r]}  
\Phi^{{\bf 0}_r}
+
{\sum_{  \ell\neq {\bf 0}}}^{\prime}
\sum_r
{\zeta}^j_{S[\ell_r]}  
\Phi^{\ell_r}
\,,
}
where  in the first sum corresponding to poles at $p^2=0$
(labeled by $\ell={\bf 0}$) 
the summation is over  the indices $r$ corresponding to physical 
eigenvectors 
${\zeta}^j_{S[{\bf 0}_r]}$,  satisfying (for each generator $\TS_\al$) the 
following condition    
\eq{\label{Eq:scalars-phys}
\lim_{q^2\to 0} 
\left\{
\ze_{S[{\bf 0}_r]}^{\,\, \rmt} 
\left[
\mathds{1} - M^2{}^{\prime}_{\!\!\!\!S\, } ({q^2})
\right]
\TS_\al v
\right\}
=0\,.
}
As before, the prime over the second sum in \refer{Eq:phys-scalar-field} 
indicates that the summation is restricted to the poles located on the real 
axis. In our conventions the canonically normalized free scalar field 
$\Phi^{\ell_r}$ of mass $m_{S(\ell)}$  has the form 
\eq{\label{Eq:Phi}
\Phi^{\ell_r}(x)
=
\int\!\frac{{\rm d}^3\ve{k}}{(2\pi)^3\,2\sqrt{m_{S(\ell)}^2+\ve{k}^2}}\,
\left\{
 \exp(-i\, \bar{k} x)\, a^{\ell_r}(\ve{k})
 +
 \exp(+i\, \bar{k} x)\, a^{\ell_r}(\ve{k})^\dagger
\right\}
\,,
}
with $\bar{k}=(\bar{k}^\mu)\equiv(\sqrt{m_{S(\ell)}^2+\ve{k}^2},\ \ve{k})$, and
\eqs{\nn
\left[a^{\ell\phantom{'}_{\!\!r\phantom{'}}}\!\!(\ve{k}),\ a^{\ell'_{r'}}\!(\ve{q})^\dagger\right]_{-}
&=&
\de_{{\ell\!\!\phantom{'}} \ell'} \, 
\de_{{r\!\!\phantom{'}} r'} \, 
2 \sqrt{m_{S(\ell)}^2+\ve{k}^2} \,(2\pi)^3 \delta^{(3)}(\ve{k}-\ve{q})\,.
}
The time-ordered propagator of the complete asymptotic
scalar field \refer{Eq:AsymField-NOWE} 
\eq{\nn
\widetilde{\vG}^{lj}\!(k,-k)=
\widetilde{\vG}^{lj}\!(k,-k)_{\rm ph}+\widetilde{\vG}^{lj}\!(k,-k)_{\rm unph}
\,,
}
(where $\widetilde{\vG}^{lj}\!(k,-k)_{\rm ph}$ is defined analogously to 
\refer{Eq:PropUnph-Formal}) matches then the form
\refer{Eq:G2-scalar:pole-part} that the complete scalar fields  propagator
takes near its poles on the real axis. In particular, the states 
of physical massless scalars in the pseudo-Fock space are 
orthogonal to the states of would-be Goldstone modes. 

It should be stressed that the asymptotic fields 
$(\Psi^{I})=(\vphi^i,\ \vA^{\al}_{\mu},\ \vh_{\be})$ do obey the consistency 
conditions \refer{Eq:TrInvCond1}-\refer{Eq:TrInvCond2} and that 
almost all of their time-ordered propagators $\vG^{IJ}(x,y)$, obtained 
using the general formula
\refer{Eq:PropJawGeneral}, 
indeed exactly reproduce  
the appropriate expressions $G^{IJ}(x,y)_{\rm pole}$ listed in Eqs. 
\refer{Eq:G2:hA:pole-part}-\refer{Eq:G2-vector:pole-part}. 
The sole exception is the propagator of  the 
time components of  the vector fields:
\eq{\nn
\widetilde{\vG}^{\be\de}_{00}(q,-q)
=\widetilde{G}^{\be\de}_{00}(q,-q)_{\rm pole}
-i\,\cR^{\be\de}
-i
{\sum_{\la\neq {\bf 0} }}^{\prime}
\frac{1}{m_{V(\la)}^{2}}
\sum_r
\ze^\be_{V[\la_r]} 
\ze^\de_{V[\la_r]}\,.
}
The difference affects only the non-pole parts and therefore is 
irrelevant for the structure of asymptotic states. 

This completes the construction 
(in the Landau gauge) of the space of 
asymptotic states  in general gauge theories with an  
arbitrary mixing of fields. 
It is however worthwhile to show that, also in the presence of generic 
mixing, the unphysical asymptotic states  do have the structure  
discussed in \cite{Bec,KugoOjima}
which is required for unitarity 
of the $S$-operator restricted to the subspace of 
physical states.   
 \kLLLL 
To this end,  let us, following \cite{Bec}, introduce 
the generator of the  BRST transformations acting on the 
asymptotic fields (compare the formulae \refer{Eq:BRS}) \kRRR
\koment{str.MPW97,96}  
\eqs{\label{Eq:BRS-asym}
& i\big[Q_{BRST},\ \vphi^j\big]_{-}
=B(0)^{j}_{\ \ga}\, \vom^{\ga}\,,
\qquad\qquad\qquad
& i\big[Q_{BRST},\ \vpsi^a\big]_{+}=0\,,
\nonumber\\[5pt]
& i\big[Q_{BRST},\ \vA^\al_\mu\big]_{-}
=\Om(0)^{\al}_{\ \ga}\, \partial_\mu\vom^{\gamma}\,,
\qquad\qquad\ \  \
& i\big[Q_{BRST},\ \vom^\alpha\big]_{+}=0\,,
\nonumber\\[5pt]
& i\big[Q_{BRST},\ \overline{\vom}_\al^{\phantom{\al}}\big]_{+} =\vh_\alpha,
\qquad\qquad\qquad\qquad\quad
& i\big[Q_{BRST},\ \vh_\alpha^{\phantom{\al}}\big]_{-}=0\,,
}
with  the asymptotic (anti)ghosts fields  having the forms
\koment{str.MPW47} 
\eqs{\label{Eq:om-asymp}
\anti{\vom}_\be(x)
&=&
-i
\int\!\frac{{\rm d}^3\ve{k}}{(2\pi)^3\,2|\ve{k}|}\,
\left\{
 \exp(-i\, \bar{k} x)\, \anti{b}_\be(\ve{k})
 +
 \exp(+i\, \bar{k} x)\, \anti{b}_\be(\ve{k})^\dagger
\right\}
\,,\nn\\[5pt]
{\vom}^\be(x)
&=&
\int\!\frac{{\rm d}^3\ve{k}}{(2\pi)^3\,2|\ve{k}|}\,
\left\{
 \exp(-i\, \bar{k} x)\, \anti{d}^\be(\ve{k})
 +
 \exp(+i\, \bar{k} x)\, \anti{d}^\be(\ve{k})^\dagger
\right\}
\,.
}
The only non-vanishing anticommutators of the operators
$\anti{b}_\be(\ve{k})$, $\anti{d}_\be(\ve{k})$, etc. 
are \koment{str.MPW51,52}
\eqs{\label{Eq:RelKom-anti-b-d^dagger}
\big[\anti{d}^\al(\ve{q}),\ \anti{b}_{\be}(\ve{k})^\dagger\big]_{+}
&=&
\big[\anti{b}_{\be}(\ve{k}),\ \anti{d}^\al(\ve{q})^\dagger\big]_{+}^\dagger
=
-i [\Om(0)^{-1}]^{\al}_{\ \be}\, 
2 |\ve{k}| \,(2\pi)^3 \delta^{(3)}(\ve{k}-\ve{q})\,.
\nn\\
}
This ensures that the time-ordered propagator 
of the ghost fields \refer{Eq:om-asymp} matches  the expression 
\refer{Eq:G2:gh-antigh:pole-part}.\footnote{We note that $\Om(0)$ is 
real,  because 
Feynman integrals contributing 
to $\Om(q^2)$ cannot acquire 
imaginary parts for $q^2=0$. \kUUUU
}
\kSSS 
The charge $Q_{BRST}$ is a pseudoHermitian and nilpotent operator. It is easy 
to check that it can be represented as \koment{str.MPW98,99}
\eqs{\label{Eq:Q-BRS-asymp}
Q_{BRST}
&=&
-i\,\Om(0)^{\al}_{\ \be}\,
\int\!\frac{{\rm d}^3\ve{k}}{(2\pi)^3\,2|\ve{k}|}\,
\left\{
\anti{d}^\be(\ve{k})^\dagger\,  {b}_\al(\ve{k})
 -
{b}_\al(\ve{k})^\dagger\,  \anti{d}^\be(\ve{k})
 \right\}
\,,
}
and thus anticommutation relations \refer{Eq:RelKom-anti-b-d^dagger}, 
as well as  $Q_{BRST}$, have the standard Kugo-Ojima form \cite{KugoOjima} 
(up to a redefinition of $\anti{d}^\be(\ve{k})$). 
We have \koment{str.MPW97}
\eqs{\label{Eq:imQ}
b_\al(\ve{k})^\dagger |{\rm 0}\rangle
&=&Q_{BRST}\,\anti{b}_\al(\ve{k})^\dagger|{\rm 0}\rangle \,, 
\nn\\[5pt]
\anti{d}^\ga(\ve{k})^\dagger |{\rm 0}\rangle 
&=&
i\,[\Om(0)^{-1}]^{\ga}_{\ \al} \, 
   Q_{BRST}\,{d}^\al(\ve{k})^\dagger|{\rm 0}\rangle \,,
}
and  \koment{str.MPW98}
\eqs{\nn
&Q_{BRST}\,b_\al(\ve{k})^\dagger |{\rm 0}\rangle =0 \,,
\nn\\[5pt]
&Q_{BRST}\,\anti{d}^\ga(\ve{k})^\dagger |{\rm 0}\rangle =0\,, 
}
which shows that the unphysical states form quartet representations 
of $Q_{BRST}$. 
One has, therefore, the following decomposition 
\cite{KugoOjima} \koment{MPW101}
\eq{\label{Eq:ker}
\ker Q_{BRST}=\sF_{\rm ph} \oplus {\rm im}\, Q_{BRST}\,,
}
in which the subspace $\sF_{\rm ph}$ is obtained by the action on the 
vacuum state $|0\rangle$ of (products of) creation operators 
$a^{\ell_r}(\ve{k})^\dagger$   and $a^{\la_r}_\ch(\ve{k})^\dagger$ 
appearing in Eqs. \refer{Eq:Phi} and \refer{Eq:bA}, as well as their spin 1/2 
counterparts (``physical particles").\footnote{
In particular, the states  
$a^{{\bf 0}_r}(\ve{k})^\dagger|{\rm 0}\rangle $
corresponding to massless eigenvectors $\ze_{S[{\bf 0}_r]}$ satisfying 
the condition \refer{Eq:scalars-phys} belong to $\sF_{\rm ph}$. 
In contrast, states created/annihilated  by the 
$\vphi_{\rm unph}^j$ part (explicitly 
given by \refer{Eq:AsymField-NOWE-unph})  of 
\refer{Eq:AsymField-NOWE} are unphysical 
would-be Goldstone modes; they are ``confined" in the sense of 
Kugo-Ojima quartet mechanism \cite{KugoOjima}.  
} 
The decomposition \refer{Eq:ker} is obvious  
in  the subspace of one-particle states; 
by constructing an 
appropriate family of projection operators \cite{KugoOjima}, one can prove 
its validity in 
the entire pseudo-Fock space $\sF$. 
In particular,  
the scalar product restricted to $\ker Q_{BRST}$ is positive 
semidefinite (elements of ${\rm im}\, Q_{BRST}$ have a vanishing norm). \kTTT

Finally, 
$Q_{BRST}$ commutes \cite{Bec} with the pseudounitary $S$-operator 
\eq{\nn 
S=\,\,:\!
\exp\left\{
-\int\,\volel{x}\,\Psi^{J}(x)
\int\,\volel{y}\,\Gamma_{JK}(x,y)\,
\derf{}{J_K(y)}
\right\}
\!:\, \exp(i\, W[J])\bigg|_{J=0}\,,
}
in which 
$(\Psi^{J})
=(\vphi^j,\ \vA^{\al}_{\mu},\ \vh_{\be},\ \vpsi^a,\ \vom^\al,\ \anti{\vom}_\al)$ 
now runs over all asymptotic fields (including ghosts).\footnote{The 
commutativity with $S$ is a consequence of the Zinn-Justin identity 
\refer{eqn:ZJidBasic}; this is why one has to include the 
$B(0)$ and $\Om(0)$ 
factors in the definition \refer{Eq:BRS-asym} of $Q_{BRST}$ \cite{Bec}. }
Hence, $\ker Q_{BRST}$ is an invariant subspace for $S$ and the amplitudes 
between 
the states belonging to $\sF_{\rm ph}$ are consistent with unitarity 
\cite{Bec,KugoOjima}.  \kYYYY

\section{Conclusions}\label{Sec:Concl}
We have shown how the asymptotic approach of Lehmann, Symanzik and Zimmermann 
to calculating $S$-matrix elements extends to general gauge theories,
treated in the Landau gauge, in the presence of arbitrary mixing of
vector (and scalar) fields. The developed formalism covers both exact and
spontaneouly
broken gauge symmetries and takes into account complication arising
if there are Goldstone bosons associated with spontaneously broken
global symetries. The pseudo-Fock space of asymptotic states
following from the structure of the poles at real values of the momentum
variable $p^2$ (corresponding to stable particles) of the matrix
propagators
of vector and scalar fields has been explicitly constructed.
Its BRST-cohomological structure ensures unitarity of the
$S$-operator restricted to the subspace of physical asymptotic states 
in the presence of a generic mixing. 

On the practical side, a simple prescription, formulated entirely in terms
of eigenvectors of certain matrices, for computing ``square-rooted
residues"
$\zeta$ of poles of the matrix propagators has been given. It can be
viewed as a straightforward generalization of the procedure used to
identify fields which are ``mass eigenstates'' in tree
level calculations and can be efficiently used also in numerical or
automatized analytical calculations.

These general results, obtained by analysing the relevant set of
Slavnov-Taylor identities, have been supplemented by the ready-to use
one-loop formulae for self-energies of vector and scalar fields valid in
any renormalizable gauge theory, and the formulated practical prescriptions
have been illustrated on two interesting examples of field mixing.

While the prescription for the $\zeta$ factors of the vector fields given
in this paper is valid only in the Landau gauge, it can be generalized to
other $R_\xi$ gauges, as will be shown in a separate publication.

Finally,
although in some reasonings restrictions were made to the perturbative
approach (mainly to guarantee
the existence of inverses of certain matrices), 
most of the results should remain valid 
outside the perturbative expansion as well.

\vspace{0.2cm}
\noindent{\bf Acknowledgments:} 
I am grateful to the Anonymous Referee for 
suggested improvements of the text.   
I also thank to \mbox{K. Meissner} for 
careful reading of the first version of the paper, and 
suggestions which allowed me to clarify the presentation of results.  
 \kJJJJ

\appendix
\section{Results in the Singlet Majoron Model}\label{Sec-App:ResMajoron}

In this appendix we list some one-loop results pertaining to
the singlet Majoron extension of the SM. 
Our conventions are described in Sec. \ref{Sec:SubSec:Examples:Maj}.   
\kSSSS  

The one-loop corrections to the vacuum expectation values 
($w_{H}\equiv v_{H(0)}$ and $w_{\cphi}\equiv v_{\cphi(0)}$ denote the tree-level 
VEVs) can be obtained using the general formula \refer{Eq:TadCond} and 
have the form: 
\eqs{\label{Eq:App:vH_1}
 v_{H(1)}
&=&
\frac{\la_2}{(4 \pi )^2 w_H^3 \left(\la_1 \la_2-\la_3^2\right)  } 
\Bigg\{
2 \sum_{i=1}^3 f_1\!\left(m_{N_i},m_{\nu _i}\right)
+6\sum_{quarks} m_q^2\, a^R({m_q})+
\nn\\[4pt]
&{}&\quad 
+
2\sum_{\ell=e\mu\tau}
m_\ell^2\, a^R(m_\ell)
-3 m_W^2 \left[a^R(m_W)+\frac{2}{3}m_W^2\right]
+\nn\\[4pt]
&{}&\quad 
-\frac{3}{2} m_Z^2 \left[a^R(m_Z)+\frac{2}{3}m_Z^2\right]
+\nn\\[4pt]
&{}&\quad 
+\bigg[
\frac{m_{\text{I}}^2\, a^R(m_{\text{I}})}
  {4 \left(m_{\text{I}}^2-m_{\text{II}}^2\right)
     \left(m_{\text{I}}^2+m_{\text{II}}^2-2 \lambda _1 w_H^2\right)}
\times\nn\\[4pt]
&{}&\qquad 
\qquad \times
\Big(3
   m_{\text{II}}^4-10 \lambda _1 m_{\text{II}}^2 w_H^2+8 \lambda _1 \left(\lambda _1-\lambda _3\right) w_H^4
+\nn\\[4pt]
&{}&\qquad
\qquad \ \ \ \ \ 
+2 m_{\text{I}}^2 (m_{\text{II}}^2-2 (\lambda _1-\lambda _3) w_H^2)
\Big)
+\nn\\[4pt]
&{}&\quad
\ \ \ \ \ +(m_{\rm I}\leftrightarrow m_{\rm II})
\bigg]
\Bigg\}
\,,
}
\eqs{\label{Eq:App:vcphi_1}
 v_{\cphi(1)}
&=&
\frac{\lambda _3}{(4 \pi )^2 \left(\lambda _1 \lambda _2-\lambda _3^2\right) 
w_H^2 w_{\varphi }} 
\bigg\{
-2 \sum_{i=1}^3 f_2\!\left(m_{N_i},m_{\nu_i}\right)
-6\sum_{quarks} m_q^2\, a^R({m_q})
+\nn\\[4pt]
&{}&
\qquad\qquad\qquad\quad
-2 \sum_{\ell=e\mu\tau} m_\ell^2\, a^R({m_\ell})
+3 m_W^2 \left[a^R\left(m_W\right)+\frac{2 m_W^2}{3}\right]
+\nn\\[4pt]
&{}&
\qquad\qquad\qquad\quad
+\frac{3}{2} m_Z^2 \left[a^R\left(m_Z\right)+\frac{2 m_Z^2}{3}\right]
\bigg\}
+\nn\\[4pt]
&{}&
-
\bigg\{
\frac{a^R(m_{\text{I}})}{2 (4 \pi )^2
   \lambda _2 m_{\text{II}}^2 \left(m_{\text{I}}^2-m_{\text{II}}^2\right) w_{\varphi }}
\Big[
\lambda _2 \left(m_{\text{I}}^2-2 \lambda _1 w_H^2\right) \left(m_{\text{II}}^2+4 \lambda _1 w_H^2\right)
+\nn\\[4pt]
&{}&
\qquad\qquad\qquad\qquad
+2 \lambda _3 \left(m_{\text{II}}^2-2 \lambda _1 w_H^2\right) \left(m_{\text{I}}^2+m_{\text{II}}^2-2 \lambda _1 w_H^2\right)
\Big] 
+\nn\\[4pt]
&{}&
\qquad\!\! +
(m_{\rm I}\leftrightarrow m_{\rm II})
\bigg\}\,.
}
The symbol $(m_{\rm I}\leftrightarrow m_{\rm II})$ used in the 
above formulae denotes 
a term 
obtained by interchanging the two masses,
$m_I$ and $m_{II}$, in the preceding one. 
The functions $f_1$ and $f_2$ expressing contributions of the neutrinos read 
\eqs{\nn
f_1(m_N,m_\nu)
&=&
\frac{1}{2 \left(m_N+m_{\nu }\right) \lambda _2 w_{\varphi }^2} \times
\nn\\[4pt]
&{}&
\times 
\big\{
\left(\lambda _3 \left(m_{\nu }-m_N\right) w_H^2+2 \lambda _2 m_{\nu } w_{\varphi }^2\right) m_N^2\, a^R\left(m_N\right)
+
\nn\\[4pt]
&{}&
\ \ \ \ +\left(\lambda _3 \left(m_N-m_{\nu }\right) w_H^2+2 \lambda _2 m_N w_{\varphi }^2\right) m_{\nu }^2\, a^R\left(m_{\nu }\right)
\big\}\,,
\nn}
\eqs{\nn
f_2(m_N,m_\nu)
&=&
\frac{1}{2 \left(m_N+m_{\nu }\right) \lambda _3 w_{\varphi }^2}
\times 
\nn\\[4pt]
&{}&
\times 
\big\{
\left(\lambda _1 \left(m_{\nu }-m_N\right) w_H^2+2 \lambda _3 m_{\nu } w_{\varphi }^2\right) m_N^2\, a^R\left(m_N\right)
+
\nn\\[4pt]
&{}&
\ \ \ \ +
\left(\lambda _1 \left(m_N-m_{\nu }\right) w_H^2+2 \lambda _3 m_N w_{\varphi }^2\right) m_{\nu }^2\, a^R\left(m_{\nu }\right)
\big\}\,.
\nn}
The factors $\mathbf{a}$, $\mathbf{b}$ and $\mathbf{c}$ parametrizing
the matrix \refer{Eq:ScProd}, obtained 
from the general expressions \refer{Eq:MV-and-MS} and \refer{Eq:De_S}, 
read 
\eqs{\label{Eq:App:ScProd-a}
\va&=&
\sum_{i=1}^3 
f_{11}(m_{N_i},m_{\nu_i})
-\frac{1}{4 \pi ^2 w_H^2} 
\left\{3 \sum_{quarks}{m_q^2} \ln\frac{m_q}{\bar{\mu}}
+
\sum_{\ell=e\mu\tau}{m_\ell^2} \ln\frac{m_\ell}{\bar{\mu}}
\right\}
+\frac{\lambda _1}{16 \pi ^2}+
\nn\\[4mm]
&{}&
+
\frac{3}{8 \pi ^2 w_H^2}
\left\{
2 m_W^2\! \left[\ln\!\left( \frac{m_W}{\bar{\mu}}\right)-\frac{5}{12}\right]
+m_Z^2 \! \left[\ln\!\left( \frac{m_Z}{\bar{\mu}}\right)-\frac{5}{12}\right]
\right\}
+\nn\\[4mm]
&{}&
-\frac{3\,m_{\text{I}}^2\, m_Z^2
   \left(m_{\text{II}}^2-2 \lambda _1 w_H^2\right) }
     {8\pi ^2 w_H^2 \left(m_{\text{I}}^2-m_{\text{II}}^2\right)   
       \left(m_{\text{I}}^2-m_Z^2\right) }
\ln\!\left(\frac{m_{\text{I}}}{m_Z}\right)+
\nn\\[4mm]
&{}&
+\frac{3\, m_{\text{II}}^2\, m_Z^2 \left(m_{\text{I}}^2-2 \lambda _1 w_H^2\right) }{8 \pi ^2 w_H^2 \left(m_{\text{I}}^2-m_{\text{II}}^2\right) \left(m_{\text{II}}^2-m_Z^2\right) }
\ln\!\left(\frac{m_{\text{II}}}{m_Z}\right)\,,
}
\eqs{\label{Eq:App:ScProd-c}
\vc&=&
\sum_{i=1}^3 
f_{22}(m_{N_i},m_{\nu_i})
+\frac{\lambda _2}{16 \pi ^2}\,,
}
\eqs{\label{Eq:App:ScProd-b}
\vb&=&
\sum_{i=1}^3 
f_{12}(m_{N_i},m_{\nu_i})
\,.
}
The functions $f_{11}$, $f_{22}$ and $f_{12}$ which represent
 contributions of the neutrino loops are given by
\eqs{\nn
f_{11}(m_{N},m_{\nu})
&=&
\frac{1}{(4 \pi )^2 w_H^2} 
\bigg\{
\frac{m_N m_{\nu } \left(m_N^2-4 m_N m_{\nu }+m_{\nu }^2\right)}{\left(m_N+m_{\nu }\right)^2}+
\nn\\[4pt]
&{}&\hspace*{-30 pt}
-4 m_N m_{\nu } \ln\!\left(\frac{m_N}{\bar\mu}\right)+
\frac{4 m_N m_{\nu }^2 \left(m_{\nu }^3-2 m_N^3\right) 
}{\left(m_N-m_{\nu }\right) \left(m_N+m_{\nu }\right)^3}
\log\! \left(\frac{m_{\nu }}{m_N}\right)
\bigg\},
\ \ \ \ \ 
\nn}
\eqs{\nn
f_{22}(m_{N},m_{\nu})
&=&
\frac{1}{(4\pi)^2 w_{\varphi }^2}
\bigg\{
\frac{m_{\nu } m_N \left(m_{\nu }^2-4 m_{\nu } m_N+m_N^2\right)}
     {\left(m_{\nu}+m_N\right){}^2}
+\nn\\[4pt]
&{}&
+
\frac{2 m_N^2 \left(m_{\nu }^4+4 m_{\nu }^2 m_N^2-2 m_{\nu }^3 m_N-m_N^4\right)}
     {\left(m_N-m_{\nu }\right) \left(m_{\nu }+m_N\right)^3}
\ln\! \left(\frac{m_N}{\bar\mu}\right)
+\nn\\[4pt]
&{}&
+\frac{2 m_{\nu }^2 \left(m_{\nu }^4+2 m_{\nu } m_N^3-4 m_{\nu }^2  m_N^2-m_N^4
      \right)}
      {\left(m_N-m_{\nu }\right)\left(m_{\nu }+m_N\right)^3}
\ln\! \left(\frac{m_{\nu }}{{\bar\mu}}\right)
\bigg\},
\nn}
\eqs{\nn
f_{12}(m_{N},m_{\nu})
&=&
\frac{1}{(4\pi)^2 w_H\, w_{\varphi }}
\bigg\{
\frac{4 m_{\nu }^2 m_N^2 \left(m_{\nu }^2-m_{\nu } m_N+m_N^2\right)}
     {\left(m_N-m_{\nu }\right) \left(m_{\nu }+m_N\right)^3}
 \ln\!\left(\frac{m_{\nu }}{m_N}\right)
+\nn\\[4pt]
&{}&\qquad\qquad\qquad
-\frac{m_{\nu } m_N \left(m_{\nu }^2-4 m_{\nu } m_N+m_N^2\right)}
      {\left(m_{\nu }+m_N\right)^2}
\bigg\}.
\nn}

\end{document}